\documentclass[prb,twocolumn,amsmath,amssymb,floatfix,superscriptaddress,longbibliography,nofootinbib,reprint]{revtex4-2}
\usepackage{graphicx}
\usepackage{bm}

\usepackage{hyperref}

\usepackage{braket,bbold,color,enumerate}
\usepackage{float}
\usepackage{amsmath,amssymb,xcolor}
\usepackage{amsfonts}
\usepackage{amsmath}
\usepackage{mathtools}
\usepackage{dsfont}
\usepackage[capitalize]{cleveref}
\usepackage{enumitem}

\newcommand{\cdag}{c^{\dag}}

\newcommand{\im}{\mathrm{i}}
\newcommand{\id}{\mathbb{1}}

\newcommand{\Chat}{\hat{\mathcal{C}}}
\newcommand{\Shat}{\hat{\mathcal{S}}}

\newcommand{\Ical}{\mathcal{I}}
\newcommand{\ei}[1]{e^{\mathrm{i}#1}}
\newcommand{\emi}[1]{e^{-\mathrm{i}#1}}
\newcommand{\sgn}{\operatorname{sgn}}

\begin{document}

\title{Noise-Induced Topology in Chiral Quantum Walks}

\author{Alexandre Chaduteau}
\affiliation{Blackett Laboratory, Imperial College London, London SW7 2AZ, United Kingdom}
\author{Derek K. K. Lee}
\affiliation{Blackett Laboratory, Imperial College London, London SW7 2AZ, United Kingdom}
\author{Frank Schindler}
\affiliation{Blackett Laboratory, Imperial College London, London SW7 2AZ, United Kingdom}
\author{Abhinav Prem}
\affiliation{Physics Program, Bard College, 30 Campus Road, Annandale-on-Hudson, New York 12504, USA}
\date{\today}

\begin{abstract}
Noise typically destroys quantum coherence and washes out transport phenomena that rely on interference. Chiral quantum walks are a case in point: a particle hopping on a lattice threaded by a magnetic flux can spread asymmetrically due to interference between paths, an asymmetry that noise is expected to erase. Remarkably, we find that the most ubiquitous noise---random fluctuations of on-site energies (dephasing)---instead renders this asymmetry topologically robust. The density profile develops a skewness that grows in time while its center stays fixed, and in a finite chain the late-time relaxation modes pile up at one edge: a skin effect lacking the drift of conventional non-Hermitian skin effects. With multiple sites per unit cell, the dephasing strength alone can cause a topological phase transition, reversing the asymmetry. On a flux ladder with noisy rung hopping, the noise causes a drift of rung coherences, a quantum ratchet with no classical counterpart. Underlying these effects, the noise-averaged dynamics is an interacting (generically non-integrable) Lindbladian that remains analytically tractable. The modes governing late-time relaxation are bound states of the noise-induced interaction and form a band with a point gap, absent both without noise and under postselection, whose winding number fixes the direction of the asymmetry. These effects are within reach of several experimental platforms, including synthetic lattices, superconducting qubits, and optical lattices.
\end{abstract}

\maketitle

\section{Introduction}
\label{sec:intro}
Environmental noise is largely detrimental to quantum coherence and entanglement. That noise need not be purely antagonistic is nonetheless an idea that has been rapidly gaining purchase, particularly in many-body quantum systems. Indeed, decoherence can even give rise to phases with no pure-state counterpart, including intrinsically mixed-state topological orders~\cite{sohal2025imto,ellison2025imto,KPY26} and intrinsically average symmetry-protected topological phases~\cite{ma2023aspt,ma2025avg}. These developments highlight how noise can work in concert with coherent quantum processes to generate novel patterns of entanglement in decohered ground states or steady states of open quantum dynamics. It is natural to ask whether noise can play a similar role in \textit{dynamics}, endowing transport and relaxation in a many-body open quantum system with topologically protected features that are impossible under coherent dynamics alone. The destructive effects of noise on transport are well documented. For instance, it generically turns ballistic spreading into diffusion, as in the Haken--Strobl model of exciton transport~\cite{haken1973exactly, PhysRevE.47.67, KenkreReineker1982, Marais2013, Plenio2008, Znidaric_2010_1}. However, there are well-known circumstances where noise helps, such as in the environment-assisted transport of excitons, demonstrated with engineered dephasing in photonic, trapped-ion, and superconducting networks~\cite{Biggerstaff2016, Caruso2016, PhysRevLett.122.050501, potocnik2018studying}. In such circumstances, noise disrupts the interference that localizes excitations in disordered networks~\cite{Rebentrost2009}; it can also broaden energy levels so that detuned sites come into resonance~\cite{Plenio2008}. In either case, noise assists by removing a coherent obstacle to transport.
\begin{figure*}[t!]
\includegraphics[width=2\columnwidth]{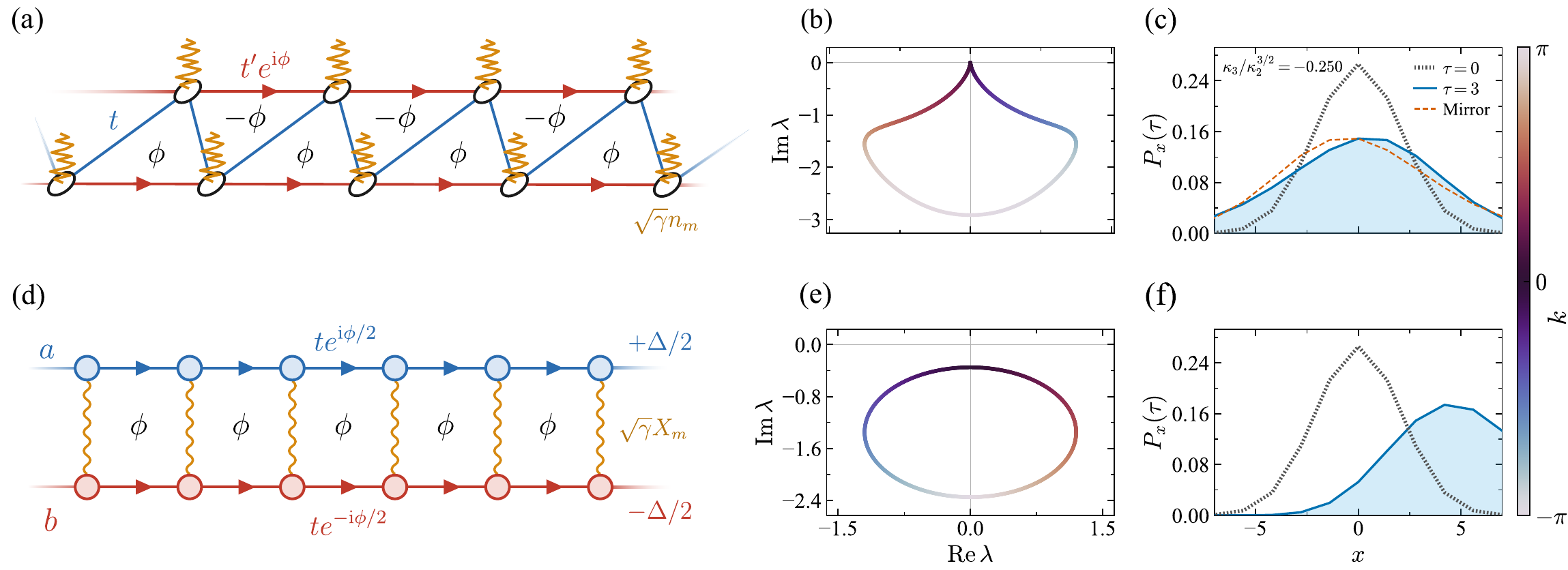}
\caption{Engineering many-body topology from noise in lattice setups with a flux. (a) The dephased flux chain, realized in a zigzag geometry with a flux through each triangular plaquette [Eq.~\eqref{eq:dephased_flux_chain}]. All next-nearest-neighbor (n.n.n.) bonds carry the same phase $\phi$, so every three-site loop encloses the same flux. Drawn in the plane, consecutive triangles have opposite orientation, so the flux alternates as $\pm\phi$ with zero net flux~\cite{Jotzu2014}. A uniform planar flux~\cite{PhysRevX.8.031045} would preserve inversion along the chain and give no asymmetry [Sec.~\ref{sec:bound_states}] for equal row energies. On-site dephasing noise is realized e.g. as white-noise stochastic fluctuations of the on-site energy [see Sec.~\ref{sec:experiment}]. (b) In periodic boundary conditions (PBC), this setup gives point-gap topology in the generator $\Chat$ of correlation-matrix dynamics. The corresponding eigenmodes are bound states induced by the noise; they are also hydrodynamic modes with low decay rate near the origin. Taylor expanding $\lambda(k)$ near the origin determines the type of late-time asymmetric behavior in the density profile. The point gap has a cusp, so the asymmetry manifests only in higher-order cumulants, such as the skewness $\kappa_3$. (c) Time evolution of an initial Gaussian density profile. The late-time asymmetry is seen here as a skewness. The parameters used here are as in Fig.~\ref{fig:dephased_flux_chain}. (d) The dephased flux ladder, with two legs $a$ and $b$ with nearest-neighbor (n.n.) hoppings and a flux [Eq.~\eqref{eq:dephased_flux_ladder}]. The two legs are also detuned by $\Delta$. The only coupling between the legs is a noisy (incoherent) hopping. (e) The corresponding point gap has nonzero linear coefficient $c_1$ in the expansion of $\lambda(k)$. (f) The corresponding profile of rung coherences generically drifts; with particle-number conservation, this can only happen for coherences, not densities.}
\label{fig:concept}
\end{figure*}

Chiral quantum walks offer a testing ground for transport amidst both noise and quantum interference. These walks are generated by hoppings whose complex amplitudes break time-reversal symmetry~\cite{Zimboras2013, PhysRevA.93.042302}, which naturally occurs for a charged particle hopping on a lattice in a magnetic field: it picks up Aharonov--Bohm phases that enter as complex (Peierls) hopping amplitudes~\cite{Hofstadter1976}. Synthetic versions of these fluxes have been engineered for neutral atoms~\cite{Jaksch_2003, Dalibard2011, PhysRevLett.111.185301, PhysRevLett.111.185302, Atala2014, Mancini2015, An2017} and for photons in superconducting circuits~\cite{Roushan2017, rosen2024synthetic}. Interference between paths enclosing a flux makes transport direction-dependent. One therefore expects noise to destroy such interference effects. Noisy quantum walks can be modeled via a stochastic Schr\"odinger equation~\cite{quantum_brownian_circuits, Jacobs01092006}, whose noise-averaged dynamics is described by a Lindbladian~\cite{lindblad1976generators, gorini1976completely}. For random fluctuations of on-site energies, the jump operators are simply the local densities.

In this work, we show that unbiased dephasing noise can render the chirality of this walk topologically robust. Moreover, with multiple sites per unit cell, the dephasing strength alone can cause topological phase transitions. Since the jump operators are quadratic and Hermitian, the Lindbladian is interacting and generates non-Gaussian states~\cite{Barthel_2022, PhysRevLett.129.120401, chaduteau2026}; nonetheless, the equations of motion for correlations obey a closed hierarchy~\cite{Zunkovic_2014, Barthel_2022}, allowing an analytical treatment. We show that the generator of the two-point correlation dynamics can host a point gap that is absent both without noise and when postselecting on no quantum jumps. That generator's eigenmodes are bound states of the interaction induced by the noise. The point-gap topology here is therefore emergent: it arises from the noise-induced interaction, rather than being built into the Hamiltonian or the jump operators. Such a point gap requires breaking both inversion and time-reversal symmetry, which a suitable flux pattern provides. Particle-number conservation makes these modes hydrodynamic, with a decay rate that vanishes at long wavelengths, and forbids any drift of the density. Instead, we show that a nonzero winding number about the steady state is equivalent to an odd moment of the density profile growing linearly in time, with a sign fixed by the winding. In a finite-sized system, the hydrodynamic modes accumulate at one edge, representing a skin effect without drift.

We illustrate these general results in three simple lattice models: (i) In a dephased flux chain, dephasing makes the chirality topologically robust; the density profile develops a skewness $\kappa_3$ that grows linearly in time while its center stays fixed. (ii) With two sites per unit cell, the noise strength alone reverses the winding number and the nonreciprocity. This transition is missed by semiclassical kinetic theory~\cite{Cercignani1988} and requires a quantum treatment. (iii) In a dephased flux ladder with noisy rung hoppings, the noise creates a drift of rung coherences, whereas the legs decouple without noise: this represents a quantum ratchet~\cite{RevModPhys.81.387, REIMANN200257} without any classical counterpart. These effects manifest for both bosons and fermions, and we discuss feasible experimental realizations across multiple platforms: synthetic lattices~\cite{An2017, An2017NC, PhysRevLett.120.040407, PhysRevLett.129.070401}, arrays of superconducting qubits~\cite{Roushan2017, rosen2024synthetic}, and optical lattices~\cite{Pichler2010, bouganne2020anomalous, PhysRevLett.115.140402, PhysRevX.7.011034}.
 
\subsection*{Relation to Prior Work}
Unlike effective non-Hermitian Hamiltonians, which describe dynamics conditioned on the absence of quantum jumps, Lindbladians describe the unconditioned dynamics of physical open many-body systems. Lindbladians can nonetheless display point-gap topology, whose hallmark is the non-Hermitian skin effect~\cite{ashida2020review, Bergholtz2021, PhysRevX.8.031079, kawabata_38_fold_way, lieu2020tenfold, Chaduteau_2025, PhysRevLett.123.170401, PhysRevLett.127.070402, original_hatano_nelson, PhysRevLett.121.086803, PhysRevLett.124.086801, PhysRevLett.125.126402}. Such skin effects have been observed in photonic quantum walks, ultracold atoms, and on a digital quantum computer~\cite{Xiao2020, Weidemann2020, PhysRevLett.129.070401, Shen2025}. Most analytical treatments use Lindbladians with linear jumps~\cite{lieu2020tenfold,PhysRevLett.123.170401, PhysRevLett.127.070402, mcdonald_big_thirdqtsn_paper, yang2022lse, Chaduteau_2025}, which are non-interacting and for which the topology resides in the single-particle damping matrix, with nonreciprocity built either into the effective Hamiltonian or in the jumps. 
In the many-body setting, symmetry classes have been catalogued~\cite{PhysRevX.13.031019, Kawabata2023}, but topological phenomena remain largely limited to specific examples~\cite{PhysRevB.108.155114, chaduteau2026}. In these examples, the skin effect also arises from a nonreciprocity built into the generator, such as asymmetric hopping, gain, or loss. It typically manifests as a drift~\cite{PhysRevResearch.1.023013, wsmq-kmq9}. Dephasing has no such nonreciprocity built in: its jump operators are Hermitian and conserve particle number, and the evolution conditioned on no quantum jumps merely damps all modes at the same rate.

We also contrast our transport study with topological steady states engineered by dissipation~\cite{diehl2015top, Diehl2011, bardyn2013top}, mixed-state topological phases~\cite{sohal2025imto, ellison2025imto, KPY26, ma2023aspt, ma2025avg}, and  dephasing-driven transitions from topological to trivial quantum states~\cite{8vtp-khnq}. Incoherent hopping of particles has been studied analytically~\cite{PhysRevB.71.214302, PhysRevLett.123.080601, haken1972coupled, PhysRevLett.117.137202, Znidaric_2010_1, PhysRevLett.131.053401}, as have models with coherent hopping that breaks time reversal~\cite{Zimboras2013, PhysRevA.93.042302}, and environment-assisted transport~\cite{haken1973exactly, Rebentrost2009, Plenio2008, Caruso2009, PhysRevLett.108.020602}. However, these works do not identify any point-gap topology. In stochastic classical processes, non-Hermitian topology has been identified in the single-particle generators~\cite{li2026classificationtopologicalphasesmatter, PhysRevLett.132.046602, Murugan2017, Dasbiswas2018}. Nonreciprocal transport has also been engineered by combining an Aharonov--Bohm flux with particle loss~\cite{PhysRevLett.124.070402}, and the persistence of the skin effect under dephasing has been studied in walks that are already non-Hermitian due to loss~\cite{Longhi2024, 2f5t-gc98}. Prior to this work, interaction-induced skin effects had been studied numerically for two-body loss~\cite{PhysRevB.108.155114} and in interacting postselected Hamiltonians~\cite{qrh6-vx64}. In Ref.~\onlinecite{chaduteau2026} the present authors found topology in the correlation dynamics for noise coupling to both local densities and currents, where the single-particle generator of the no-jump dynamics already has a winding number. In the present work, the single-particle generator is trivial, so the topology is purely noise-induced, and it arises for simple dephasing (the most common form of noise). We also derive analytical transport coefficients and link the spectrum to late-time hydrodynamics.

The remainder of the text is organized as follows: in Sec.~\ref{sec:general_setting}, we review the stochastic Schrödinger equation and its continuous-time Lindblad version for quadratic Hermitian jumps; we then outline the general recipe for noise-induced topology, and link it to late-time transport. In Sec.~\ref{sec:models}, we highlight these general lessons through explicit models.
We discuss experimental realizations in Sec.~\ref{sec:experiment} and conclude in Sec.~\ref{sec:discussion}.

\section{Setting and general results}
\label{sec:general_setting} 

In this Section, we introduce stochastic quantum walks whose noise-averaged dynamics is a Lindbladian with a quadratic Hermitian Hamiltonian and quadratic Hermitian jumps. We derive the correlation-matrix dynamics as a closed equation of motion. This Section contains our central results, which do not depend on the specific models of Sec.~\ref{sec:models}. First, the quantum jumps act as an attractive interaction between the ket and bra indices of the correlation matrix, and the resulting bound states can carry point-gap topology that is absent from the no-jump dynamics [Secs.~\ref{subsec:setup} and~\ref{sec:bound_states}]. Second, we identify that such a noise-induced point gap requires breaking both inversion and time reversal, which for on-site dephasing and a single band requires hoppings of at least two different ranges [Sec.~\ref{sec:bound_states}]. Third, with several sites per unit cell, interband coherences can allow the noise strength alone to change the winding number, a transition beyond semiclassical kinetic theory [Sec.~\ref{sec:bound_states}]. Finally, when particle number is conserved, the density cannot drift, and a nonzero winding number about the steady state is equivalent to an odd cumulant of the density profile growing linearly in time, with a sign fixed by the winding number [Sec.~\ref{subsec:latetime}].

\subsection{Lindbladian dynamics with Hermitian quadratic jumps}
\label{subsec:setup}

We consider bosons/fermions with annihilation/creation operators $c_i/\cdag_i,\; i=1\ldots N$ for an integer number $N$ of modes. These obey canonical commutation (for bosons) or anti-commutation (for fermions) relations. We consider quadratic, particle-number conserving (Hermitian) Hamiltonians $H = \sum_{i,j=1}^N\mathcal{H}_{ij}\cdag_ic_j$, where $\mathcal{H}$ is a Hermitian $N\times N$ matrix. Similarly, we define a family of quadratic Hermitian operators $K_m$, indexed by $m$, of the form $K_m = \sum_{ij}(\mathcal{K}_m)_{ij}\cdag_ic_j$. We consider time ($\tau$) evolution of a pure state $\ket{\psi}$ under the stochastic Schrödinger equation~\cite{Jacobs01092006}
\begin{equation}\label{eq:SSE}
    \mathrm{d}\ket{\psi}=\bigg[-\im H-\frac{1}{2}\sum_m K_m^2\bigg]\ket{\psi}\mathrm{d}\tau  - \im\sum_m K_m\mathrm{d}W_m\ket{\psi}.
\end{equation}
Here, $\mathrm{d}W_m$ are independent Wiener increments. They implement noise: every trajectory $\ket{\psi_\tau}$ is different. Denoting the expectation value with $\mathbb{E}$, the Brownian (white noise) assumption means that (1) $\mathbb{E}[W_m(\tau)] = 0$, and (2) $\mathbb{E}[\mathrm{d}W_m\mathrm{d}W_n] = \delta_{mn} \mathrm{d}\tau$, which is the It\^o rule~\cite{breuer2002theory}. Averaging the above stochastic equation yields a Lindblad equation, written here as
\begin{equation}\label{eq:lindblad_quadratic_jumps}
\begin{aligned}
   \im\partial_\tau\rho &= \mathcal{L} \rho = [H,\rho]-\frac{\im}{2}\sum_m[K_m,[K_m,\rho]],
\end{aligned}
\end{equation}
where $\rho = \mathbb{E}\left[\ket{\psi}\bra{\psi}\right]$ denotes the density matrix of the system. Such equations arise when describing simple models of decoherence, such as quantum Brownian motion~\cite{PhysRevA.94.042123, CALDEIRA1983587, PhysRevLett.46.211}. Typical jumps $K_m$ include dephasing~\cite{Znidaric_2010_1, Znidaric_2010_2, 8vtp-khnq} via densities $n_m=\cdag_m c_m$, as well as currents $j_{m}=\im (\cdag_mc_{m+1}-\cdag_{m+1}c_m)$, in which case $m=1 \dots N$ denotes a spatial index.

Quadratic jumps render the Lindbladian quartic in the bosons/fermions, so the model is interacting and its dynamics do not preserve Gaussianity~\cite{Barthel_2022}. Nevertheless, the Heisenberg equation of motion for the correlation matrix $C_{ij}=\mathrm{Tr}(\cdag_ic_j\rho)$ remains closed and does not include higher order correlators~\cite{chaduteau2026,Barthel_2022} [see also Appendix~\ref{appendix:CM_derivation}],
\begin{equation}\label{eq:CM_EOM}
    \im\frac{\mathrm{d}C}{\mathrm{d}\tau} = C\cdot \Theta^T - \Theta^* \cdot C + \im \sum_m \mathcal{K}_m^T \cdot C \cdot \mathcal{K}_m^T,
\end{equation}
where the dot indicates matrix multiplication, and we introduced the matrix $\Theta_{ij} = \mathcal{H}_{ij} - \frac{\im}{2}\sum_m(\mathcal{K}^2_m)_{ij}$. $\Theta$ is generally non-Hermitian. Remarkably, Eq.~\eqref{eq:CM_EOM} is linear in $C$ and hence remains polynomially solvable despite describing dynamics for an interacting open system~\cite{Zunkovic_2014, Barthel_2022}. We can vectorize the correlation matrix as $\ket{C} = \sum_{i,j} C_{ij} \ket{j} \otimes \ket{i}$, in which case Eq.~\eqref{eq:CM_EOM} becomes $\im \, \mathrm d|C\rangle/\mathrm d\tau=\Chat|C\rangle$, with 
\begin{equation}\label{eq:vectorised_Chat}
    \Chat=\Theta\otimes\mathbb 1-\mathbb 1\otimes\Theta^*+\im\sum_m \mathcal{K}_m \otimes \mathcal{K}_m^T
\end{equation}
following the procedure of Ref.~\onlinecite{chaduteau2026}. This is generically a non-Hermitian matrix. Eigenmodes then evolve as $C(\tau)=e^{-\im\lambda\tau}C$, with $\mathrm{Im}\,\lambda\le0$ since the dynamics are completely positive and trace preserving. The adjoint Lindbladian maps quadratic operators onto quadratic operators: eigenvalues of $\Chat$ are thus also eigenvalues of the many-body Lindbladian $\mathcal{L}$.

The generator $\Chat$ acts on two indices, the bra $i$ and the ket $j$ of $C_{ij}$, which we can picture as the two legs of a ladder. Absent the last term in Eq.~\eqref{eq:CM_EOM}, the two legs evolve independently. The spectrum of $\Chat$ is then $\{\lambda_i-\lambda_j^*\}$, with $\lambda_i$ the eigenvalues of $\Theta$. This is known as the continuum of eigenvalues~\cite{Hislop1996, chaduteau2026}. This continuum governs the postselected dynamics, conditioned on no quantum jumps~\cite{Daley04032014}, and for on-site dephasing it has no point gap. The last term in Eq.~\eqref{eq:CM_EOM}, generated by the quantum jumps, couples the two legs and plays the role of an interaction. When this interaction is attractive, it binds ket and bra into pairs [see Appendix~\ref{appendix:late_time_expansions}], whose eigenvalues form the so-called discrete spectrum~\cite{Hislop1996, chaduteau2026}. For on-site dephasing, the problem can alternatively be mapped onto a Hubbard model with an imaginary interaction~\cite{PhysRevLett.117.137202}, in which the ket and bra legs play the role of two spin species.

Let us now turn to the symmetries of this family of models. Translational invariance maps $K_m$ to $K_{m+1}$. Under periodic boundary conditions (PBC), this gives a weak symmetry for the Lindbladian~\cite{Buca_2012, PhysRevX.6.041031}, which allows us to block diagonalize $\Chat$. The jump term couples the ket/bra translations, giving a single block label which we call the ``momentum'' $k \in \frac{2\pi}{N}\mathbb{Z}$. Generally, $\Chat(k)$ is an infinite-dimensional matrix~\cite{chaduteau2026, PhysRevX.15.021016}.
Hermiticity of the density matrix imposes Hermiticity on the correlation matrix: $C^\dagger = C$. In turn, this provides a symmetry of $\Chat$ such that
\begin{equation}
\label{eq:momentum_C_ab_K_mymmetry}
U(k)\Chat(k)U^\dagger(k)=-\Chat^*(-k),
\end{equation}
where $U_{ab}(k) = \delta_{a,-b}\ei{kb}$~\cite{chaduteau2026}. This forces the spectrum of $\Chat$ to be symmetric under $\lambda(k) \to -\lambda^*(-k)$. This SWAP symmetry is of the (daggered) particle-hole type, placing $\Chat$ in class $D^\dagger$ of the non-Hermitian topological classification~\cite{kawabata_38_fold_way, PhysRevX.13.031019}. In one dimension, this class admits a $\mathbb{Z}$ point gap classification, diagnosed by a winding number\footnote{As $\Chat(k)$ is infinite-dimensional, the definition of the winding number here should formally be regularized. For our purposes however, it suffices to truncate $\Chat(k)$ to $|a| < R$ for $R$ larger than a few lattice sites.}
\begin{equation}
\label{eq:winding_number_definition}
    \nu(\lambda) = \frac{1}{2\pi \im} \int_{0}^{2\pi}\mathrm{d}k \frac{\partial}{\partial k}\log \det \left[\Chat(k) - \lambda\id\right],
\end{equation}
where $\lambda \in \mathbb{C}$.

We call a point gap \textit{noise-induced} if it exists in $\Chat$ but not in $\Theta$; such a point gap generally originates from the jump term in Eq.~\eqref{eq:CM_EOM}. Only a subset of generic non-Hermitian perturbations in class $D^\dagger$ preserves the closed hierarchy of equations of motion. Since restricting the allowed perturbations can only remove deformation paths, $\nu$ remains a topological invariant in this setting: generators with different winding numbers cannot be connected without closing the point gap. Moreover, every integer can be realized, since $\nu$ is additive under stacking decoupled chains and negative invariants are also realizable [see Sec.~\ref{sec:models}]. The winding number remains a $\mathbb{Z}$-valued invariant within our restricted setting, although we do not exclude that this setting supports additional invariants.

\subsection{Bound states and symmetry breaking}
\label{sec:bound_states}

We now ask when the jump term produces a point gap, focusing primarily on the most experimentally relevant case of on-site dephasing, with jump operators $K_m=\sqrt\gamma\,n_m$. Since $\sum_m\mathcal K_m^2=\gamma\id$, the single-particle matrix $\Theta=\mathcal H-\frac{\im\gamma}{2}\id$ is a Hermitian matrix shifted by an imaginary constant and cannot have a point gap. Any point gap of $\Chat$ must therefore be induced purely by the jump term in Eq.~\eqref{eq:CM_EOM}.

This setup admits a description as an impurity problem. Let us write the correlation matrix as $C_{i,i+a}$ in terms of the relative index $a=j-i$, and let $h_s=\mathcal H_{i,i+s}$. Fourier transforming in $i$ via $C_{k,a} := \sum_i C_{i,i+a}\ei{ki}$, Eq.~\eqref{eq:CM_EOM} becomes [see Appendix~\ref{appendix:late_time_expansions}]
\begin{equation}
\label{eq:impurity_problem}
\im\dot C_{k,a}=\sum_sT_s(k)\,C_{k,a+s}-\im\gamma\,(1-\delta_{a,0})\,C_{k,a},
\end{equation}
where $T_s(k)=-2\im\,h_s\sin\tfrac{sk}{2}$. Eq.~\eqref{eq:impurity_problem} describes a fictitious particle hopping on the relative index $a$ between ket and bra, damped uniformly at rate $\gamma$ everywhere except at $a=0$, where the two coincide. This undamped central site acts as an attractive impurity, so we generically expect bound states localized near $a=0$~\cite{chaduteau2026, PhysRevLett.117.137202}, i.e. modes in which ket and bra remain closely bound together.

Bound states by themselves, however, do not guarantee a point gap in the spectrum of $\Chat$. Consider first a unique bound-state band $\lambda(k)$. SWAP symmetry enforces the spectrum $\{\lambda(k)\}$ of $\Chat(k)$ to be symmetric under $\lambda(k) \to -\lambda^*(-k)$. The real part is odd, while the imaginary part is even in $k$. For a point gap, we additionally require $\lambda(k)$ to have nonzero real part: $\mathrm{Re}\,\lambda(k)\neq0$. Two kinds of symmetry forbid this, as they map $\Chat(k)$ to $\Chat(-k)$: 
\begin{enumerate}[label=(\arabic*)]
    \item Inversion symmetry, imposing eigenvalues to be symmetric under $\lambda(k) \to \lambda(-k)$
    \item Time-reversal symmetry, i.e. real $\mathcal{H}$ and $\mathcal{K}_m$, which gives a transposition symmetry $\Chat^T(k)=\Chat(-k)$, and hence the same spectral constraint.
\end{enumerate}
Combined with SWAP, either one forces $\mathrm{Re}\,\lambda(k) = 0$.

A noise-induced point gap therefore requires breaking both inversion and time reversal, although their product may remain a symmetry. More generally, either symmetry gives $\det[\Chat(k)-\lambda\id]=\det[\Chat(-k)-\lambda\id]$, so that $\nu(\lambda)=0$ for every reference point, regardless of the number of bands. These symmetries can be broken either by the jumps or by the Hamiltonian. In Ref.~\onlinecite{chaduteau2026}, noise-induced topology was implemented by breaking inversion via the jumps. Here, motivated by simple experimental realizations, we use on-site dephasing, so that $\Theta$ stays trivial, and break both symmetries via the Hamiltonian instead.

For a single-band $\mathcal{H}$ with a dispersion $E(k)$, inversion and time reversal each impose $E(k) = E(-k)$ up to a shift $k \to k + k_0$. Point gaps require nonzero odd moments of the group velocity $v(k) = \partial_k E(k)$, which is only possible for a dispersion that breaks inversion and time reversal. Writing $h_s = |h_s|\ei{\theta_s}$, a shift $k \to k + k_0$ shifts $\theta_s \to \theta_s + sk_0$, so the phase of a single hopping range can always be removed and cannot give rise to chirality. With at least two ranges, the phases cannot be removed together. The flux-threaded models of Sec.~\ref{sec:models} are built on this principle.

With multiple sites per unit cell, the noise strength alone can change the winding number. In this case, the impurity of Eq.~\eqref{eq:impurity_problem} has one entry per site density of the unit cell. Between dephasing events, interband coherences contribute to the transport coefficients. Their frequencies set time scales independent of $1/\gamma$, which semiclassical kinetic (Boltzmann) theory~\cite{Cercignani1988}, built from band velocities and scattering rates, does not contain. Consequently, the winding number $\nu$ can change with $\gamma$ alone, giving a topological transition which is driven purely by the noise strength and which is beyond kinetic theory. This is discussed in detail in Sec.~\ref{subsec:two_band_chain}.

\subsection{Linking the spectrum to late-time dynamics}
\label{subsec:latetime}

At late times, the dynamics only depends on the least-damped band(s) of $\Chat(k)$, closest to the origin [Fig.~\ref{fig:concept}]. We focus here on a single such band and treat the general case in Appendix~\ref{appendix:multi_band}.

Hermitian jumps make $C \propto \id$ a null eigenvector of $\Chat$, so that $\lambda=0$ is always an eigenvalue, corresponding to the steady state(s)\footnote{The many-body Lindbladian has one steady state per $U(1)$ symmetry sector since we assume a strong $U(1)$ symmetry here. These all map to a correlation matrix proportional to the identity, $C \propto \id$. We assume that $\Chat$ has a one-dimensional kernel generated by $\id$, which holds when particle number is the only (quadratic) conserved quantity. This steady state commutes with $\mathcal{H}$ and each $\mathcal{K}_m$.}. We assume that $C \propto \id$ is the only null eigenvector. Since $\lambda = 0$ cannot be an exceptional point [Appendix~\ref{appendix:multi_band}], only a single band $\lambda(k)$ then passes through the origin. SWAP symmetry then enforces $\lambda(k)=-\lambda^*(-k)$. Expanding around $k=0$, we have
\begin{equation}
\label{eq:eigenvalue_expansion}
\begin{aligned}
    \lambda(k) &= \sum_{n \text{ odd}}c_nk^n -\im \sum_{n \text{ even}}D_nk^n,
\end{aligned}
\end{equation}
where $c_n,D_n$ are real coefficients, and $n \in \mathbb{N}^*$. This specific form is enforced by the SWAP symmetry.

At late times, a generic density perturbation $n_m(\tau) = \langle \cdag_m c_m\rangle(\tau)$ can be expanded in terms of momentum modes as
\begin{equation}
    n_m(\tau) = \int_{-\pi}^\pi \frac{\mathrm{d}k}{2\pi}\, e^{-\im km}e^{-\im \lambda(k)\tau}\, n(k,0),
\end{equation}
where $n(k,0)$ is the initial momentum-space profile. Normalizing to a distribution $P_m(\tau) = n_m(\tau)/\sum_m n_m(\tau)$, we then define density cumulants~\cite{GardinerZoller2004} $\kappa_n$ through
\begin{equation}\label{eq:cumulants_definition}
\log \left[\sum_m P_m(\tau)\,e^{\im km}\right]=\sum_n\kappa_n(\tau)\frac{(\im k)^n}{n!}.
\end{equation}
One can expand the above, differentiate in $\tau$, and compare coefficients with Eq.~\eqref{eq:eigenvalue_expansion} to find
\begin{equation}\label{eq:cumulants}
\begin{aligned}
\dot{\kappa}_1 &= -c_1,\\ 
\dot{\kappa}_2 &= 2D_2, \\
\dot{\kappa}_3 &= 6c_3, \ldots
\end{aligned}
\end{equation}
Here, dots are shorthand for derivatives in time $\tau$. Cumulants are more natural variables than moments in this context for two reasons. Firstly, cumulants are additive under the convolution generated by $e^{-\im \lambda(k)\tau}$ in $n_m(\tau)$, making the $\dot{\kappa}_n$ directly map to a single Taylor coefficient of $\lambda(k)$. Secondly, moments $\mu_n$ mix orders, e.g. $\mu_4 = \kappa_4 + 3\kappa^2_2$, so that $\mu_4$ also has $\tau^2$ growth from diffusive spreading. This obscures the higher-order asymmetries we wish to isolate.

Note that particle-number conservation forbids drift. With a strong $U(1)$ symmetry, the steady state $\rho_S \propto \id$~\cite{PhysRevA.87.012108} carries no current, so the density cannot drift and $c_1=0$. The band through the origin then has a cusp there, and any asymmetry in transport must appear in higher odd cumulants.

The winding number fixes the direction of this asymmetry, directly linking it to point-gap topology. Let $n_*$ be the lowest odd integer with $c_{n_*}\neq0$. Then, the winding number of Eq.~\eqref{eq:winding_number_definition} around reference energies $-\im\epsilon$, just below the steady state, is
\begin{equation}
\label{eq:winding_sign_rule}
\nu(-\im\epsilon)=-\sgn c_{n_*},\qquad \sgn\dot\kappa_{n_*}=(-1)^{(n_*-1)/2}\nu,
\end{equation}
while $\nu=0$ if all odd coefficients vanish. A nonzero winding number near the origin is thus equivalent to an odd cumulant of the density profile that grows linearly in time, with its sign set by $\nu$.

This equivalence holds for multiple bands, provided only one of them touches the origin, and for any initial state with nonzero particle number [Appendix~\ref{appendix:multi_band}]. At large enough dephasing, the relevant point gap takes the form of a loop that touches the origin, and its winding cannot change unless the leading odd coefficient passes through zero, i.e. through a topological transition at which the asymmetry reverses. The same holds for reference points just below isolated bands that obey $\lambda(k) = -\lambda^*(-k)$, as in the flux ladder of Sec.~\ref{sec:models}. The winding number can also be defined without translation invariance, through a twisted boundary condition; using this, we show in Appendix~\ref{appendix:multi_band} that, for the dephased flux chain, static on-site disorder reduces the asymmetry without changing its sign [see also Sec.~\ref{sec:models}].

In practice, we diagnose the asymmetry by checking when inversion symmetry, $\Ical: i \to -i$, is broken in the equations of motion. If inversion symmetry is respected, then the spectrum is symmetric under $\lambda(k) \to \lambda(-k)$, and every odd $\dot\kappa_n$ vanishes. In the following sections, we will probe transport by initializing the system with a Gaussian density profile in the middle of the chain; see Appendix~\ref{appendix:multi_band} for a discussion of other initializations.

\section{Models}
\label{sec:models}

In this Section, we illustrate the general principles of noise-induced topology established in the previous section in simple lattice models threaded by a magnetic flux, and identify their consequences for transport. Throughout, we abbreviate nearest-neighbor and next-nearest-neighbor hoppings as n.n. and n.n.n., respectively.

\subsection{A dephased flux chain}
\label{subsec:flux_chain}

As said above, for particle-number conserving jumps $K_m$, the Lindbladian has a strong $U(1)$ symmetry and the density cannot drift, $\dot{\kappa}_1 = 0$ [Sec.~\ref{subsec:latetime}]. Any noise-induced topology in $\Chat$ must therefore manifest in higher-order cumulants, with a point gap that is necessarily cusped at the origin. Following Sec.~\ref{sec:bound_states}, it also requires breaking both inversion and time reversal, which we do through the Hamiltonian.

A minimal model doing so is a tight-binding chain with Peierls phases, subject to on-site dephasing~\cite{Zimboras2013, haken1973exactly, PhysRevLett.117.137202, PhysRevX.8.031045} [see also Fig.~\ref{fig:concept}(a)]
\begin{equation}\label{eq:dephased_flux_chain}
\begin{aligned}
H&=\sum_m\Big[t\,c^\dagger_{m+1}c_m+t'\ei{\phi}c^\dagger_{m+2}c_m+\text{h.c.}\Big],\\ K_m&=\sqrt\gamma\,n_m,
\end{aligned}
\end{equation}
where $t, t'$, and $\gamma$ are all real and positive. We set the phase $\phi$ to $\pi/2$ unless stated otherwise; values in the range $(0,\pi)$ do not qualitatively change our results. This is a variant of Haken--Strobl models~\cite{haken1973exactly, PhysRevE.47.67, PhysRevLett.131.053401} for incoherent transport of excitons. Its $\gamma = 0$ limit, a time-reversal-breaking hopping Hamiltonian, has been studied both experimentally and theoretically as a chiral quantum walk~\cite{Zimboras2013, PhysRevX.8.031045, An2017}. In this limit, the particle spreads ballistically, with a profile set by the distribution of group velocities $v(k)=\partial_kE(k)$, whose third moment $\langle v^3\rangle=-12t^2t'\sin\phi$ is already nonzero. The Lindbladian spectrum is then real, however, and there is no point-gap topology.

\begin{figure}
    \centering
    \includegraphics[width=1.02\linewidth]{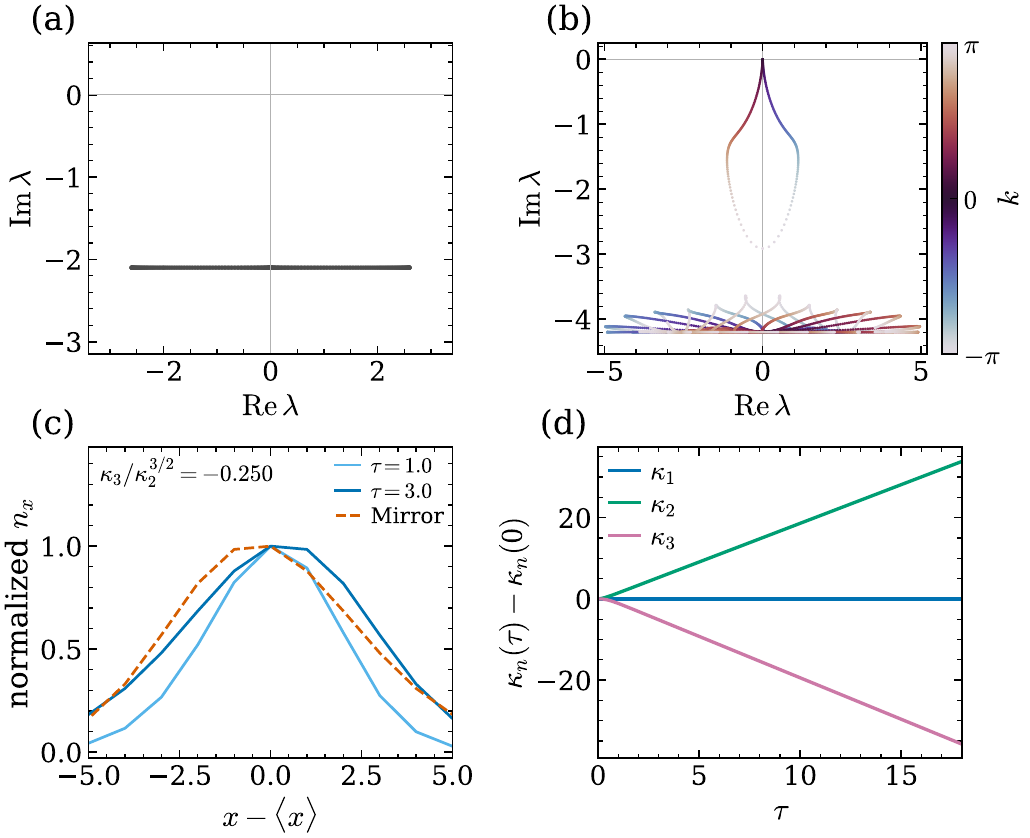}
    \caption{ Dephased flux chain [Eq.~\eqref{eq:dephased_flux_chain}] with $t=1$, $t'=0.5$ and $\phi=\pi/2$, as well as on-site dephasing strength $\gamma=4.2$. (a) Spectrum of the single-particle generator $\Theta$, which is flat at imaginary part $-\gamma/2$. (b) Spectrum of the correlation matrix dynamics generator $\Chat$. It displays a point gap around the origin (winding number $\nu = +1$), atop a continuum of eigenvalues, understood from $\Theta$. (c) Time evolution of an initial Gaussian density profile. Various curves show the profile at different times. The profile spreads diffusively and develops an asymmetry, measured by the skewness $\kappa_3$. The direction of asymmetry is dictated by the winding number $\nu$ [Eq.~\eqref{eq:winding_number_definition}]. The dashed profile indicates the inversion-mapped profile at $\tau=3$, to visualize the asymmetry. (d) Time evolution of the cumulants. The mean $\kappa_1$ remains fixed, i.e. there is no drift. The variance $\kappa_2$ and skewness $\kappa_3$ follow Eq.~\eqref{eq:analytical_dephased_flux_chain_cumulants}.}
    \label{fig:dephased_flux_chain}
\end{figure}

With dephasing, the situation changes. For generic $\gamma$, $\sum_m \mathcal{K}^2_m = \gamma \id$, so $\Theta$ has no point gap [Fig.~\ref{fig:dephased_flux_chain}(a)], whereas $\Chat$ can host one [Fig.~\ref{fig:dephased_flux_chain}(b)], which is therefore induced purely by the jump term in Eq.~\eqref{eq:CM_EOM}. The corresponding bound states form the late-time band of $\Chat$. They already exist for $t'=0$~\cite{chaduteau2026, PhysRevLett.117.137202}, but the point gap only opens once the flux $\phi$ is present. This flux cannot be gauged away. This mechanism is analogous to Holstein's three-site problem~\cite{Holstein1961} for the hopping Hall effect.

The leading coefficients, which we compute exactly in Appendix~\ref{appendix:late_time_expansions}, are
\begin{equation}
\label{eq:analytical_dephased_flux_chain_cumulants}
    D_2=\frac{2t^2+8t'^2}{\gamma},\qquad c_3=-\frac{12\,t^2t'\sin\phi}{\gamma^2},
\end{equation}
so the asymmetry first appears as a skewness. Numerics confirm this [Fig.~\ref{fig:dephased_flux_chain}(c) and (d)]: the mean $\kappa_1$ remains fixed, while $\kappa_2$ and $\kappa_3$ grow at the predicted rates. The asymmetry disappears for $\phi = 0, \pi$, and for $\gamma \gg t, t'$, the point gap of $\Chat$ shrinks and the density dynamics tends to a classical random walk. 

Thus, the asymmetry manifests in the skewness of the density profile. The flux chain also hosts a topological phase transition at $\phi = 0$ (and $\phi=\pi$), across which the point gap closes and the winding number reverses, as reflected in $c_3\propto\sin\phi$. At $\gamma=0$, the Hamiltonian walk already reverses chirality at $\phi=0$; dephasing turns this reversal into a topological phase transition. The dephased chiral walk is thus a minimal platform for noise-induced topology.

Although $\kappa_3$ grows linearly, the standardized skewness $\tilde\kappa_3\equiv\kappa_3/\kappa_2^{3/2}$ decays, as for any diffusive process, since the profile becomes Gaussian:
\begin{equation}
\label{eq:standardised_skewness}
\tilde\kappa_3(\tau) \approx -\frac{36\sin\phi}{\sqrt2}\,
\frac{f}{(2+8f^2)^{3/2}}\,(\gamma\tau)^{-1/2},
\end{equation}
where $f = t'/t$. Its magnitude is largest, $|\tilde\kappa_3|\approx\sqrt3\,(\gamma\tau)^{-1/2}$, at $\phi=\pi/2$ and $f=1/(2\sqrt2)$. Since Eq.~\eqref{eq:standardised_skewness} depends only on $\gamma\tau$, the dephasing sets the timescale but not the size of the asymmetry.

The asymmetry is most pronounced in the tails of the profile. For the parameters of Fig.~\ref{fig:dephased_flux_chain} and a particle initially on a single site, the probability of finding the particle at $x \geq 20$ at $\tau=20/t$ is five times smaller than at $x \leq -20$. More generally, at fixed speed $u=x/\tau$, the log-ratio of opposite tails, $\ln[n_x(\tau)/n_{-x}(\tau)]$, grows linearly in time [Fig.~\ref{fig:fig_tails} of Appendix~\ref{appendix:late_time_expansions}], with a sign set by $\nu$ for small $u$.

The point-gap topology implies a non-Hermitian skin effect, i.e. exponential localization of the eigenvectors of $\Chat$ under open boundary conditions (OBC) [Appendix~\ref{appendix:OBC}], which establishes a bulk--boundary correspondence for the hydrodynamic modes~\cite{chaduteau2026}. Unlike conventional skin effects, which typically accompany a drift, in our case the skin effect occurs with $\dot\kappa_1=0$. Since the steady state remains uniform~\cite{PhysRevA.87.012108}, the skin effect only affects the relaxation~\cite{PhysRevLett.123.170401}.

We contrast this with Ref.~\onlinecite{chaduteau2026}, whose model considers correlated density-current jumps $K_m = An_m + Bj_m$, with the single-particle generator $\Theta$ also point-gapped. In the present case, the topology resides solely in $\Chat$. The same construction can push the asymmetry to higher orders. As we detail in Appendix~\ref{appendix:higher_orders}, longer-range hoppings in $H$, still with on-site dephasing, give noise-induced topology whose asymmetry first appears in higher odd cumulants. Experimentally, this requires fine tuning since nonzero hoppings at lower ranges restore the lower-order cumulants. 

Finally, we turn to disorder, which can arise both from $H$ and from the jumps $K_m$. In Appendix~\ref{appendix:multi_band}, we show that, by the $U(1)$ symmetry, any nonzero winding number near the origin is equivalent to a late-time asymmetry in the transport, also in the presence of multiple bands. Replacing the momentum by a twisted boundary condition, we further find that the winding number of the dephased flux chain survives static on-site disorder, which reduces the asymmetry but crucially does not change its sign.

\subsection{Topological transition driven by dephasing}\label{subsec:two_band_chain}
\begin{figure}
    \centering
    \includegraphics[width=1.01\linewidth]{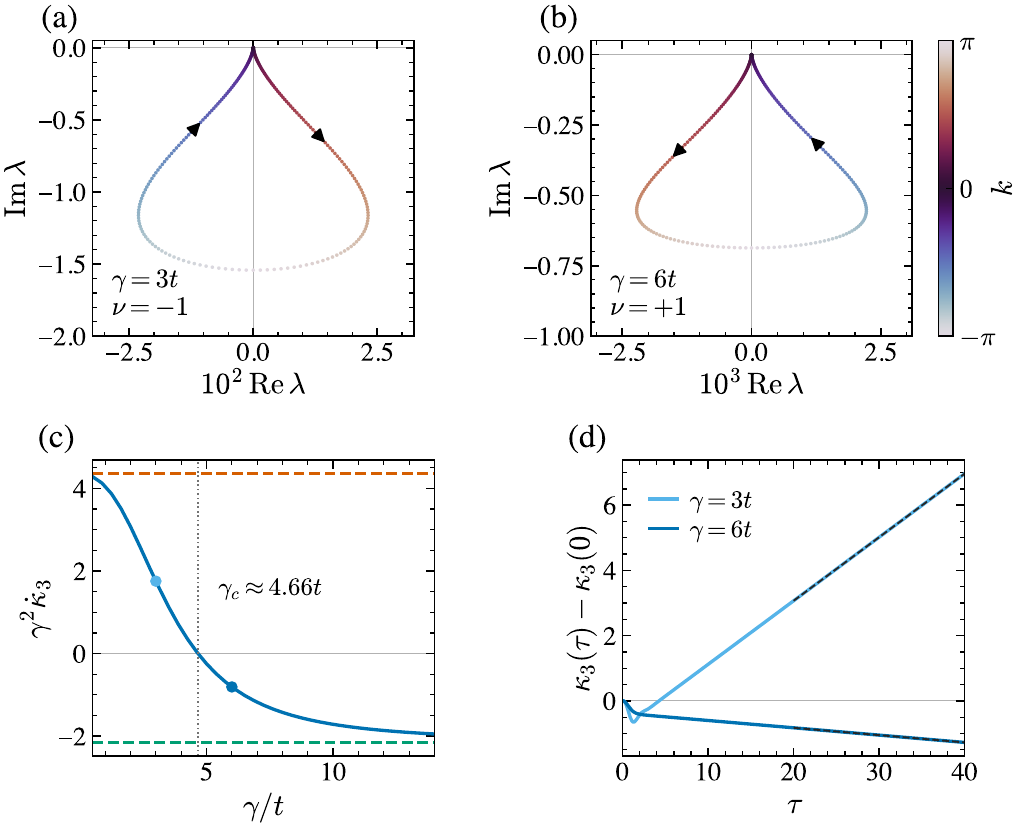}
    \caption{Topological phase transition driven by dephasing strength alone [Eq.~\eqref{eq:dephasing_transition_model}]. We take $\delta=0.7$ and $t'=t/2$. (a), (b) PBC spectra of $\Chat(k)$ near the origin for $\gamma = 3t$ (below the critical $\gamma_c \approx 4.66t$) and $\gamma = 6t$ (above $\gamma_c$). The hydrodynamic band is colored by $k$, with arrows indicating increasing $k$, and the other eigenvalues are in gray. The real axis is rescaled by $10^2$ and $10^3$, respectively. The two loops have opposite winding numbers. (c) The numerical skewness growth rate $\gamma^2 \dot{\kappa}_3$ switches sign at $\gamma_c$, as does the winding number. Both semiclassical kinetic theory and Holstein's mechanism (dashed) give constants at low and high dephasing rate limits, respectively. (d) Real-space evolution of $\kappa_3$ from a Gaussian profile at $\gamma=3t$ and $\gamma=6t$. Dashed lines show predicted slopes $6c_3$.}
    \label{fig:dephasing_transition_model}
\end{figure}
With one site per unit cell, as in the dephased flux chain, the skewness growth is fixed by the velocity distribution, $c_3 = \langle v^3 \rangle / \gamma^2$. There, a topological phase transition can only occur by changing the Hamiltonian, e.g. via $\phi$. With several sites per unit cell, however, interband coherences also contribute to the transport coefficients [Sec.~\ref{sec:bound_states}], and the noise strength alone can then drive a topological transition.

We realize this with two additions to the dephased flux chain of Eq.~\eqref{eq:dephased_flux_chain} at $\phi=\pi/2$:
\begin{equation}\label{eq:dephasing_transition_model}
    \begin{aligned}
    H = \sum_m\Big[&t\big(1+\delta(-1)^m\big)c^\dagger_{m+1}c_m\\
    &+t' \big(\im+(-1)^m\big)c^\dagger_{m+2}c_m+\text{h.c.}\Big],
    \end{aligned}
\end{equation}
with on-site dephasing $K_m=\sqrt\gamma\,n_m$. The n.n. hopping is dimerized by $\delta$, and a real range-$2$ n.n.n. hopping that alternates in sign doubles the unit cell. The n.n.n. hoppings have Peierls phases alternating between $\pi/4$ and $3\pi/4$. Both inversion and time reversal are broken.

Now, $\Chat(k)$ has two bound-state bands. The hydrodynamic band forms a loop that touches the origin, while the other forms a separate loop gapped from the origin, which only contributes transients and affects neither the winding number near the origin nor the rates $\dot{\kappa}_n$. Compared with $\delta=0$, the extra dimerization increases $\gamma_c$, so that the hydrodynamic loop is isolated from the continuum on both sides of the transition [Fig.~\ref{fig:dephasing_transition_model}(a), (b)]. For the parameters in Fig.~\ref{fig:dephasing_transition_model}, at $\gamma_c \approx 4.66t$ the point gap of the hydrodynamic band closes, and $c_3$ vanishes [Fig.~\ref{fig:dephasing_transition_model}(c)]. For $\gamma < \gamma_c$, the winding number near the origin is $\nu = -1$, with $\dot\kappa_3=+0.19t$ at $\gamma=3t$. Above $\gamma_c$ it reverses to $\nu = +1$, the value of the flux chain, with $\dot\kappa_3=-0.022t$ at $\gamma=6t$. Real-space evolution of Eq.~\eqref{eq:CM_EOM} confirms both rates [Fig.~\ref{fig:dephasing_transition_model}(d)]. This is a topological transition driven by the noise strength alone, which, for on-site dephasing, requires several sites per unit cell.

The reason for this transition is seen most clearly in the two limits of the dephasing strength. At weak dephasing, interband coherences average out between dephasing events, and the ``particle'' hopping in the correlation matrix [Eq.~\eqref{eq:impurity_problem}] streams with band velocities. This is the regime of semiclassical kinetic theory, built from band velocities and scattering rates, in which $\gamma^2\dot\kappa_3$ tends to a constant. At strong dephasing, $\gamma\gg t$, coherences between sites decay before they can propagate, and the ``particle'' performs a classical random walk between sites, as in the Haken--Strobl model~\cite{haken1973exactly}. The flux then enters at next order, through Holstein's three-site mechanism~\cite{Holstein1961}, and $\gamma^2\dot\kappa_3$ again tends to a constant. The two limits involve different mechanisms, band velocities versus hopping around flux triangles, so they need not share a sign. When they differ, $c_3$, and hence $\nu$, must change sign at an intermediate $\gamma_c$.

Kinetic theory contains no time scale besides $1/\gamma$, so the sign of $c_3$ it predicts cannot change with $\gamma$. Indeed, the kinetic prediction $\gamma^2\dot\kappa_3$ is constant and agrees with the exact result only for $\gamma\ll\gamma_c$ [Fig.~\ref{fig:dephasing_transition_model}(c)]. Kinetic theory therefore misses the transition, whereas the winding number captures it accurately.

Only the staggered range-$2$ hopping is essential: without it, we find no transition. Without $\delta$, the winding direction and the skewness do reverse at some $\gamma_c$, but the hydrodynamic band merges with the continuum and does not form an isolated point gap. Finally, the amplitude of $\dot{\kappa}_3$ is weaker here than in the flux chain, so resolving its sign will require more noise samples.

\subsection{Flux and dephasing on a ladder}
\label{subsec:ladder}

Particle-number conservation forbids any drift of the density: in the language of Brownian motors~\cite{REIMANN200257, RevModPhys.81.387}, there is no density ratchet. This does not, however, restrict noise-induced topology to higher-order cumulants, since coherences are not conserved and can drift. To isolate them, we use a symmetry that separates the (conserved) density sector from the (non-conserved) coherence sector. This is most easily realized by introducing two sublattices $a$ and $b$, and introducing noise that is odd under the parity flip $b \to -b$ (equivalently, $a \to -a$).

Consider a dephased flux ladder with noisy rung hopping,
\begin{equation}
\label{eq:dephased_flux_ladder}
\begin{aligned}
H=&\sum_m\Big[t\,\ei{\phi/2}a^\dagger_{m+1}a_m+t\,\emi{\phi/2}b^\dagger_{m+1}b_m+\text{h.c.}\Big]\\
&+\frac{\Delta}{2}\sum_m\big(n^a_m-n^b_m\big),\\
K_m=&\sqrt\gamma\,(a^\dagger_mb_m+b^\dagger_ma_m).
\end{aligned}
\end{equation}
The two sublattice hoppings involve opposite complex Peierls phases, and they are detuned by $\Delta \in \mathbb{R}$. The matrices $\mathcal{K}_m$ involve the first-quantized operator $X_m = \ket{a_m}\bra{b_m} + \ket{b_m}\bra{a_m}$, whose square projects onto rung $m$. As a result, $\Theta = \mathcal{H} - \frac{\im}{2}\gamma \id$ is again topologically trivial.

The parity flip $b \to -b$ maps $K_m \to -K_m$, giving a weak $\mathbb{Z}_2$ symmetry of the Lindbladian. This in turn induces a $\mathbb{Z}_2$ symmetry $\Shat$ with $\Shat \Chat \Shat^{-1} = \Chat$, allowing block-diagonalization into two sectors: a parity-even sector of densities $C_{aa}$, $C_{bb}$, and a parity-odd sector of coherences $C_{ab}$, $C_{ba}$. By the strong $U(1)$ symmetry, the even sector must conserve density, while the odd sector need not.
\begin{figure}
    \centering
    \includegraphics[width=1.02\linewidth]{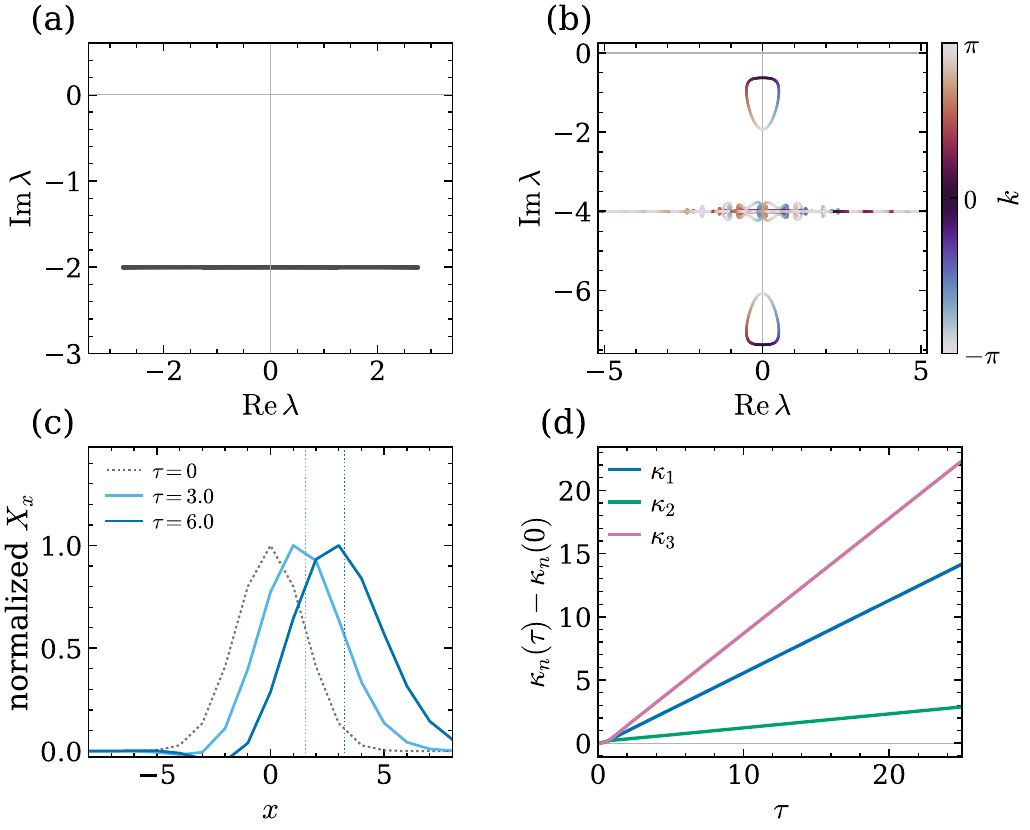}
    \caption{Dephased flux ladder [Eq.~\eqref{eq:dephased_flux_ladder}] with $t=1$, $\phi=3\pi/10$, $\Delta=1.5$ and rung tunneling noise strength $\gamma=4$. (a) Spectrum of $\Theta$, which is flat, as in the dephased flux chain. (b) Spectrum of $\Chat$ in the odd $\mathbb{Z}_2$ sector. Two point gaps are open, indicating asymmetric transport along the ladder. The even-parity sector does not have such point gaps, indicating symmetric diffusion along the ladder. (c) Normalized rung coherence profile $X_x=\langle a^\dagger_xb_x+b^\dagger_xa_x\rangle$ at $\tau=0,3,6$; dotted lines mark $\langle x\rangle$, which drifts at rate $-c_1$ [Eq.~\eqref{eq:dephased_flux_ladder_drift}]. (d) Evolution of the cumulants $\kappa_n$, which grow linearly.}
    \label{fig:dephased_flux_ladder}
\end{figure}

In the odd sector, the noise drives a drift of rung coherences. In Appendix~\ref{appendix:late_time_expansions}, we find the drift coefficient
\begin{equation}
\label{eq:dephased_flux_ladder_drift}
c_1 = -\frac{8\,t^2\,\Delta\sin\phi}{\sqrt{\gamma^4+64\,t^2\Delta^2\sin^2(\phi/2)}} \, ,
\end{equation}
which matches numerics [Fig.~\ref{fig:dephased_flux_ladder}]. The drift requires both flux and detuning, since $c_1$ is odd in $\Delta$ and in $\sin\phi$, and its direction reverses with the sign of either. Such a drift of coherences has no counterpart in a classical walk of densities: it is a quantum ratchet, enabled by the incoherent jumps. At $\gamma=0$, the legs decouple and their Peierls phases can be gauged away; it is the noisy rungs that close plaquettes whose flux $\phi$ cannot be gauged away. Unlike in the chain, the odd-parity band is gapped from the origin, so the drift of the coherences occurs on top of a decay at a rate $\Gamma_0$ [see Appendix~\ref{appendix:late_time_expansions}].

The point gap has a direct real-space consequence under OBC [Appendix~\ref{appendix:OBC}]: rung coherences localize exponentially at one edge and are depleted at the other. The winding number selects the edge; inversion, which maps $\phi \to -\phi$, exchanges the two edges. The accumulation disappears for $\Delta=0$ and for $\phi\in\{0,\pi\}$. Coherence and density skin effects in the non-interacting case were studied for non-unitary circuits with measurement and feedback in Ref.~\onlinecite{chahine2026independentdensitycoherenceskin}.

\section{Experimental realizations}
\label{sec:experiment}

Observing the direct link we have established between bound-state topology and late-time hydrodynamics requires no engineered dissipation: a flux pattern that breaks both inversion and time reversal, combined with on-site dephasing (random fluctuations of the on-site energies), is enough. Below, we describe a protocol and a measurable observable, along with the assumptions the effect relies on, before turning to specific platforms; of these, momentum-state lattices currently appear to offer the most direct route.

\subsection{Protocol and observable}
\label{subsec:exp_protocol}

Equation~\eqref{eq:SSE} is the It\^o form of unitary evolution under the stochastic Hamiltonian
\begin{equation}
H(\tau)=H+\sum_m\xi_m(\tau)K_m \, ,
\end{equation}
where $\xi_m$ is Gaussian white noise with $\mathbb E[\xi_m(\tau)\xi_n(\tau')]=\delta_{mn}\delta(\tau-\tau')$~\cite{PhysRevLett.118.140403}. For $K_m=\sqrt\gamma\,n_m$, this is a random modulation of the on-site energies, $\epsilon_m(\tau)=\sqrt\gamma\,\xi_m(\tau)$. Each experimental run then corresponds to one noise realization, and averaging over runs yields the dephasing of Eq.~\eqref{eq:lindblad_quadratic_jumps}. Since Eq.~\eqref{eq:CM_EOM} holds for both bosons and fermions [see Appendix~\ref{appendix:CM_derivation}], this protocol applies to a wide range of platforms.

Focusing on the dephased flux chain, a concrete observable is the linear growth rate of the third cumulant, $\dot\kappa_3=6c_3$ [Eqs.~\eqref{eq:cumulants} and~\eqref{eq:analytical_dephased_flux_chain_cumulants}], extracted from a fit of $\kappa_3(\tau)$. Unlike the standardized skewness, which decays as $(\gamma\tau)^{-1/2}$ [Eq.~\eqref{eq:standardised_skewness}], this rate stays constant at late times, and it is the same for any initial state with nonzero particle number [Appendix~\ref{appendix:multi_band}]. Any localized initial profile therefore works, a Gaussian density profile being a natural choice. Since inversion maps $\phi\to-\phi$ and reverses the sign of $c_3$, the difference $\kappa_3(\tau;\phi)-\kappa_3(\tau;-\phi)$ moreover cancels any skewness due to an imperfectly symmetric initial state or to systematic imperfections that are even in $\phi$.

To estimate the statistics needed to resolve the sign of the asymmetry, consider the parameters of Fig.~\ref{fig:dephased_flux_chain} and a particle initially on a single site, for which $\tilde\kappa_3\approx-0.22$ at $\tau=12/t$. For $M$ independent position samples, the standard error of the sample skewness is close to $\sqrt{6/M}$, so resolving $|\tilde\kappa_3|\approx0.2$ to $3\sigma$ requires $M\gtrsim1.5\times10^3$ samples per time step. At this time, the profile has a standard deviation of about five sites, so chains of $\sim 30$ sites suffice.

\subsection{Assumptions and robustness}
\label{subsec:exp_requirements}

Beyond the flux pattern, our analysis rests on a few assumptions. Since most experimental noise is colored, we assume that the correlation time of the noise is short compared with the other timescales, so that a Markovian description applies. We also assume that the noise on different sites is uncorrelated, since spatial correlations weaken the damping of short-range coherences $C_{i,i+a}$ and modify the impurity relation of Eq.~\eqref{eq:impurity_problem}. Interactions between particles, e.g. quartic terms in $H$, break the closure of the equations of motion and should remain weak, as should site-dependent loss, which nontrivially modifies $\Theta$ and hence $\Chat$.

The effect is, on the other hand, robust to several common imperfections. Residual decoherence, such as on-site dephasing intrinsic to the platform, only adds to $\gamma$. Uniform particle loss, which is likely to be present, can be modeled by adding linear jumps $J_m=\sqrt\Gamma\,c_m$ to the Lindbladian. These shift $\Theta$, and hence the spectrum of $\Chat$, down the imaginary axis without changing the jump term in Eq.~\eqref{eq:CM_EOM}, so the noise-induced point gap retains its winding number; the correlation matrix $C(\tau)$ simply acquires a factor $e^{-\Gamma\tau}$, which cancels in normalized cumulants. Finally, weak static disorder reduces the asymmetry without changing its sign [Appendix~\ref{appendix:multi_band}].

\subsection{Platforms}
\label{subsec:exp_platforms}

\emph{Momentum-state lattices.} In a Bose--Einstein condensate, Bragg transitions couple an array of discrete atomic momentum states $p_m=2m\hbar k$ and mimic n.n. tunneling~\cite{PhysRevA.92.043606,PhysRevA.93.051602}, while multifrequency laser beams also drive second-order Bragg transitions that provide n.n.n. tunneling. Because the kinetic energy is quadratic in momentum, each link between momentum states has its own resonance frequency and is addressed by its own frequency component, whose amplitude, phase, and detuning set the hopping, the Peierls phase, and the on-site energy. Since every link can thus be programmed individually, this platform is particularly well suited to our models. Zigzag chains with programmable n.n.n. Peierls phases~\cite{PhysRevX.8.031045} and flux ladders~\cite{An2017} have already been realized, and time-dependent disorder~\cite{An2017NC} as well as engineered loss and non-Hermitian skin effects~\cite{Lapp2019,PhysRevLett.129.070401} have been demonstrated on the same platform. Ref.~\onlinecite{PhysRevX.8.031045} used alternating phases $(-1)^m\phi$, i.e. a uniform planar flux, whereas Eq.~\eqref{eq:dephased_flux_chain} requires equal phases on all n.n.n. links; this only changes the phases of the corresponding Bragg components. The chain of Sec.~\ref{subsec:two_band_chain} additionally requires alternating amplitudes on the n.n. links and alternating real parts on the n.n.n. links, again set by the corresponding Bragg components. On-site dephasing, $K_m=\sqrt\gamma\,n_m$, is implemented by modulating the Bragg detunings so that the site energies $\epsilon_m(\tau)$ are independent white noises, while independent amplitude modulation of each rung component implements the noisy rungs of Eq.~\eqref{eq:dephased_flux_ladder}. For a given noise realization, a single time-of-flight image records the full density profile, and averaging over realizations gives the Lindblad dynamics; for the ladder, a short resonant rung pulse before imaging maps the rung coherence onto a population imbalance between the legs. The main constraints are the lattice size and residual interactions, which are long-ranged in the synthetic dimension~\cite{PhysRevLett.120.040407} but remain weak in a dilute condensate.

\emph{Superconducting qubits.} An array of transmons hosting a single excitation realizes a single-particle quantum walk, with complex hoppings obtained from parametric modulation of qubits or couplers~\cite{Roushan2017, rosen2024synthetic} and on-site dephasing from white frequency noise injected into each qubit's flux line~\cite{potocnik2018studying}. Single-shot readout of all qubits gives one sample of the excitation position per run, and since each run can use a new noise realization, the run average is exactly the Lindblad average. The main obstacle is geometric: the flux chain requires triangular plaquettes arranged in a zigzag~\cite{Roushan2017}, whereas current processors are mostly arranged in square grids~\cite{Karamlou2022}. Variations in the $T_1$ time between qubits act as site-dependent loss and should be kept small over the course of the measurement.

\emph{Optical lattices.} Spin-polarized fermions do not interact at low energy, so they realize Eq.~\eqref{eq:CM_EOM} with many particles per shot, and quantum gas microscopes resolve the density profile site by site~\cite{Cheuk2015,Parsons2015}. Peierls phases on the bonds of triangular and zigzag lattices follow from elliptical or circular lattice shaking~\cite{PhysRevLett.108.225304,Jotzu2014}, and near-resonant light scattering in a deep lattice dephases the atoms, with jump operators given by the local densities~\cite{Pichler2010,bouganne2020anomalous,PhysRevLett.115.140402,PhysRevX.7.011034}. Two caveats apply. First, the noise from scattered photons is correlated over a length comparable to an optical wavelength~\cite{PhysRevA.82.013615}, so neighboring sites do not dephase independently and the impurity of Eq.~\eqref{eq:impurity_problem} is modified, as discussed above; these correlations should be kept small. Second, photon recoil causes interband excitations, which must remain small over the course of the measurement. Fermionic flux ladders have also been realized along a synthetic dimension, with $\Delta$ a Raman detuning~\cite{Mancini2015}. There, however, a single global beam drives all rungs, so modulating it produces one collective jump, $K=\sqrt\gamma\sum_m X_m$, rather than the independent rung noise of Eq.~\eqref{eq:dephased_flux_ladder}, and does not realize our model; site-resolved control of the rung amplitude would be needed.

In addition to these platforms, engineered noise has been demonstrated in trapped ions~\cite{PhysRevLett.122.050501} and in photonic platforms~\cite{Biggerstaff2016,Caruso2016}; we leave detailed proposals for these to future work.

\section{Discussion}
\label{sec:discussion}

We have shown that noise can induce nontrivial topology in the relaxation dynamics of an interacting open quantum system. On-site dephasing, the most ubiquitous type of noise, does not wash out the asymmetry of chiral quantum walks, but instead renders it topological, inducing point-gap topology and a skin effect. This topology is induced solely by the jump terms and is absent under postselection. Moreover, for quadratic Hermitian jumps, the equations of motion for correlations close, so the models we consider remain analytically and polynomially soluble. The jump term acts as an interaction, and the corresponding eigenstates of the point gap are bound states of this interaction. In contrast to the conventional link between drift and the skin effect~\cite{PhysRevResearch.1.023013, wsmq-kmq9}, when density is conserved these effects can only manifest in higher-order cumulants of the density profile. Thus, a zero drift does not imply the absence of a skin effect, as illustrated by the dephased flux chain [Eq.~\eqref{eq:dephased_flux_chain}]. However, drift due to the bound-state point gap is still possible in sectors that are not conserved, as illustrated by the dephased flux ladder [Eq.~\eqref{eq:dephased_flux_ladder}]. At late times, the asymmetry is largest in the tails of the density profile.

For on-site dephasing and one site per unit cell, the leading transport coefficients $D_2$ and $c_{n_*}$ coincide with those of a classical velocity-jump process. The quantum input is the velocity distribution, whose odd moments originate from the flux. With multiple sites per unit cell, interband coherences enter at the same order, and can reverse the transport and the winding number as the noise strength alone is varied [Sec.~\ref{subsec:two_band_chain}]. The ladder model, with noisy rung hopping, is instead a true quantum ratchet with no classical counterpart~\cite{REIMANN200257}.

In our case of number-conserving noise, we have shown that a nonzero winding number near the origin is equivalent to a topologically robust late-time asymmetry in the transport. This holds generically, provided a single band touches the origin, and survives static disorder [Appendix~\ref{appendix:multi_band}]. While the winding number dictates the direction of asymmetric transport, it is blind to whether the late-time transport is ballistic, diffusive, or of a different kind. In addition, for $t'=0$, the Lindbladian of the dephased chain maps exactly to a Hubbard model with imaginary interaction~\cite{PhysRevLett.117.137202}, which is integrable via a Bethe ansatz. Turning on $t'$ adds next-nearest-neighbor hoppings to this Hubbard model, which is not expected to be integrable~\cite{PhysRevE.88.012108}. Thus, integrability is separate from the closure of the equations of motion we use here.

Moreover, the point-gap topology here is interaction-induced. While the Lindbladian is interacting, the topology resides in the two-point sector, where the relevant interaction acts between the two indices of the correlation matrix, the ket and the bra [see Sec.~\ref{sec:bound_states}]. The relevant modes are bound states of this effective two-body problem, with coupling set by the dephasing rate $\gamma$. Noisy quantum walks therefore provide a simple testbed for interaction-induced non-Hermitian topology~\cite{PhysRevB.108.155114, qrh6-vx64, PhysRevLett.129.180401, chaduteau2026}, requiring neither strong two-body interactions nor postselection. Nonetheless, it is worth stating what this topology does not provide. Since densities relax to a uniform steady state, the averaged dynamics cannot amplify a signal or pump particles and, unlike drive- or gain-based non-Hermitian devices~\cite{PhysRevX.5.021025, Wanjura2020, mcdonald2020exponentially}, cannot serve as an isolator or a directional amplifier. Instead, the value of this topology is diagnostic and conceptual. First, non-Hermitian topology arises here from a Hermitian Hamiltonian and Hermitian jump operators, in non-Gaussian states. Second, the sign of a growing odd cumulant or of the tail asymmetry is a protected signature of broken inversion and time reversal in a noisy, incoherent system, obtained without applying a bias. Higher cumulants of this kind are measurable: the third cumulant of current fluctuations has been resolved in mesoscopic tunnel junctions~\cite{PhysRevLett.91.196601,PhysRevLett.95.176601}, whereas here $\kappa_3$ follows directly from position samples [Sec.~\ref{sec:experiment}].

We have also discussed a variety of experimental platforms on which these effects could be observed. Apart from experiment, our work opens up several promising directions. In two spatial dimensions, crystalline symmetries such as $C_3$ replace inversion symmetry and can constrain which odd cumulants may grow, suggesting a richer set of noise-induced transport classes. Breaking time reversal there also allows a transverse, Hall-like component, analogous to Holstein's hopping Hall effect~\cite{Holstein1961} and to the odd diffusivity of chiral active matter~\cite{Hargus2021}. It also remains open whether the restricted class of Lindbladians with closed equations of motion supports invariants beyond the winding number. Finally, non-Markovian effects, such as those studied in Ref.~\onlinecite{PhysRevLett.122.050501}, would be interesting to investigate, both for theoretical interest and for their experimental relevance.

\begin{acknowledgments}
We thank D. Huse, B. Lian, S. Chirame, and A. Gupta for insightful discussions. A.C. acknowledges support from Imperial College London via a President’s PhD Scholarship. This work was supported by a UKRI Future Leaders Fellowship MR/Y017331/1. A.P. is supported by the U.S. Department of Energy, Office of Science, Office of Advanced Scientific Computing Research, through the Exploratory Research for Extreme Scale Science (EXPRESS) program under Award No. DE-SC0026216. The scientific content of this work was conceived by the authors without the use of generative
AI. The authors acknowledge the use of generative AI (Claude Opus 5.5) for literature searches and preliminary numerical analysis. All content, claims, and conclusions have been reviewed and verified by the authors to ensure accuracy and originality. The authors take full responsibility for the contents of the paper.
\end{acknowledgments}

\bibliography{refs}

\begin{thebibliography}{126}%
\makeatletter
\providecommand \@ifxundefined [1]{%
 \@ifx{#1\undefined}
}%
\providecommand \@ifnum [1]{%
 \ifnum #1\expandafter \@firstoftwo
 \else \expandafter \@secondoftwo
 \fi
}%
\providecommand \@ifx [1]{%
 \ifx #1\expandafter \@firstoftwo
 \else \expandafter \@secondoftwo
 \fi
}%
\providecommand \natexlab [1]{#1}%
\providecommand \enquote  [1]{``#1''}%
\providecommand \bibnamefont  [1]{#1}%
\providecommand \bibfnamefont [1]{#1}%
\providecommand \citenamefont [1]{#1}%
\providecommand \href@noop [0]{\@secondoftwo}%
\providecommand \href [0]{\begingroup \@sanitize@url \@href}%
\providecommand \@href[1]{\@@startlink{#1}\@@href}%
\providecommand \@@href[1]{\endgroup#1\@@endlink}%
\providecommand \@sanitize@url [0]{\catcode `\\12\catcode `\$12\catcode `\&12\catcode `\#12\catcode `\^12\catcode `\_12\catcode `\%12\relax}%
\providecommand \@@startlink[1]{}%
\providecommand \@@endlink[0]{}%
\providecommand \url  [0]{\begingroup\@sanitize@url \@url }%
\providecommand \@url [1]{\endgroup\@href {#1}{\urlprefix }}%
\providecommand \urlprefix  [0]{URL }%
\providecommand \Eprint [0]{\href }%
\providecommand \doibase [0]{https://doi.org/}%
\providecommand \selectlanguage [0]{\@gobble}%
\providecommand \bibinfo  [0]{\@secondoftwo}%
\providecommand \bibfield  [0]{\@secondoftwo}%
\providecommand \translation [1]{[#1]}%
\providecommand \BibitemOpen [0]{}%
\providecommand \bibitemStop [0]{}%
\providecommand \bibitemNoStop [0]{.\EOS\space}%
\providecommand \EOS [0]{\spacefactor3000\relax}%
\providecommand \BibitemShut  [1]{\csname bibitem#1\endcsname}%
\let\auto@bib@innerbib\@empty
\bibitem [{\citenamefont {Sohal}\ and\ \citenamefont {Prem}(2025)}]{sohal2025imto}%
  \BibitemOpen
  \bibfield  {author} {\bibinfo {author} {\bibfnamefont {R.}~\bibnamefont {Sohal}}\ and\ \bibinfo {author} {\bibfnamefont {A.}~\bibnamefont {Prem}},\ }\bibfield  {title} {\bibinfo {title} {Noisy approach to intrinsically mixed-state topological order},\ }\href {https://doi.org/10.1103/PRXQuantum.6.010313} {\bibfield  {journal} {\bibinfo  {journal} {PRX Quantum}\ }\textbf {\bibinfo {volume} {6}},\ \bibinfo {pages} {010313} (\bibinfo {year} {2025})}\BibitemShut {NoStop}%
\bibitem [{\citenamefont {Ellison}\ and\ \citenamefont {Cheng}(2025)}]{ellison2025imto}%
  \BibitemOpen
  \bibfield  {author} {\bibinfo {author} {\bibfnamefont {T.~D.}\ \bibnamefont {Ellison}}\ and\ \bibinfo {author} {\bibfnamefont {M.}~\bibnamefont {Cheng}},\ }\bibfield  {title} {\bibinfo {title} {Toward a classification of mixed-state topological orders in two dimensions},\ }\href {https://doi.org/10.1103/PRXQuantum.6.010315} {\bibfield  {journal} {\bibinfo  {journal} {PRX Quantum}\ }\textbf {\bibinfo {volume} {6}},\ \bibinfo {pages} {010315} (\bibinfo {year} {2025})}\BibitemShut {NoStop}%
\bibitem [{\citenamefont {{Kobayashi}}\ \emph {et~al.}(2026)\citenamefont {{Kobayashi}}, \citenamefont {{Prem}},\ and\ \citenamefont {{Yu}}}]{KPY26}%
  \BibitemOpen
  \bibfield  {author} {\bibinfo {author} {\bibfnamefont {R.}~\bibnamefont {{Kobayashi}}}, \bibinfo {author} {\bibfnamefont {A.}~\bibnamefont {{Prem}}},\ and\ \bibinfo {author} {\bibfnamefont {M.}~\bibnamefont {{Yu}}},\ }\bibfield  {title} {\bibinfo {title} {{Realizing Braided Fusion 2-Categories as (3+1)D Mixed-State Topological Orders}},\ }\href@noop {} {\bibfield  {journal} {\bibinfo  {journal} {arXiv e-prints}\ } (\bibinfo {year} {2026})},\ \Eprint {https://arxiv.org/abs/2610.03864} {arXiv:2610.03864 [quant-ph]} \BibitemShut {NoStop}%
\bibitem [{\citenamefont {Ma}\ and\ \citenamefont {Wang}(2023)}]{ma2023aspt}%
  \BibitemOpen
  \bibfield  {author} {\bibinfo {author} {\bibfnamefont {R.}~\bibnamefont {Ma}}\ and\ \bibinfo {author} {\bibfnamefont {C.}~\bibnamefont {Wang}},\ }\bibfield  {title} {\bibinfo {title} {Average symmetry-protected topological phases},\ }\href {https://doi.org/10.1103/PhysRevX.13.031016} {\bibfield  {journal} {\bibinfo  {journal} {Phys. Rev. X}\ }\textbf {\bibinfo {volume} {13}},\ \bibinfo {pages} {031016} (\bibinfo {year} {2023})}\BibitemShut {NoStop}%
\bibitem [{\citenamefont {Ma}\ \emph {et~al.}(2025)\citenamefont {Ma}, \citenamefont {Zhang}, \citenamefont {Bi}, \citenamefont {Cheng},\ and\ \citenamefont {Wang}}]{ma2025avg}%
  \BibitemOpen
  \bibfield  {author} {\bibinfo {author} {\bibfnamefont {R.}~\bibnamefont {Ma}}, \bibinfo {author} {\bibfnamefont {J.-H.}\ \bibnamefont {Zhang}}, \bibinfo {author} {\bibfnamefont {Z.}~\bibnamefont {Bi}}, \bibinfo {author} {\bibfnamefont {M.}~\bibnamefont {Cheng}},\ and\ \bibinfo {author} {\bibfnamefont {C.}~\bibnamefont {Wang}},\ }\bibfield  {title} {\bibinfo {title} {Topological phases with average symmetries: The decohered, the disordered, and the intrinsic},\ }\href {https://doi.org/10.1103/PhysRevX.15.021062} {\bibfield  {journal} {\bibinfo  {journal} {Phys. Rev. X}\ }\textbf {\bibinfo {volume} {15}},\ \bibinfo {pages} {021062} (\bibinfo {year} {2025})}\BibitemShut {NoStop}%
\bibitem [{\citenamefont {Haken}\ and\ \citenamefont {Strobl}(1973)}]{haken1973exactly}%
  \BibitemOpen
  \bibfield  {author} {\bibinfo {author} {\bibfnamefont {H.}~\bibnamefont {Haken}}\ and\ \bibinfo {author} {\bibfnamefont {G.}~\bibnamefont {Strobl}},\ }\bibfield  {title} {\bibinfo {title} {An exactly solvable model for coherent and incoherent exciton motion},\ }\href {https://doi.org/10.1007/BF01399723} {\bibfield  {journal} {\bibinfo  {journal} {Z. Phys.}\ }\textbf {\bibinfo {volume} {262}},\ \bibinfo {pages} {135} (\bibinfo {year} {1973})}\BibitemShut {NoStop}%
\bibitem [{\citenamefont {Rips}(1993)}]{PhysRevE.47.67}%
  \BibitemOpen
  \bibfield  {author} {\bibinfo {author} {\bibfnamefont {I.}~\bibnamefont {Rips}},\ }\bibfield  {title} {\bibinfo {title} {Stochastic models of exciton transport: The {H}aken-{S}trobl model},\ }\href {https://doi.org/10.1103/PhysRevE.47.67} {\bibfield  {journal} {\bibinfo  {journal} {Phys. Rev. E}\ }\textbf {\bibinfo {volume} {47}},\ \bibinfo {pages} {67} (\bibinfo {year} {1993})}\BibitemShut {NoStop}%
\bibitem [{\citenamefont {Kenkre}\ and\ \citenamefont {Reineker}(1982)}]{KenkreReineker1982}%
  \BibitemOpen
  \bibfield  {author} {\bibinfo {author} {\bibfnamefont {V.~M.}\ \bibnamefont {Kenkre}}\ and\ \bibinfo {author} {\bibfnamefont {P.}~\bibnamefont {Reineker}},\ }\href {https://doi.org/10.1007/BFb0041435} {\emph {\bibinfo {title} {Exciton Dynamics in Molecular Crystals and Aggregates}}},\ \bibinfo {series} {Springer Tracts in Modern Physics}, Vol.~\bibinfo {volume} {94}\ (\bibinfo  {publisher} {Springer},\ \bibinfo {address} {Berlin, Heidelberg},\ \bibinfo {year} {1982})\BibitemShut {NoStop}%
\bibitem [{\citenamefont {Marais}\ \emph {et~al.}(2013)\citenamefont {Marais}, \citenamefont {Sinayskiy}, \citenamefont {Kay}, \citenamefont {Petruccione},\ and\ \citenamefont {Ekert}}]{Marais2013}%
  \BibitemOpen
  \bibfield  {author} {\bibinfo {author} {\bibfnamefont {A.}~\bibnamefont {Marais}}, \bibinfo {author} {\bibfnamefont {I.}~\bibnamefont {Sinayskiy}}, \bibinfo {author} {\bibfnamefont {A.}~\bibnamefont {Kay}}, \bibinfo {author} {\bibfnamefont {F.}~\bibnamefont {Petruccione}},\ and\ \bibinfo {author} {\bibfnamefont {A.}~\bibnamefont {Ekert}},\ }\bibfield  {title} {\bibinfo {title} {Decoherence-assisted transport in quantum networks},\ }\href {https://doi.org/10.1088/1367-2630/15/1/013038} {\bibfield  {journal} {\bibinfo  {journal} {New J. Phys.}\ }\textbf {\bibinfo {volume} {15}},\ \bibinfo {pages} {013038} (\bibinfo {year} {2013})}\BibitemShut {NoStop}%
\bibitem [{\citenamefont {Plenio}\ and\ \citenamefont {Huelga}(2008)}]{Plenio2008}%
  \BibitemOpen
  \bibfield  {author} {\bibinfo {author} {\bibfnamefont {M.~B.}\ \bibnamefont {Plenio}}\ and\ \bibinfo {author} {\bibfnamefont {S.~F.}\ \bibnamefont {Huelga}},\ }\bibfield  {title} {\bibinfo {title} {Dephasing-assisted transport: Quantum networks and biomolecules},\ }\href {https://doi.org/10.1088/1367-2630/10/11/113019} {\bibfield  {journal} {\bibinfo  {journal} {New J. Phys.}\ }\textbf {\bibinfo {volume} {10}},\ \bibinfo {pages} {113019} (\bibinfo {year} {2008})}\BibitemShut {NoStop}%
\bibitem [{\citenamefont {{\v{Z}}nidari{\v{c}}}(2010{\natexlab{a}})}]{Znidaric_2010_1}%
  \BibitemOpen
  \bibfield  {author} {\bibinfo {author} {\bibfnamefont {M.}~\bibnamefont {{\v{Z}}nidari{\v{c}}}},\ }\bibfield  {title} {\bibinfo {title} {Dephasing-induced diffusive transport in the anisotropic {H}eisenberg model},\ }\href {https://doi.org/10.1088/1367-2630/12/4/043001} {\bibfield  {journal} {\bibinfo  {journal} {New J. Phys.}\ }\textbf {\bibinfo {volume} {12}},\ \bibinfo {pages} {043001} (\bibinfo {year} {2010}{\natexlab{a}})}\BibitemShut {NoStop}%
\bibitem [{\citenamefont {Biggerstaff}\ \emph {et~al.}(2016)\citenamefont {Biggerstaff}, \citenamefont {Heilmann}, \citenamefont {Zecevik}, \citenamefont {Gr{\"a}fe}, \citenamefont {Broome}, \citenamefont {Fedrizzi}, \citenamefont {Nolte}, \citenamefont {Szameit}, \citenamefont {White},\ and\ \citenamefont {Kassal}}]{Biggerstaff2016}%
  \BibitemOpen
  \bibfield  {author} {\bibinfo {author} {\bibfnamefont {D.~N.}\ \bibnamefont {Biggerstaff}}, \bibinfo {author} {\bibfnamefont {R.}~\bibnamefont {Heilmann}}, \bibinfo {author} {\bibfnamefont {A.~A.}\ \bibnamefont {Zecevik}}, \bibinfo {author} {\bibfnamefont {M.}~\bibnamefont {Gr{\"a}fe}}, \bibinfo {author} {\bibfnamefont {M.~A.}\ \bibnamefont {Broome}}, \bibinfo {author} {\bibfnamefont {A.}~\bibnamefont {Fedrizzi}}, \bibinfo {author} {\bibfnamefont {S.}~\bibnamefont {Nolte}}, \bibinfo {author} {\bibfnamefont {A.}~\bibnamefont {Szameit}}, \bibinfo {author} {\bibfnamefont {A.~G.}\ \bibnamefont {White}},\ and\ \bibinfo {author} {\bibfnamefont {I.}~\bibnamefont {Kassal}},\ }\bibfield  {title} {\bibinfo {title} {Enhancing coherent transport in a photonic network using controllable decoherence},\ }\href {https://doi.org/10.1038/ncomms11282} {\bibfield  {journal} {\bibinfo  {journal} {Nat. Commun.}\ }\textbf {\bibinfo {volume} {7}},\ \bibinfo {pages} {11282} (\bibinfo {year} {2016})}\BibitemShut {NoStop}%
\bibitem [{\citenamefont {Caruso}\ \emph {et~al.}(2016)\citenamefont {Caruso}, \citenamefont {Crespi}, \citenamefont {Ciriolo}, \citenamefont {Sciarrino},\ and\ \citenamefont {Osellame}}]{Caruso2016}%
  \BibitemOpen
  \bibfield  {author} {\bibinfo {author} {\bibfnamefont {F.}~\bibnamefont {Caruso}}, \bibinfo {author} {\bibfnamefont {A.}~\bibnamefont {Crespi}}, \bibinfo {author} {\bibfnamefont {A.~G.}\ \bibnamefont {Ciriolo}}, \bibinfo {author} {\bibfnamefont {F.}~\bibnamefont {Sciarrino}},\ and\ \bibinfo {author} {\bibfnamefont {R.}~\bibnamefont {Osellame}},\ }\bibfield  {title} {\bibinfo {title} {Fast escape of a quantum walker from an integrated photonic maze},\ }\href {https://doi.org/10.1038/ncomms11682} {\bibfield  {journal} {\bibinfo  {journal} {Nat. Commun.}\ }\textbf {\bibinfo {volume} {7}},\ \bibinfo {pages} {11682} (\bibinfo {year} {2016})}\BibitemShut {NoStop}%
\bibitem [{\citenamefont {Maier}\ \emph {et~al.}(2019)\citenamefont {Maier}, \citenamefont {Brydges}, \citenamefont {Jurcevic}, \citenamefont {Trautmann}, \citenamefont {Hempel}, \citenamefont {Lanyon}, \citenamefont {Hauke}, \citenamefont {Blatt},\ and\ \citenamefont {Roos}}]{PhysRevLett.122.050501}%
  \BibitemOpen
  \bibfield  {author} {\bibinfo {author} {\bibfnamefont {C.}~\bibnamefont {Maier}}, \bibinfo {author} {\bibfnamefont {T.}~\bibnamefont {Brydges}}, \bibinfo {author} {\bibfnamefont {P.}~\bibnamefont {Jurcevic}}, \bibinfo {author} {\bibfnamefont {N.}~\bibnamefont {Trautmann}}, \bibinfo {author} {\bibfnamefont {C.}~\bibnamefont {Hempel}}, \bibinfo {author} {\bibfnamefont {B.~P.}\ \bibnamefont {Lanyon}}, \bibinfo {author} {\bibfnamefont {P.}~\bibnamefont {Hauke}}, \bibinfo {author} {\bibfnamefont {R.}~\bibnamefont {Blatt}},\ and\ \bibinfo {author} {\bibfnamefont {C.~F.}\ \bibnamefont {Roos}},\ }\bibfield  {title} {\bibinfo {title} {Environment-assisted quantum transport in a 10-qubit network},\ }\href {https://doi.org/10.1103/PhysRevLett.122.050501} {\bibfield  {journal} {\bibinfo  {journal} {Phys. Rev. Lett.}\ }\textbf {\bibinfo {volume} {122}},\ \bibinfo {pages} {050501} (\bibinfo {year} {2019})}\BibitemShut {NoStop}%
\bibitem [{\citenamefont {Poto{\v{c}}nik}\ \emph {et~al.}(2018)\citenamefont {Poto{\v{c}}nik}, \citenamefont {Bargerbos}, \citenamefont {Schr{\"o}der}, \citenamefont {Khan}, \citenamefont {Collodo}, \citenamefont {Gasparinetti}, \citenamefont {Salath{\'e}}, \citenamefont {Creatore}, \citenamefont {Eichler}, \citenamefont {T{\"u}reci}, \citenamefont {Chin},\ and\ \citenamefont {Wallraff}}]{potocnik2018studying}%
  \BibitemOpen
  \bibfield  {author} {\bibinfo {author} {\bibfnamefont {A.}~\bibnamefont {Poto{\v{c}}nik}}, \bibinfo {author} {\bibfnamefont {A.}~\bibnamefont {Bargerbos}}, \bibinfo {author} {\bibfnamefont {F.~A. Y.~N.}\ \bibnamefont {Schr{\"o}der}}, \bibinfo {author} {\bibfnamefont {S.~A.}\ \bibnamefont {Khan}}, \bibinfo {author} {\bibfnamefont {M.~C.}\ \bibnamefont {Collodo}}, \bibinfo {author} {\bibfnamefont {S.}~\bibnamefont {Gasparinetti}}, \bibinfo {author} {\bibfnamefont {Y.}~\bibnamefont {Salath{\'e}}}, \bibinfo {author} {\bibfnamefont {C.}~\bibnamefont {Creatore}}, \bibinfo {author} {\bibfnamefont {C.}~\bibnamefont {Eichler}}, \bibinfo {author} {\bibfnamefont {H.~E.}\ \bibnamefont {T{\"u}reci}}, \bibinfo {author} {\bibfnamefont {A.~W.}\ \bibnamefont {Chin}},\ and\ \bibinfo {author} {\bibfnamefont {A.}~\bibnamefont {Wallraff}},\ }\bibfield  {title} {\bibinfo {title} {Studying light-harvesting models with superconducting circuits},\ }\href {https://doi.org/10.1038/s41467-018-03312-x} {\bibfield  {journal} {\bibinfo
  {journal} {Nat. Commun.}\ }\textbf {\bibinfo {volume} {9}},\ \bibinfo {pages} {904} (\bibinfo {year} {2018})}\BibitemShut {NoStop}%
\bibitem [{\citenamefont {Rebentrost}\ \emph {et~al.}(2009)\citenamefont {Rebentrost}, \citenamefont {Mohseni}, \citenamefont {Kassal}, \citenamefont {Lloyd},\ and\ \citenamefont {Aspuru-Guzik}}]{Rebentrost2009}%
  \BibitemOpen
  \bibfield  {author} {\bibinfo {author} {\bibfnamefont {P.}~\bibnamefont {Rebentrost}}, \bibinfo {author} {\bibfnamefont {M.}~\bibnamefont {Mohseni}}, \bibinfo {author} {\bibfnamefont {I.}~\bibnamefont {Kassal}}, \bibinfo {author} {\bibfnamefont {S.}~\bibnamefont {Lloyd}},\ and\ \bibinfo {author} {\bibfnamefont {A.}~\bibnamefont {Aspuru-Guzik}},\ }\bibfield  {title} {\bibinfo {title} {Environment-assisted quantum transport},\ }\href {https://doi.org/10.1088/1367-2630/11/3/033003} {\bibfield  {journal} {\bibinfo  {journal} {New J. Phys.}\ }\textbf {\bibinfo {volume} {11}},\ \bibinfo {pages} {033003} (\bibinfo {year} {2009})}\BibitemShut {NoStop}%
\bibitem [{\citenamefont {Jotzu}\ \emph {et~al.}(2014)\citenamefont {Jotzu}, \citenamefont {Messer}, \citenamefont {Desbuquois}, \citenamefont {Lebrat}, \citenamefont {Uehlinger}, \citenamefont {Greif},\ and\ \citenamefont {Esslinger}}]{Jotzu2014}%
  \BibitemOpen
  \bibfield  {author} {\bibinfo {author} {\bibfnamefont {G.}~\bibnamefont {Jotzu}}, \bibinfo {author} {\bibfnamefont {M.}~\bibnamefont {Messer}}, \bibinfo {author} {\bibfnamefont {R.}~\bibnamefont {Desbuquois}}, \bibinfo {author} {\bibfnamefont {M.}~\bibnamefont {Lebrat}}, \bibinfo {author} {\bibfnamefont {T.}~\bibnamefont {Uehlinger}}, \bibinfo {author} {\bibfnamefont {D.}~\bibnamefont {Greif}},\ and\ \bibinfo {author} {\bibfnamefont {T.}~\bibnamefont {Esslinger}},\ }\bibfield  {title} {\bibinfo {title} {Experimental realization of the topological {H}aldane model with ultracold fermions},\ }\href {https://doi.org/10.1038/nature13915} {\bibfield  {journal} {\bibinfo  {journal} {Nature}\ }\textbf {\bibinfo {volume} {515}},\ \bibinfo {pages} {237} (\bibinfo {year} {2014})}\BibitemShut {NoStop}%
\bibitem [{\citenamefont {An}\ \emph {et~al.}(2018{\natexlab{a}})\citenamefont {An}, \citenamefont {Meier},\ and\ \citenamefont {Gadway}}]{PhysRevX.8.031045}%
  \BibitemOpen
  \bibfield  {author} {\bibinfo {author} {\bibfnamefont {F.~A.}\ \bibnamefont {An}}, \bibinfo {author} {\bibfnamefont {E.~J.}\ \bibnamefont {Meier}},\ and\ \bibinfo {author} {\bibfnamefont {B.}~\bibnamefont {Gadway}},\ }\bibfield  {title} {\bibinfo {title} {Engineering a flux-dependent mobility edge in disordered zigzag chains},\ }\href {https://doi.org/10.1103/PhysRevX.8.031045} {\bibfield  {journal} {\bibinfo  {journal} {Phys. Rev. X}\ }\textbf {\bibinfo {volume} {8}},\ \bibinfo {pages} {031045} (\bibinfo {year} {2018}{\natexlab{a}})}\BibitemShut {NoStop}%
\bibitem [{\citenamefont {Zimbor{\'a}s}\ \emph {et~al.}(2013)\citenamefont {Zimbor{\'a}s}, \citenamefont {Faccin}, \citenamefont {K{\'a}d{\'a}r}, \citenamefont {Whitfield}, \citenamefont {Lanyon},\ and\ \citenamefont {Biamonte}}]{Zimboras2013}%
  \BibitemOpen
  \bibfield  {author} {\bibinfo {author} {\bibfnamefont {Z.}~\bibnamefont {Zimbor{\'a}s}}, \bibinfo {author} {\bibfnamefont {M.}~\bibnamefont {Faccin}}, \bibinfo {author} {\bibfnamefont {Z.}~\bibnamefont {K{\'a}d{\'a}r}}, \bibinfo {author} {\bibfnamefont {J.~D.}\ \bibnamefont {Whitfield}}, \bibinfo {author} {\bibfnamefont {B.~P.}\ \bibnamefont {Lanyon}},\ and\ \bibinfo {author} {\bibfnamefont {J.}~\bibnamefont {Biamonte}},\ }\bibfield  {title} {\bibinfo {title} {Quantum transport enhancement by time-reversal symmetry breaking},\ }\href {https://doi.org/10.1038/srep02361} {\bibfield  {journal} {\bibinfo  {journal} {Sci. Rep.}\ }\textbf {\bibinfo {volume} {3}},\ \bibinfo {pages} {2361} (\bibinfo {year} {2013})}\BibitemShut {NoStop}%
\bibitem [{\citenamefont {Lu}\ \emph {et~al.}(2016)\citenamefont {Lu}, \citenamefont {Biamonte}, \citenamefont {Li}, \citenamefont {Li}, \citenamefont {Johnson}, \citenamefont {Bergholm}, \citenamefont {Faccin}, \citenamefont {Zimbor{\'a}s}, \citenamefont {Laflamme}, \citenamefont {Baugh},\ and\ \citenamefont {Lloyd}}]{PhysRevA.93.042302}%
  \BibitemOpen
  \bibfield  {author} {\bibinfo {author} {\bibfnamefont {D.}~\bibnamefont {Lu}}, \bibinfo {author} {\bibfnamefont {J.~D.}\ \bibnamefont {Biamonte}}, \bibinfo {author} {\bibfnamefont {J.}~\bibnamefont {Li}}, \bibinfo {author} {\bibfnamefont {H.}~\bibnamefont {Li}}, \bibinfo {author} {\bibfnamefont {T.~H.}\ \bibnamefont {Johnson}}, \bibinfo {author} {\bibfnamefont {V.}~\bibnamefont {Bergholm}}, \bibinfo {author} {\bibfnamefont {M.}~\bibnamefont {Faccin}}, \bibinfo {author} {\bibfnamefont {Z.}~\bibnamefont {Zimbor{\'a}s}}, \bibinfo {author} {\bibfnamefont {R.}~\bibnamefont {Laflamme}}, \bibinfo {author} {\bibfnamefont {J.}~\bibnamefont {Baugh}},\ and\ \bibinfo {author} {\bibfnamefont {S.}~\bibnamefont {Lloyd}},\ }\bibfield  {title} {\bibinfo {title} {Chiral quantum walks},\ }\href {https://doi.org/10.1103/PhysRevA.93.042302} {\bibfield  {journal} {\bibinfo  {journal} {Phys. Rev. A}\ }\textbf {\bibinfo {volume} {93}},\ \bibinfo {pages} {042302} (\bibinfo {year} {2016})}\BibitemShut {NoStop}%
\bibitem [{\citenamefont {{H}ofstadter}(1976)}]{Hofstadter1976}%
  \BibitemOpen
  \bibfield  {author} {\bibinfo {author} {\bibfnamefont {D.~R.}\ \bibnamefont {{H}ofstadter}},\ }\bibfield  {title} {\bibinfo {title} {Energy levels and wave functions of {B}loch electrons in rational and irrational magnetic fields},\ }\href {https://doi.org/10.1103/PhysRevB.14.2239} {\bibfield  {journal} {\bibinfo  {journal} {Phys. Rev. B}\ }\textbf {\bibinfo {volume} {14}},\ \bibinfo {pages} {2239} (\bibinfo {year} {1976})}\BibitemShut {NoStop}%
\bibitem [{\citenamefont {Jaksch}\ and\ \citenamefont {Zoller}(2003)}]{Jaksch_2003}%
  \BibitemOpen
  \bibfield  {author} {\bibinfo {author} {\bibfnamefont {D.}~\bibnamefont {Jaksch}}\ and\ \bibinfo {author} {\bibfnamefont {P.}~\bibnamefont {Zoller}},\ }\bibfield  {title} {\bibinfo {title} {Creation of effective magnetic fields in optical lattices: the {H}ofstadter butterfly for cold neutral atoms},\ }\href {https://doi.org/10.1088/1367-2630/5/1/356} {\bibfield  {journal} {\bibinfo  {journal} {New J. Phys.}\ }\textbf {\bibinfo {volume} {5}},\ \bibinfo {pages} {56} (\bibinfo {year} {2003})}\BibitemShut {NoStop}%
\bibitem [{\citenamefont {Dalibard}\ \emph {et~al.}(2011)\citenamefont {Dalibard}, \citenamefont {Gerbier}, \citenamefont {Juzeli\={u}nas},\ and\ \citenamefont {\"Ohberg}}]{Dalibard2011}%
  \BibitemOpen
  \bibfield  {author} {\bibinfo {author} {\bibfnamefont {J.}~\bibnamefont {Dalibard}}, \bibinfo {author} {\bibfnamefont {F.}~\bibnamefont {Gerbier}}, \bibinfo {author} {\bibfnamefont {G.}~\bibnamefont {Juzeli\={u}nas}},\ and\ \bibinfo {author} {\bibfnamefont {P.}~\bibnamefont {\"Ohberg}},\ }\bibfield  {title} {\bibinfo {title} {Colloquium: Artificial gauge potentials for neutral atoms},\ }\href {https://doi.org/10.1103/RevModPhys.83.1523} {\bibfield  {journal} {\bibinfo  {journal} {Rev. Mod. Phys.}\ }\textbf {\bibinfo {volume} {83}},\ \bibinfo {pages} {1523} (\bibinfo {year} {2011})}\BibitemShut {NoStop}%
\bibitem [{\citenamefont {Aidelsburger}\ \emph {et~al.}(2013)\citenamefont {Aidelsburger}, \citenamefont {Atala}, \citenamefont {Lohse}, \citenamefont {Barreiro}, \citenamefont {Paredes},\ and\ \citenamefont {Bloch}}]{PhysRevLett.111.185301}%
  \BibitemOpen
  \bibfield  {author} {\bibinfo {author} {\bibfnamefont {M.}~\bibnamefont {Aidelsburger}}, \bibinfo {author} {\bibfnamefont {M.}~\bibnamefont {Atala}}, \bibinfo {author} {\bibfnamefont {M.}~\bibnamefont {Lohse}}, \bibinfo {author} {\bibfnamefont {J.~T.}\ \bibnamefont {Barreiro}}, \bibinfo {author} {\bibfnamefont {B.}~\bibnamefont {Paredes}},\ and\ \bibinfo {author} {\bibfnamefont {I.}~\bibnamefont {Bloch}},\ }\bibfield  {title} {\bibinfo {title} {Realization of the {H}ofstadter hamiltonian with ultracold atoms in optical lattices},\ }\href {https://doi.org/10.1103/PhysRevLett.111.185301} {\bibfield  {journal} {\bibinfo  {journal} {Phys. Rev. Lett.}\ }\textbf {\bibinfo {volume} {111}},\ \bibinfo {pages} {185301} (\bibinfo {year} {2013})}\BibitemShut {NoStop}%
\bibitem [{\citenamefont {Miyake}\ \emph {et~al.}(2013)\citenamefont {Miyake}, \citenamefont {Siviloglou}, \citenamefont {Kennedy}, \citenamefont {Burton},\ and\ \citenamefont {Ketterle}}]{PhysRevLett.111.185302}%
  \BibitemOpen
  \bibfield  {author} {\bibinfo {author} {\bibfnamefont {H.}~\bibnamefont {Miyake}}, \bibinfo {author} {\bibfnamefont {G.~A.}\ \bibnamefont {Siviloglou}}, \bibinfo {author} {\bibfnamefont {C.~J.}\ \bibnamefont {Kennedy}}, \bibinfo {author} {\bibfnamefont {W.~C.}\ \bibnamefont {Burton}},\ and\ \bibinfo {author} {\bibfnamefont {W.}~\bibnamefont {Ketterle}},\ }\bibfield  {title} {\bibinfo {title} {Realizing the {H}arper hamiltonian with laser-assisted tunneling in optical lattices},\ }\href {https://doi.org/10.1103/PhysRevLett.111.185302} {\bibfield  {journal} {\bibinfo  {journal} {Phys. Rev. Lett.}\ }\textbf {\bibinfo {volume} {111}},\ \bibinfo {pages} {185302} (\bibinfo {year} {2013})}\BibitemShut {NoStop}%
\bibitem [{\citenamefont {Atala}\ \emph {et~al.}(2014)\citenamefont {Atala}, \citenamefont {Aidelsburger}, \citenamefont {Lohse}, \citenamefont {Barreiro}, \citenamefont {Paredes},\ and\ \citenamefont {Bloch}}]{Atala2014}%
  \BibitemOpen
  \bibfield  {author} {\bibinfo {author} {\bibfnamefont {M.}~\bibnamefont {Atala}}, \bibinfo {author} {\bibfnamefont {M.}~\bibnamefont {Aidelsburger}}, \bibinfo {author} {\bibfnamefont {M.}~\bibnamefont {Lohse}}, \bibinfo {author} {\bibfnamefont {J.~T.}\ \bibnamefont {Barreiro}}, \bibinfo {author} {\bibfnamefont {B.}~\bibnamefont {Paredes}},\ and\ \bibinfo {author} {\bibfnamefont {I.}~\bibnamefont {Bloch}},\ }\bibfield  {title} {\bibinfo {title} {Observation of chiral currents with ultracold atoms in bosonic ladders},\ }\href@noop {} {\bibfield  {journal} {\bibinfo  {journal} {Nat. Phys.}\ }\textbf {\bibinfo {volume} {10}},\ \bibinfo {pages} {588} (\bibinfo {year} {2014})}\BibitemShut {NoStop}%
\bibitem [{\citenamefont {Mancini}\ \emph {et~al.}(2015)\citenamefont {Mancini}, \citenamefont {Pagano}, \citenamefont {Cappellini}, \citenamefont {Livi}, \citenamefont {Rider}, \citenamefont {Catani}, \citenamefont {Sias}, \citenamefont {Zoller}, \citenamefont {Inguscio}, \citenamefont {Dalmonte},\ and\ \citenamefont {Fallani}}]{Mancini2015}%
  \BibitemOpen
  \bibfield  {author} {\bibinfo {author} {\bibfnamefont {M.}~\bibnamefont {Mancini}}, \bibinfo {author} {\bibfnamefont {G.}~\bibnamefont {Pagano}}, \bibinfo {author} {\bibfnamefont {G.}~\bibnamefont {Cappellini}}, \bibinfo {author} {\bibfnamefont {L.}~\bibnamefont {Livi}}, \bibinfo {author} {\bibfnamefont {M.}~\bibnamefont {Rider}}, \bibinfo {author} {\bibfnamefont {J.}~\bibnamefont {Catani}}, \bibinfo {author} {\bibfnamefont {C.}~\bibnamefont {Sias}}, \bibinfo {author} {\bibfnamefont {P.}~\bibnamefont {Zoller}}, \bibinfo {author} {\bibfnamefont {M.}~\bibnamefont {Inguscio}}, \bibinfo {author} {\bibfnamefont {M.}~\bibnamefont {Dalmonte}},\ and\ \bibinfo {author} {\bibfnamefont {L.}~\bibnamefont {Fallani}},\ }\bibfield  {title} {\bibinfo {title} {Observation of chiral edge states with neutral fermions in synthetic {H}all ribbons},\ }\href {https://doi.org/10.1126/science.aaa8736} {\bibfield  {journal} {\bibinfo  {journal} {Science}\ }\textbf {\bibinfo {volume} {349}},\ \bibinfo {pages} {1510} (\bibinfo {year}
  {2015})}\BibitemShut {NoStop}%
\bibitem [{\citenamefont {An}\ \emph {et~al.}(2017{\natexlab{a}})\citenamefont {An}, \citenamefont {Meier},\ and\ \citenamefont {Gadway}}]{An2017}%
  \BibitemOpen
  \bibfield  {author} {\bibinfo {author} {\bibfnamefont {F.~A.}\ \bibnamefont {An}}, \bibinfo {author} {\bibfnamefont {E.~J.}\ \bibnamefont {Meier}},\ and\ \bibinfo {author} {\bibfnamefont {B.}~\bibnamefont {Gadway}},\ }\bibfield  {title} {\bibinfo {title} {Direct observation of chiral currents and magnetic reflection in atomic flux lattices},\ }\href {https://doi.org/10.1126/sciadv.1602685} {\bibfield  {journal} {\bibinfo  {journal} {Sci. Adv.}\ }\textbf {\bibinfo {volume} {3}},\ \bibinfo {pages} {e1602685} (\bibinfo {year} {2017}{\natexlab{a}})}\BibitemShut {NoStop}%
\bibitem [{\citenamefont {Roushan}\ \emph {et~al.}(2017)\citenamefont {Roushan}, \citenamefont {Neill}, \citenamefont {Megrant}, \citenamefont {Chen}, \citenamefont {Babbush}, \citenamefont {Barends}, \citenamefont {Campbell}, \citenamefont {Chen}, \citenamefont {Chiaro}, \citenamefont {Dunsworth}, \citenamefont {Fowler}, \citenamefont {Jeffrey}, \citenamefont {Kelly}, \citenamefont {Lucero}, \citenamefont {Mutus}, \citenamefont {O'Malley}, \citenamefont {Neeley}, \citenamefont {Quintana}, \citenamefont {Sank}, \citenamefont {Vainsencher}, \citenamefont {Wenner}, \citenamefont {White}, \citenamefont {Kapit}, \citenamefont {Neven},\ and\ \citenamefont {Martinis}}]{Roushan2017}%
  \BibitemOpen
  \bibfield  {author} {\bibinfo {author} {\bibfnamefont {P.}~\bibnamefont {Roushan}}, \bibinfo {author} {\bibfnamefont {C.}~\bibnamefont {Neill}}, \bibinfo {author} {\bibfnamefont {A.}~\bibnamefont {Megrant}}, \bibinfo {author} {\bibfnamefont {Y.}~\bibnamefont {Chen}}, \bibinfo {author} {\bibfnamefont {R.}~\bibnamefont {Babbush}}, \bibinfo {author} {\bibfnamefont {R.}~\bibnamefont {Barends}}, \bibinfo {author} {\bibfnamefont {B.}~\bibnamefont {Campbell}}, \bibinfo {author} {\bibfnamefont {Z.}~\bibnamefont {Chen}}, \bibinfo {author} {\bibfnamefont {B.}~\bibnamefont {Chiaro}}, \bibinfo {author} {\bibfnamefont {A.}~\bibnamefont {Dunsworth}}, \bibinfo {author} {\bibfnamefont {A.}~\bibnamefont {Fowler}}, \bibinfo {author} {\bibfnamefont {E.}~\bibnamefont {Jeffrey}}, \bibinfo {author} {\bibfnamefont {J.}~\bibnamefont {Kelly}}, \bibinfo {author} {\bibfnamefont {E.}~\bibnamefont {Lucero}}, \bibinfo {author} {\bibfnamefont {J.}~\bibnamefont {Mutus}}, \bibinfo {author} {\bibfnamefont {P.~J.~J.}\ \bibnamefont
  {O'Malley}}, \bibinfo {author} {\bibfnamefont {M.}~\bibnamefont {Neeley}}, \bibinfo {author} {\bibfnamefont {C.}~\bibnamefont {Quintana}}, \bibinfo {author} {\bibfnamefont {D.}~\bibnamefont {Sank}}, \bibinfo {author} {\bibfnamefont {A.}~\bibnamefont {Vainsencher}}, \bibinfo {author} {\bibfnamefont {J.}~\bibnamefont {Wenner}}, \bibinfo {author} {\bibfnamefont {T.}~\bibnamefont {White}}, \bibinfo {author} {\bibfnamefont {E.}~\bibnamefont {Kapit}}, \bibinfo {author} {\bibfnamefont {H.}~\bibnamefont {Neven}},\ and\ \bibinfo {author} {\bibfnamefont {J.~M.}\ \bibnamefont {Martinis}},\ }\bibfield  {title} {\bibinfo {title} {Chiral ground-state currents of interacting photons in a synthetic magnetic field},\ }\href {https://doi.org/10.1038/nphys3930} {\bibfield  {journal} {\bibinfo  {journal} {Nat. Phys.}\ }\textbf {\bibinfo {volume} {13}},\ \bibinfo {pages} {146} (\bibinfo {year} {2017})}\BibitemShut {NoStop}%
\bibitem [{\citenamefont {Rosen}\ \emph {et~al.}(2024)\citenamefont {Rosen}, \citenamefont {Muschinske}, \citenamefont {Barrett}, \citenamefont {Chatterjee}, \citenamefont {Hays}, \citenamefont {DeMarco}, \citenamefont {Karamlou}, \citenamefont {Rower}, \citenamefont {Das}, \citenamefont {Kim}, \citenamefont {Niedzielski}, \citenamefont {Schuldt}, \citenamefont {Serniak}, \citenamefont {Schwartz}, \citenamefont {Yoder}, \citenamefont {Grover},\ and\ \citenamefont {Oliver}}]{rosen2024synthetic}%
  \BibitemOpen
  \bibfield  {author} {\bibinfo {author} {\bibfnamefont {I.~T.}\ \bibnamefont {Rosen}}, \bibinfo {author} {\bibfnamefont {S.}~\bibnamefont {Muschinske}}, \bibinfo {author} {\bibfnamefont {C.~N.}\ \bibnamefont {Barrett}}, \bibinfo {author} {\bibfnamefont {A.}~\bibnamefont {Chatterjee}}, \bibinfo {author} {\bibfnamefont {M.}~\bibnamefont {Hays}}, \bibinfo {author} {\bibfnamefont {M.~A.}\ \bibnamefont {DeMarco}}, \bibinfo {author} {\bibfnamefont {A.~H.}\ \bibnamefont {Karamlou}}, \bibinfo {author} {\bibfnamefont {D.~A.}\ \bibnamefont {Rower}}, \bibinfo {author} {\bibfnamefont {R.}~\bibnamefont {Das}}, \bibinfo {author} {\bibfnamefont {D.~K.}\ \bibnamefont {Kim}}, \bibinfo {author} {\bibfnamefont {B.~M.}\ \bibnamefont {Niedzielski}}, \bibinfo {author} {\bibfnamefont {M.}~\bibnamefont {Schuldt}}, \bibinfo {author} {\bibfnamefont {K.}~\bibnamefont {Serniak}}, \bibinfo {author} {\bibfnamefont {M.~E.}\ \bibnamefont {Schwartz}}, \bibinfo {author} {\bibfnamefont {J.~L.}\ \bibnamefont {Yoder}}, \bibinfo {author}
  {\bibfnamefont {J.~A.}\ \bibnamefont {Grover}},\ and\ \bibinfo {author} {\bibfnamefont {W.~D.}\ \bibnamefont {Oliver}},\ }\bibfield  {title} {\bibinfo {title} {A synthetic magnetic vector potential in a 2{D} superconducting qubit array},\ }\href {https://doi.org/10.1038/s41567-024-02661-3} {\bibfield  {journal} {\bibinfo  {journal} {Nat. Phys.}\ }\textbf {\bibinfo {volume} {20}},\ \bibinfo {pages} {1881} (\bibinfo {year} {2024})}\BibitemShut {NoStop}%
\bibitem [{\citenamefont {Fisher}\ \emph {et~al.}(2023)\citenamefont {Fisher}, \citenamefont {Khemani}, \citenamefont {Nahum},\ and\ \citenamefont {Vijay}}]{quantum_brownian_circuits}%
  \BibitemOpen
  \bibfield  {author} {\bibinfo {author} {\bibfnamefont {M.~P.}\ \bibnamefont {Fisher}}, \bibinfo {author} {\bibfnamefont {V.}~\bibnamefont {Khemani}}, \bibinfo {author} {\bibfnamefont {A.}~\bibnamefont {Nahum}},\ and\ \bibinfo {author} {\bibfnamefont {S.}~\bibnamefont {Vijay}},\ }\bibfield  {title} {\bibinfo {title} {Random quantum circuits},\ }\href {https://doi.org/10.1146/annurev-conmatphys-031720-030658} {\bibfield  {journal} {\bibinfo  {journal} {Annu. Rev. Condens. Matter Phys.}\ }\textbf {\bibinfo {volume} {14}},\ \bibinfo {pages} {335} (\bibinfo {year} {2023})}\BibitemShut {NoStop}%
\bibitem [{\citenamefont {Jacobs}\ and\ \citenamefont {Steck}(2006)}]{Jacobs01092006}%
  \BibitemOpen
  \bibfield  {author} {\bibinfo {author} {\bibfnamefont {K.}~\bibnamefont {Jacobs}}\ and\ \bibinfo {author} {\bibfnamefont {D.~A.}\ \bibnamefont {Steck}},\ }\bibfield  {title} {\bibinfo {title} {A straightforward introduction to continuous quantum measurement},\ }\href {https://doi.org/10.1080/00107510601101934} {\bibfield  {journal} {\bibinfo  {journal} {Contemp. Phys.}\ }\textbf {\bibinfo {volume} {47}},\ \bibinfo {pages} {279} (\bibinfo {year} {2006})}\BibitemShut {NoStop}%
\bibitem [{\citenamefont {Lindblad}(1976)}]{lindblad1976generators}%
  \BibitemOpen
  \bibfield  {author} {\bibinfo {author} {\bibfnamefont {G.}~\bibnamefont {Lindblad}},\ }\bibfield  {title} {\bibinfo {title} {On the generators of quantum dynamical semigroups},\ }\href {https://doi.org/10.1007/BF01608499} {\bibfield  {journal} {\bibinfo  {journal} {Commun. Math. Phys.}\ }\textbf {\bibinfo {volume} {48}},\ \bibinfo {pages} {119} (\bibinfo {year} {1976})}\BibitemShut {NoStop}%
\bibitem [{\citenamefont {Gorini}\ \emph {et~al.}(1976)\citenamefont {Gorini}, \citenamefont {Kossakowski},\ and\ \citenamefont {Sudarshan}}]{gorini1976completely}%
  \BibitemOpen
  \bibfield  {author} {\bibinfo {author} {\bibfnamefont {V.}~\bibnamefont {Gorini}}, \bibinfo {author} {\bibfnamefont {A.}~\bibnamefont {Kossakowski}},\ and\ \bibinfo {author} {\bibfnamefont {E.~C.~G.}\ \bibnamefont {Sudarshan}},\ }\bibfield  {title} {\bibinfo {title} {Completely positive dynamical semigroups of {$N$}-level systems},\ }\href {https://doi.org/10.1063/1.522979} {\bibfield  {journal} {\bibinfo  {journal} {J. Math. Phys.}\ }\textbf {\bibinfo {volume} {17}},\ \bibinfo {pages} {821} (\bibinfo {year} {1976})}\BibitemShut {NoStop}%
\bibitem [{\citenamefont {Barthel}\ and\ \citenamefont {Zhang}(2022)}]{Barthel_2022}%
  \BibitemOpen
  \bibfield  {author} {\bibinfo {author} {\bibfnamefont {T.}~\bibnamefont {Barthel}}\ and\ \bibinfo {author} {\bibfnamefont {Y.}~\bibnamefont {Zhang}},\ }\bibfield  {title} {\bibinfo {title} {Solving quasi-free and quadratic {L}indblad master equations for open fermionic and bosonic systems},\ }\href {https://doi.org/10.1088/1742-5468/ac8e5c} {\bibfield  {journal} {\bibinfo  {journal} {J. Stat. Mech.: Theory Exp.}\ }\textbf {\bibinfo {volume} {2022}}\bibinfo  {number} { (11)},\ \bibinfo {pages} {113101}}\BibitemShut {NoStop}%
\bibitem [{\citenamefont {Zhang}\ and\ \citenamefont {Barthel}(2022)}]{PhysRevLett.129.120401}%
  \BibitemOpen
\bibfield  {number} {  }\bibfield  {author} {\bibinfo {author} {\bibfnamefont {Y.}~\bibnamefont {Zhang}}\ and\ \bibinfo {author} {\bibfnamefont {T.}~\bibnamefont {Barthel}},\ }\bibfield  {title} {\bibinfo {title} {Criticality and phase classification for quadratic open quantum many-body systems},\ }\href {https://doi.org/10.1103/PhysRevLett.129.120401} {\bibfield  {journal} {\bibinfo  {journal} {Phys. Rev. Lett.}\ }\textbf {\bibinfo {volume} {129}},\ \bibinfo {pages} {120401} (\bibinfo {year} {2022})}\BibitemShut {NoStop}%
\bibitem [{\citenamefont {Chaduteau}\ \emph {et~al.}(2026{\natexlab{a}})\citenamefont {Chaduteau}, \citenamefont {Lee}, \citenamefont {Schindler},\ and\ \citenamefont {Prem}}]{chaduteau2026}%
  \BibitemOpen
  \bibfield  {author} {\bibinfo {author} {\bibfnamefont {A.}~\bibnamefont {Chaduteau}}, \bibinfo {author} {\bibfnamefont {D.~K.~K.}\ \bibnamefont {Lee}}, \bibinfo {author} {\bibfnamefont {F.}~\bibnamefont {Schindler}},\ and\ \bibinfo {author} {\bibfnamefont {A.}~\bibnamefont {Prem}},\ }\href {https://arxiv.org/abs/2607.07801} {\bibinfo {title} {Topology from decoherence}} (\bibinfo {year} {2026}{\natexlab{a}}),\ \Eprint {https://arxiv.org/abs/2607.07801} {arXiv:2607.07801 [quant-ph]} \BibitemShut {NoStop}%
\bibitem [{\citenamefont {{\v{Z}}unkovi{\v{c}}}(2014)}]{Zunkovic_2014}%
  \BibitemOpen
  \bibfield  {author} {\bibinfo {author} {\bibfnamefont {B.}~\bibnamefont {{\v{Z}}unkovi{\v{c}}}},\ }\bibfield  {title} {\bibinfo {title} {Closed hierarchy of correlations in {M}arkovian open quantum systems},\ }\href {https://doi.org/10.1088/1367-2630/16/1/013042} {\bibfield  {journal} {\bibinfo  {journal} {New J. Phys.}\ }\textbf {\bibinfo {volume} {16}},\ \bibinfo {pages} {013042} (\bibinfo {year} {2014})}\BibitemShut {NoStop}%
\bibitem [{\citenamefont {Cercignani}(1988)}]{Cercignani1988}%
  \BibitemOpen
  \bibfield  {author} {\bibinfo {author} {\bibfnamefont {C.}~\bibnamefont {Cercignani}},\ }\href {https://doi.org/10.1007/978-1-4612-1039-9} {\emph {\bibinfo {title} {The {B}oltzmann Equation and Its Applications}}},\ \bibinfo {series} {Applied Mathematical Sciences}, Vol.~\bibinfo {volume} {67}\ (\bibinfo  {publisher} {Springer},\ \bibinfo {address} {New York},\ \bibinfo {year} {1988})\BibitemShut {NoStop}%
\bibitem [{\citenamefont {H{\"a}nggi}\ and\ \citenamefont {Marchesoni}(2009)}]{RevModPhys.81.387}%
  \BibitemOpen
  \bibfield  {author} {\bibinfo {author} {\bibfnamefont {P.}~\bibnamefont {H{\"a}nggi}}\ and\ \bibinfo {author} {\bibfnamefont {F.}~\bibnamefont {Marchesoni}},\ }\bibfield  {title} {\bibinfo {title} {Artificial {B}rownian motors: Controlling transport on the nanoscale},\ }\href {https://doi.org/10.1103/RevModPhys.81.387} {\bibfield  {journal} {\bibinfo  {journal} {Rev. Mod. Phys.}\ }\textbf {\bibinfo {volume} {81}},\ \bibinfo {pages} {387} (\bibinfo {year} {2009})}\BibitemShut {NoStop}%
\bibitem [{\citenamefont {Reimann}(2002)}]{REIMANN200257}%
  \BibitemOpen
  \bibfield  {author} {\bibinfo {author} {\bibfnamefont {P.}~\bibnamefont {Reimann}},\ }\bibfield  {title} {\bibinfo {title} {{B}rownian motors: noisy transport far from equilibrium},\ }\href {https://doi.org/10.1016/S0370-1573(01)00081-3} {\bibfield  {journal} {\bibinfo  {journal} {Phys. Rep.}\ }\textbf {\bibinfo {volume} {361}},\ \bibinfo {pages} {57} (\bibinfo {year} {2002})}\BibitemShut {NoStop}%
\bibitem [{\citenamefont {An}\ \emph {et~al.}(2017{\natexlab{b}})\citenamefont {An}, \citenamefont {Meier},\ and\ \citenamefont {Gadway}}]{An2017NC}%
  \BibitemOpen
  \bibfield  {author} {\bibinfo {author} {\bibfnamefont {F.~A.}\ \bibnamefont {An}}, \bibinfo {author} {\bibfnamefont {E.~J.}\ \bibnamefont {Meier}},\ and\ \bibinfo {author} {\bibfnamefont {B.}~\bibnamefont {Gadway}},\ }\bibfield  {title} {\bibinfo {title} {Diffusive and arrested transport of atoms under tailored disorder},\ }\href {https://doi.org/10.1038/s41467-017-00387-w} {\bibfield  {journal} {\bibinfo  {journal} {Nat. Commun.}\ }\textbf {\bibinfo {volume} {8}},\ \bibinfo {pages} {325} (\bibinfo {year} {2017}{\natexlab{b}})}\BibitemShut {NoStop}%
\bibitem [{\citenamefont {An}\ \emph {et~al.}(2018{\natexlab{b}})\citenamefont {An}, \citenamefont {Meier}, \citenamefont {Ang'ong'a},\ and\ \citenamefont {Gadway}}]{PhysRevLett.120.040407}%
  \BibitemOpen
  \bibfield  {author} {\bibinfo {author} {\bibfnamefont {F.~A.}\ \bibnamefont {An}}, \bibinfo {author} {\bibfnamefont {E.~J.}\ \bibnamefont {Meier}}, \bibinfo {author} {\bibfnamefont {J.}~\bibnamefont {Ang'ong'a}},\ and\ \bibinfo {author} {\bibfnamefont {B.}~\bibnamefont {Gadway}},\ }\bibfield  {title} {\bibinfo {title} {Correlated dynamics in a synthetic lattice of momentum states},\ }\href {https://doi.org/10.1103/PhysRevLett.120.040407} {\bibfield  {journal} {\bibinfo  {journal} {Phys. Rev. Lett.}\ }\textbf {\bibinfo {volume} {120}},\ \bibinfo {pages} {040407} (\bibinfo {year} {2018}{\natexlab{b}})}\BibitemShut {NoStop}%
\bibitem [{\citenamefont {Liang}\ \emph {et~al.}(2022)\citenamefont {Liang}, \citenamefont {Xie}, \citenamefont {Dong}, \citenamefont {Li}, \citenamefont {Li}, \citenamefont {Gadway}, \citenamefont {Yi},\ and\ \citenamefont {Yan}}]{PhysRevLett.129.070401}%
  \BibitemOpen
  \bibfield  {author} {\bibinfo {author} {\bibfnamefont {Q.}~\bibnamefont {Liang}}, \bibinfo {author} {\bibfnamefont {D.}~\bibnamefont {Xie}}, \bibinfo {author} {\bibfnamefont {Z.}~\bibnamefont {Dong}}, \bibinfo {author} {\bibfnamefont {H.}~\bibnamefont {Li}}, \bibinfo {author} {\bibfnamefont {H.}~\bibnamefont {Li}}, \bibinfo {author} {\bibfnamefont {B.}~\bibnamefont {Gadway}}, \bibinfo {author} {\bibfnamefont {W.}~\bibnamefont {Yi}},\ and\ \bibinfo {author} {\bibfnamefont {B.}~\bibnamefont {Yan}},\ }\bibfield  {title} {\bibinfo {title} {Dynamic signatures of non-{H}ermitian skin effect and topology in ultracold atoms},\ }\href {https://doi.org/10.1103/PhysRevLett.129.070401} {\bibfield  {journal} {\bibinfo  {journal} {Phys. Rev. Lett.}\ }\textbf {\bibinfo {volume} {129}},\ \bibinfo {pages} {070401} (\bibinfo {year} {2022})}\BibitemShut {NoStop}%
\bibitem [{\citenamefont {Pichler}\ \emph {et~al.}(2010)\citenamefont {Pichler}, \citenamefont {Daley},\ and\ \citenamefont {Zoller}}]{Pichler2010}%
  \BibitemOpen
  \bibfield  {author} {\bibinfo {author} {\bibfnamefont {H.}~\bibnamefont {Pichler}}, \bibinfo {author} {\bibfnamefont {A.~J.}\ \bibnamefont {Daley}},\ and\ \bibinfo {author} {\bibfnamefont {P.}~\bibnamefont {Zoller}},\ }\bibfield  {title} {\bibinfo {title} {Nonequilibrium dynamics of bosonic atoms in optical lattices: Decoherence of many-body states due to spontaneous emission},\ }\href {https://doi.org/10.1103/PhysRevA.82.063605} {\bibfield  {journal} {\bibinfo  {journal} {Phys. Rev. A}\ }\textbf {\bibinfo {volume} {82}},\ \bibinfo {pages} {063605} (\bibinfo {year} {2010})}\BibitemShut {NoStop}%
\bibitem [{\citenamefont {Bouganne}\ \emph {et~al.}(2020)\citenamefont {Bouganne}, \citenamefont {Bosch~Aguilera}, \citenamefont {Ghermaoui}, \citenamefont {Beugnon},\ and\ \citenamefont {Gerbier}}]{bouganne2020anomalous}%
  \BibitemOpen
  \bibfield  {author} {\bibinfo {author} {\bibfnamefont {R.}~\bibnamefont {Bouganne}}, \bibinfo {author} {\bibfnamefont {M.}~\bibnamefont {Bosch~Aguilera}}, \bibinfo {author} {\bibfnamefont {A.}~\bibnamefont {Ghermaoui}}, \bibinfo {author} {\bibfnamefont {J.}~\bibnamefont {Beugnon}},\ and\ \bibinfo {author} {\bibfnamefont {F.}~\bibnamefont {Gerbier}},\ }\bibfield  {title} {\bibinfo {title} {Anomalous decay of coherence in a dissipative many-body system},\ }\href {https://doi.org/10.1038/s41567-019-0678-2} {\bibfield  {journal} {\bibinfo  {journal} {Nat. Phys.}\ }\textbf {\bibinfo {volume} {16}},\ \bibinfo {pages} {21} (\bibinfo {year} {2020})}\BibitemShut {NoStop}%
\bibitem [{\citenamefont {Patil}\ \emph {et~al.}(2015)\citenamefont {Patil}, \citenamefont {Chakram},\ and\ \citenamefont {Vengalattore}}]{PhysRevLett.115.140402}%
  \BibitemOpen
  \bibfield  {author} {\bibinfo {author} {\bibfnamefont {Y.~S.}\ \bibnamefont {Patil}}, \bibinfo {author} {\bibfnamefont {S.}~\bibnamefont {Chakram}},\ and\ \bibinfo {author} {\bibfnamefont {M.}~\bibnamefont {Vengalattore}},\ }\bibfield  {title} {\bibinfo {title} {Measurement-induced localization of an ultracold lattice gas},\ }\href {https://doi.org/10.1103/PhysRevLett.115.140402} {\bibfield  {journal} {\bibinfo  {journal} {Phys. Rev. Lett.}\ }\textbf {\bibinfo {volume} {115}},\ \bibinfo {pages} {140402} (\bibinfo {year} {2015})}\BibitemShut {NoStop}%
\bibitem [{\citenamefont {L{\"u}schen}\ \emph {et~al.}(2017)\citenamefont {L{\"u}schen}, \citenamefont {Bordia}, \citenamefont {Hodgman}, \citenamefont {Schreiber}, \citenamefont {Sarkar}, \citenamefont {Daley}, \citenamefont {Fischer}, \citenamefont {Altman}, \citenamefont {Bloch},\ and\ \citenamefont {Schneider}}]{PhysRevX.7.011034}%
  \BibitemOpen
  \bibfield  {author} {\bibinfo {author} {\bibfnamefont {H.~P.}\ \bibnamefont {L{\"u}schen}}, \bibinfo {author} {\bibfnamefont {P.}~\bibnamefont {Bordia}}, \bibinfo {author} {\bibfnamefont {S.~S.}\ \bibnamefont {Hodgman}}, \bibinfo {author} {\bibfnamefont {M.}~\bibnamefont {Schreiber}}, \bibinfo {author} {\bibfnamefont {S.}~\bibnamefont {Sarkar}}, \bibinfo {author} {\bibfnamefont {A.~J.}\ \bibnamefont {Daley}}, \bibinfo {author} {\bibfnamefont {M.~H.}\ \bibnamefont {Fischer}}, \bibinfo {author} {\bibfnamefont {E.}~\bibnamefont {Altman}}, \bibinfo {author} {\bibfnamefont {I.}~\bibnamefont {Bloch}},\ and\ \bibinfo {author} {\bibfnamefont {U.}~\bibnamefont {Schneider}},\ }\bibfield  {title} {\bibinfo {title} {Signatures of many-body localization in a controlled open quantum system},\ }\href {https://doi.org/10.1103/PhysRevX.7.011034} {\bibfield  {journal} {\bibinfo  {journal} {Phys. Rev. X}\ }\textbf {\bibinfo {volume} {7}},\ \bibinfo {pages} {011034} (\bibinfo {year} {2017})}\BibitemShut {NoStop}%
\bibitem [{\citenamefont {Ashida}\ \emph {et~al.}(2020)\citenamefont {Ashida}, \citenamefont {Gong},\ and\ \citenamefont {Ueda}}]{ashida2020review}%
  \BibitemOpen
  \bibfield  {author} {\bibinfo {author} {\bibfnamefont {Y.}~\bibnamefont {Ashida}}, \bibinfo {author} {\bibfnamefont {Z.}~\bibnamefont {Gong}},\ and\ \bibinfo {author} {\bibfnamefont {M.}~\bibnamefont {Ueda}},\ }\bibfield  {title} {\bibinfo {title} {{N}on-{H}ermitian physics},\ }\href {https://doi.org/10.1080/00018732.2021.1876991} {\bibfield  {journal} {\bibinfo  {journal} {Adv. Phys.}\ }\textbf {\bibinfo {volume} {69}},\ \bibinfo {pages} {249} (\bibinfo {year} {2020})}\BibitemShut {NoStop}%
\bibitem [{\citenamefont {Bergholtz}\ \emph {et~al.}(2021)\citenamefont {Bergholtz}, \citenamefont {Budich},\ and\ \citenamefont {Kunst}}]{Bergholtz2021}%
  \BibitemOpen
  \bibfield  {author} {\bibinfo {author} {\bibfnamefont {E.~J.}\ \bibnamefont {Bergholtz}}, \bibinfo {author} {\bibfnamefont {J.~C.}\ \bibnamefont {Budich}},\ and\ \bibinfo {author} {\bibfnamefont {F.~K.}\ \bibnamefont {Kunst}},\ }\bibfield  {title} {\bibinfo {title} {Exceptional topology of non-{H}ermitian systems},\ }\href {https://doi.org/10.1103/RevModPhys.93.015005} {\bibfield  {journal} {\bibinfo  {journal} {Rev. Mod. Phys.}\ }\textbf {\bibinfo {volume} {93}},\ \bibinfo {pages} {015005} (\bibinfo {year} {2021})}\BibitemShut {NoStop}%
\bibitem [{\citenamefont {Gong}\ \emph {et~al.}(2018)\citenamefont {Gong}, \citenamefont {Ashida}, \citenamefont {Kawabata}, \citenamefont {Takasan}, \citenamefont {Higashikawa},\ and\ \citenamefont {Ueda}}]{PhysRevX.8.031079}%
  \BibitemOpen
  \bibfield  {author} {\bibinfo {author} {\bibfnamefont {Z.}~\bibnamefont {Gong}}, \bibinfo {author} {\bibfnamefont {Y.}~\bibnamefont {Ashida}}, \bibinfo {author} {\bibfnamefont {K.}~\bibnamefont {Kawabata}}, \bibinfo {author} {\bibfnamefont {K.}~\bibnamefont {Takasan}}, \bibinfo {author} {\bibfnamefont {S.}~\bibnamefont {Higashikawa}},\ and\ \bibinfo {author} {\bibfnamefont {M.}~\bibnamefont {Ueda}},\ }\bibfield  {title} {\bibinfo {title} {Topological phases of non-{H}ermitian systems},\ }\href {https://doi.org/10.1103/PhysRevX.8.031079} {\bibfield  {journal} {\bibinfo  {journal} {Phys. Rev. X}\ }\textbf {\bibinfo {volume} {8}},\ \bibinfo {pages} {031079} (\bibinfo {year} {2018})}\BibitemShut {NoStop}%
\bibitem [{\citenamefont {Kawabata}\ \emph {et~al.}(2019)\citenamefont {Kawabata}, \citenamefont {Shiozaki}, \citenamefont {Ueda},\ and\ \citenamefont {Sato}}]{kawabata_38_fold_way}%
  \BibitemOpen
  \bibfield  {author} {\bibinfo {author} {\bibfnamefont {K.}~\bibnamefont {Kawabata}}, \bibinfo {author} {\bibfnamefont {K.}~\bibnamefont {Shiozaki}}, \bibinfo {author} {\bibfnamefont {M.}~\bibnamefont {Ueda}},\ and\ \bibinfo {author} {\bibfnamefont {M.}~\bibnamefont {Sato}},\ }\bibfield  {title} {\bibinfo {title} {Symmetry and topology in {N}on-{H}ermitian physics},\ }\href {https://doi.org/10.1103/PhysRevX.9.041015} {\bibfield  {journal} {\bibinfo  {journal} {Phys. Rev. X}\ }\textbf {\bibinfo {volume} {9}},\ \bibinfo {pages} {041015} (\bibinfo {year} {2019})}\BibitemShut {NoStop}%
\bibitem [{\citenamefont {Lieu}\ \emph {et~al.}(2020)\citenamefont {Lieu}, \citenamefont {McGinley},\ and\ \citenamefont {Cooper}}]{lieu2020tenfold}%
  \BibitemOpen
  \bibfield  {author} {\bibinfo {author} {\bibfnamefont {S.}~\bibnamefont {Lieu}}, \bibinfo {author} {\bibfnamefont {M.}~\bibnamefont {McGinley}},\ and\ \bibinfo {author} {\bibfnamefont {N.~R.}\ \bibnamefont {Cooper}},\ }\bibfield  {title} {\bibinfo {title} {Tenfold way for quadratic {L}indbladians},\ }\href {https://doi.org/10.1103/PhysRevLett.124.040401} {\bibfield  {journal} {\bibinfo  {journal} {Phys. Rev. Lett.}\ }\textbf {\bibinfo {volume} {124}},\ \bibinfo {pages} {040401} (\bibinfo {year} {2020})}\BibitemShut {NoStop}%
\bibitem [{\citenamefont {Chaduteau}\ \emph {et~al.}(2026{\natexlab{b}})\citenamefont {Chaduteau}, \citenamefont {Lee},\ and\ \citenamefont {Schindler}}]{Chaduteau_2025}%
  \BibitemOpen
  \bibfield  {author} {\bibinfo {author} {\bibfnamefont {A.}~\bibnamefont {Chaduteau}}, \bibinfo {author} {\bibfnamefont {D.~K.~K.}\ \bibnamefont {Lee}},\ and\ \bibinfo {author} {\bibfnamefont {F.}~\bibnamefont {Schindler}},\ }\bibfield  {title} {\bibinfo {title} {{L}indbladian versus postselected {N}on-{H}ermitian topology},\ }\href {https://doi.org/10.1103/ljvt-w6hw} {\bibfield  {journal} {\bibinfo  {journal} {Phys. Rev. Lett.}\ }\textbf {\bibinfo {volume} {136}},\ \bibinfo {pages} {016603} (\bibinfo {year} {2026}{\natexlab{b}})}\BibitemShut {NoStop}%
\bibitem [{\citenamefont {Song}\ \emph {et~al.}(2019)\citenamefont {Song}, \citenamefont {Yao},\ and\ \citenamefont {Wang}}]{PhysRevLett.123.170401}%
  \BibitemOpen
  \bibfield  {author} {\bibinfo {author} {\bibfnamefont {F.}~\bibnamefont {Song}}, \bibinfo {author} {\bibfnamefont {S.}~\bibnamefont {Yao}},\ and\ \bibinfo {author} {\bibfnamefont {Z.}~\bibnamefont {Wang}},\ }\bibfield  {title} {\bibinfo {title} {{N}on-{H}ermitian skin effect and chiral damping in open quantum systems},\ }\href {https://doi.org/10.1103/PhysRevLett.123.170401} {\bibfield  {journal} {\bibinfo  {journal} {Phys. Rev. Lett.}\ }\textbf {\bibinfo {volume} {123}},\ \bibinfo {pages} {170401} (\bibinfo {year} {2019})}\BibitemShut {NoStop}%
\bibitem [{\citenamefont {Haga}\ \emph {et~al.}(2021)\citenamefont {Haga}, \citenamefont {Nakagawa}, \citenamefont {Hamazaki},\ and\ \citenamefont {Ueda}}]{PhysRevLett.127.070402}%
  \BibitemOpen
  \bibfield  {author} {\bibinfo {author} {\bibfnamefont {T.}~\bibnamefont {Haga}}, \bibinfo {author} {\bibfnamefont {M.}~\bibnamefont {Nakagawa}}, \bibinfo {author} {\bibfnamefont {R.}~\bibnamefont {Hamazaki}},\ and\ \bibinfo {author} {\bibfnamefont {M.}~\bibnamefont {Ueda}},\ }\bibfield  {title} {\bibinfo {title} {{L}iouvillian skin effect: Slowing down of relaxation processes without gap closing},\ }\href {https://doi.org/10.1103/PhysRevLett.127.070402} {\bibfield  {journal} {\bibinfo  {journal} {Phys. Rev. Lett.}\ }\textbf {\bibinfo {volume} {127}},\ \bibinfo {pages} {070402} (\bibinfo {year} {2021})}\BibitemShut {NoStop}%
\bibitem [{\citenamefont {Hatano}\ and\ \citenamefont {Nelson}(1996)}]{original_hatano_nelson}%
  \BibitemOpen
  \bibfield  {author} {\bibinfo {author} {\bibfnamefont {N.}~\bibnamefont {Hatano}}\ and\ \bibinfo {author} {\bibfnamefont {D.~R.}\ \bibnamefont {Nelson}},\ }\bibfield  {title} {\bibinfo {title} {Localization transitions in {N}on-{H}ermitian quantum mechanics},\ }\href {https://doi.org/10.1103/PhysRevLett.77.570} {\bibfield  {journal} {\bibinfo  {journal} {Phys. Rev. Lett.}\ }\textbf {\bibinfo {volume} {77}},\ \bibinfo {pages} {570} (\bibinfo {year} {1996})}\BibitemShut {NoStop}%
\bibitem [{\citenamefont {Yao}\ and\ \citenamefont {Wang}(2018)}]{PhysRevLett.121.086803}%
  \BibitemOpen
  \bibfield  {author} {\bibinfo {author} {\bibfnamefont {S.}~\bibnamefont {Yao}}\ and\ \bibinfo {author} {\bibfnamefont {Z.}~\bibnamefont {Wang}},\ }\bibfield  {title} {\bibinfo {title} {Edge states and topological invariants of non-{H}ermitian systems},\ }\href {https://doi.org/10.1103/PhysRevLett.121.086803} {\bibfield  {journal} {\bibinfo  {journal} {Phys. Rev. Lett.}\ }\textbf {\bibinfo {volume} {121}},\ \bibinfo {pages} {086803} (\bibinfo {year} {2018})}\BibitemShut {NoStop}%
\bibitem [{\citenamefont {Okuma}\ \emph {et~al.}(2020)\citenamefont {Okuma}, \citenamefont {Kawabata}, \citenamefont {Shiozaki},\ and\ \citenamefont {Sato}}]{PhysRevLett.124.086801}%
  \BibitemOpen
  \bibfield  {author} {\bibinfo {author} {\bibfnamefont {N.}~\bibnamefont {Okuma}}, \bibinfo {author} {\bibfnamefont {K.}~\bibnamefont {Kawabata}}, \bibinfo {author} {\bibfnamefont {K.}~\bibnamefont {Shiozaki}},\ and\ \bibinfo {author} {\bibfnamefont {M.}~\bibnamefont {Sato}},\ }\bibfield  {title} {\bibinfo {title} {Topological origin of {N}on-{H}ermitian skin effects},\ }\href {https://doi.org/10.1103/PhysRevLett.124.086801} {\bibfield  {journal} {\bibinfo  {journal} {Phys. Rev. Lett.}\ }\textbf {\bibinfo {volume} {124}},\ \bibinfo {pages} {086801} (\bibinfo {year} {2020})}\BibitemShut {NoStop}%
\bibitem [{\citenamefont {Zhang}\ \emph {et~al.}(2020)\citenamefont {Zhang}, \citenamefont {Yang},\ and\ \citenamefont {Fang}}]{PhysRevLett.125.126402}%
  \BibitemOpen
  \bibfield  {author} {\bibinfo {author} {\bibfnamefont {K.}~\bibnamefont {Zhang}}, \bibinfo {author} {\bibfnamefont {Z.}~\bibnamefont {Yang}},\ and\ \bibinfo {author} {\bibfnamefont {C.}~\bibnamefont {Fang}},\ }\bibfield  {title} {\bibinfo {title} {Correspondence between winding numbers and skin modes in non-{H}ermitian systems},\ }\href {https://doi.org/10.1103/PhysRevLett.125.126402} {\bibfield  {journal} {\bibinfo  {journal} {Phys. Rev. Lett.}\ }\textbf {\bibinfo {volume} {125}},\ \bibinfo {pages} {126402} (\bibinfo {year} {2020})}\BibitemShut {NoStop}%
\bibitem [{\citenamefont {Xiao}\ \emph {et~al.}(2020)\citenamefont {Xiao}, \citenamefont {Deng}, \citenamefont {Wang}, \citenamefont {Zhu}, \citenamefont {Wang}, \citenamefont {Yi},\ and\ \citenamefont {Xue}}]{Xiao2020}%
  \BibitemOpen
  \bibfield  {author} {\bibinfo {author} {\bibfnamefont {L.}~\bibnamefont {Xiao}}, \bibinfo {author} {\bibfnamefont {T.}~\bibnamefont {Deng}}, \bibinfo {author} {\bibfnamefont {K.}~\bibnamefont {Wang}}, \bibinfo {author} {\bibfnamefont {G.}~\bibnamefont {Zhu}}, \bibinfo {author} {\bibfnamefont {Z.}~\bibnamefont {Wang}}, \bibinfo {author} {\bibfnamefont {W.}~\bibnamefont {Yi}},\ and\ \bibinfo {author} {\bibfnamefont {P.}~\bibnamefont {Xue}},\ }\bibfield  {title} {\bibinfo {title} {{N}on-{H}ermitian bulk--boundary correspondence in quantum dynamics},\ }\href {https://doi.org/10.1038/s41567-020-0836-6} {\bibfield  {journal} {\bibinfo  {journal} {Nat. Phys.}\ }\textbf {\bibinfo {volume} {16}},\ \bibinfo {pages} {761} (\bibinfo {year} {2020})}\BibitemShut {NoStop}%
\bibitem [{\citenamefont {Weidemann}\ \emph {et~al.}(2020)\citenamefont {Weidemann}, \citenamefont {Kremer}, \citenamefont {Helbig}, \citenamefont {Hofmann}, \citenamefont {Stegmaier}, \citenamefont {Greiter}, \citenamefont {Thomale},\ and\ \citenamefont {Szameit}}]{Weidemann2020}%
  \BibitemOpen
  \bibfield  {author} {\bibinfo {author} {\bibfnamefont {S.}~\bibnamefont {Weidemann}}, \bibinfo {author} {\bibfnamefont {M.}~\bibnamefont {Kremer}}, \bibinfo {author} {\bibfnamefont {T.}~\bibnamefont {Helbig}}, \bibinfo {author} {\bibfnamefont {T.}~\bibnamefont {Hofmann}}, \bibinfo {author} {\bibfnamefont {A.}~\bibnamefont {Stegmaier}}, \bibinfo {author} {\bibfnamefont {M.}~\bibnamefont {Greiter}}, \bibinfo {author} {\bibfnamefont {R.}~\bibnamefont {Thomale}},\ and\ \bibinfo {author} {\bibfnamefont {A.}~\bibnamefont {Szameit}},\ }\bibfield  {title} {\bibinfo {title} {Topological funneling of light},\ }\href {https://doi.org/10.1126/science.aaz8727} {\bibfield  {journal} {\bibinfo  {journal} {Science}\ }\textbf {\bibinfo {volume} {368}},\ \bibinfo {pages} {311} (\bibinfo {year} {2020})}\BibitemShut {NoStop}%
\bibitem [{\citenamefont {Shen}\ \emph {et~al.}(2025)\citenamefont {Shen}, \citenamefont {Chen}, \citenamefont {Yang},\ and\ \citenamefont {Lee}}]{Shen2025}%
  \BibitemOpen
  \bibfield  {author} {\bibinfo {author} {\bibfnamefont {R.}~\bibnamefont {Shen}}, \bibinfo {author} {\bibfnamefont {T.}~\bibnamefont {Chen}}, \bibinfo {author} {\bibfnamefont {B.}~\bibnamefont {Yang}},\ and\ \bibinfo {author} {\bibfnamefont {C.~H.}\ \bibnamefont {Lee}},\ }\bibfield  {title} {\bibinfo {title} {Observation of the non-{H}ermitian skin effect and {F}ermi skin on a digital quantum computer},\ }\href {https://doi.org/10.1038/s41467-025-55953-4} {\bibfield  {journal} {\bibinfo  {journal} {Nat. Commun.}\ }\textbf {\bibinfo {volume} {16}},\ \bibinfo {pages} {1340} (\bibinfo {year} {2025})}\BibitemShut {NoStop}%
\bibitem [{\citenamefont {McDonald}\ and\ \citenamefont {Clerk}(2023)}]{mcdonald_big_thirdqtsn_paper}%
  \BibitemOpen
  \bibfield  {author} {\bibinfo {author} {\bibfnamefont {A.}~\bibnamefont {McDonald}}\ and\ \bibinfo {author} {\bibfnamefont {A.~A.}\ \bibnamefont {Clerk}},\ }\bibfield  {title} {\bibinfo {title} {Third quantization of open quantum systems: Dissipative symmetries and connections to phase-space and {K}eldysh field-theory formulations},\ }\href {https://doi.org/10.1103/PhysRevResearch.5.033107} {\bibfield  {journal} {\bibinfo  {journal} {Phys. Rev. Res.}\ }\textbf {\bibinfo {volume} {5}},\ \bibinfo {pages} {033107} (\bibinfo {year} {2023})}\BibitemShut {NoStop}%
\bibitem [{\citenamefont {Yang}\ \emph {et~al.}(2022)\citenamefont {Yang}, \citenamefont {Jiang},\ and\ \citenamefont {Bergholtz}}]{yang2022lse}%
  \BibitemOpen
  \bibfield  {author} {\bibinfo {author} {\bibfnamefont {F.}~\bibnamefont {Yang}}, \bibinfo {author} {\bibfnamefont {Q.-D.}\ \bibnamefont {Jiang}},\ and\ \bibinfo {author} {\bibfnamefont {E.~J.}\ \bibnamefont {Bergholtz}},\ }\bibfield  {title} {\bibinfo {title} {{L}iouvillian skin effect in an exactly solvable model},\ }\href {https://doi.org/10.1103/PhysRevResearch.4.023160} {\bibfield  {journal} {\bibinfo  {journal} {Phys. Rev. Res.}\ }\textbf {\bibinfo {volume} {4}},\ \bibinfo {pages} {023160} (\bibinfo {year} {2022})}\BibitemShut {NoStop}%
\bibitem [{\citenamefont {S\'a}\ \emph {et~al.}(2023)\citenamefont {S\'a}, \citenamefont {Ribeiro},\ and\ \citenamefont {Prosen}}]{PhysRevX.13.031019}%
  \BibitemOpen
  \bibfield  {author} {\bibinfo {author} {\bibfnamefont {L.}~\bibnamefont {S\'a}}, \bibinfo {author} {\bibfnamefont {P.}~\bibnamefont {Ribeiro}},\ and\ \bibinfo {author} {\bibfnamefont {T.}~\bibnamefont {Prosen}},\ }\bibfield  {title} {\bibinfo {title} {Symmetry classification of many-body {L}indbladians: Tenfold way and beyond},\ }\href {https://doi.org/10.1103/PhysRevX.13.031019} {\bibfield  {journal} {\bibinfo  {journal} {Phys. Rev. X}\ }\textbf {\bibinfo {volume} {13}},\ \bibinfo {pages} {031019} (\bibinfo {year} {2023})}\BibitemShut {NoStop}%
\bibitem [{\citenamefont {Kawabata}\ \emph {et~al.}(2023)\citenamefont {Kawabata}, \citenamefont {Kulkarni}, \citenamefont {Li}, \citenamefont {Numasawa},\ and\ \citenamefont {Ryu}}]{Kawabata2023}%
  \BibitemOpen
  \bibfield  {author} {\bibinfo {author} {\bibfnamefont {K.}~\bibnamefont {Kawabata}}, \bibinfo {author} {\bibfnamefont {A.}~\bibnamefont {Kulkarni}}, \bibinfo {author} {\bibfnamefont {J.}~\bibnamefont {Li}}, \bibinfo {author} {\bibfnamefont {T.}~\bibnamefont {Numasawa}},\ and\ \bibinfo {author} {\bibfnamefont {S.}~\bibnamefont {Ryu}},\ }\bibfield  {title} {\bibinfo {title} {Symmetry of open quantum systems: Classification of dissipative quantum chaos},\ }\href {https://doi.org/10.1103/PRXQuantum.4.030328} {\bibfield  {journal} {\bibinfo  {journal} {PRX Quantum}\ }\textbf {\bibinfo {volume} {4}},\ \bibinfo {pages} {030328} (\bibinfo {year} {2023})}\BibitemShut {NoStop}%
\bibitem [{\citenamefont {Hamanaka}\ \emph {et~al.}(2023)\citenamefont {Hamanaka}, \citenamefont {Yamamoto},\ and\ \citenamefont {Yoshida}}]{PhysRevB.108.155114}%
  \BibitemOpen
  \bibfield  {author} {\bibinfo {author} {\bibfnamefont {S.}~\bibnamefont {Hamanaka}}, \bibinfo {author} {\bibfnamefont {K.}~\bibnamefont {Yamamoto}},\ and\ \bibinfo {author} {\bibfnamefont {T.}~\bibnamefont {Yoshida}},\ }\bibfield  {title} {\bibinfo {title} {Interaction-induced {L}iouvillian skin effect in a fermionic chain with a two-body loss},\ }\href {https://doi.org/10.1103/PhysRevB.108.155114} {\bibfield  {journal} {\bibinfo  {journal} {Phys. Rev. B}\ }\textbf {\bibinfo {volume} {108}},\ \bibinfo {pages} {155114} (\bibinfo {year} {2023})}\BibitemShut {NoStop}%
\bibitem [{\citenamefont {Longhi}(2019)}]{PhysRevResearch.1.023013}%
  \BibitemOpen
  \bibfield  {author} {\bibinfo {author} {\bibfnamefont {S.}~\bibnamefont {Longhi}},\ }\bibfield  {title} {\bibinfo {title} {Probing non-{H}ermitian skin effect and {N}on-{B}loch phase transitions},\ }\href {https://doi.org/10.1103/PhysRevResearch.1.023013} {\bibfield  {journal} {\bibinfo  {journal} {Phys. Rev. Res.}\ }\textbf {\bibinfo {volume} {1}},\ \bibinfo {pages} {023013} (\bibinfo {year} {2019})}\BibitemShut {NoStop}%
\bibitem [{\citenamefont {Shang}\ and\ \citenamefont {Hu}(2025)}]{wsmq-kmq9}%
  \BibitemOpen
  \bibfield  {author} {\bibinfo {author} {\bibfnamefont {J.}~\bibnamefont {Shang}}\ and\ \bibinfo {author} {\bibfnamefont {H.}~\bibnamefont {Hu}},\ }\bibfield  {title} {\bibinfo {title} {Spreading dynamics in the {H}atano-{N}elson model},\ }\href {https://doi.org/10.1103/wsmq-kmq9} {\bibfield  {journal} {\bibinfo  {journal} {Phys. Rev. B}\ }\textbf {\bibinfo {volume} {112}},\ \bibinfo {pages} {014205} (\bibinfo {year} {2025})}\BibitemShut {NoStop}%
\bibitem [{\citenamefont {Budich}\ and\ \citenamefont {Diehl}(2015)}]{diehl2015top}%
  \BibitemOpen
  \bibfield  {author} {\bibinfo {author} {\bibfnamefont {J.~C.}\ \bibnamefont {Budich}}\ and\ \bibinfo {author} {\bibfnamefont {S.}~\bibnamefont {Diehl}},\ }\bibfield  {title} {\bibinfo {title} {Topology of density matrices},\ }\href {https://doi.org/10.1103/PhysRevB.91.165140} {\bibfield  {journal} {\bibinfo  {journal} {Phys. Rev. B}\ }\textbf {\bibinfo {volume} {91}},\ \bibinfo {pages} {165140} (\bibinfo {year} {2015})}\BibitemShut {NoStop}%
\bibitem [{\citenamefont {Diehl}\ \emph {et~al.}(2011)\citenamefont {Diehl}, \citenamefont {Rico}, \citenamefont {Baranov},\ and\ \citenamefont {Zoller}}]{Diehl2011}%
  \BibitemOpen
  \bibfield  {author} {\bibinfo {author} {\bibfnamefont {S.}~\bibnamefont {Diehl}}, \bibinfo {author} {\bibfnamefont {E.}~\bibnamefont {Rico}}, \bibinfo {author} {\bibfnamefont {M.~A.}\ \bibnamefont {Baranov}},\ and\ \bibinfo {author} {\bibfnamefont {P.}~\bibnamefont {Zoller}},\ }\bibfield  {title} {\bibinfo {title} {Topology by dissipation in atomic quantum wires},\ }\href {https://doi.org/10.1038/nphys2106} {\bibfield  {journal} {\bibinfo  {journal} {Nat. Phys.}\ }\textbf {\bibinfo {volume} {7}},\ \bibinfo {pages} {971} (\bibinfo {year} {2011})}\BibitemShut {NoStop}%
\bibitem [{\citenamefont {{Bardyn}}\ \emph {et~al.}(2013)\citenamefont {{Bardyn}}, \citenamefont {{Baranov}}, \citenamefont {{Kraus}}, \citenamefont {{Rico}}, \citenamefont {{{\.I}mamo{\u{g}}lu}}, \citenamefont {{Zoller}},\ and\ \citenamefont {{Diehl}}}]{bardyn2013top}%
  \BibitemOpen
  \bibfield  {author} {\bibinfo {author} {\bibfnamefont {C.-E.}\ \bibnamefont {{Bardyn}}}, \bibinfo {author} {\bibfnamefont {M.~A.}\ \bibnamefont {{Baranov}}}, \bibinfo {author} {\bibfnamefont {C.~V.}\ \bibnamefont {{Kraus}}}, \bibinfo {author} {\bibfnamefont {E.}~\bibnamefont {{Rico}}}, \bibinfo {author} {\bibfnamefont {A.}~\bibnamefont {{{\.I}mamo{\u{g}}lu}}}, \bibinfo {author} {\bibfnamefont {P.}~\bibnamefont {{Zoller}}},\ and\ \bibinfo {author} {\bibfnamefont {S.}~\bibnamefont {{Diehl}}},\ }\bibfield  {title} {\bibinfo {title} {{Topology by dissipation}},\ }\href {https://doi.org/10.1088/1367-2630/15/8/085001} {\bibfield  {journal} {\bibinfo  {journal} {New J. Phys.}\ }\textbf {\bibinfo {volume} {15}},\ \bibinfo {eid} {085001} (\bibinfo {year} {2013})}\BibitemShut {NoStop}%
\bibitem [{\citenamefont {Kiely}\ and\ \citenamefont {Xu}(2025)}]{8vtp-khnq}%
  \BibitemOpen
  \bibfield  {author} {\bibinfo {author} {\bibfnamefont {T.~G.}\ \bibnamefont {Kiely}}\ and\ \bibinfo {author} {\bibfnamefont {C.}~\bibnamefont {Xu}},\ }\bibfield  {title} {\bibinfo {title} {Phase transition of topological index driven by dephasing},\ }\href {https://doi.org/10.1103/8vtp-khnq} {\bibfield  {journal} {\bibinfo  {journal} {Phys. Rev. B}\ }\textbf {\bibinfo {volume} {112}},\ \bibinfo {pages} {L201112} (\bibinfo {year} {2025})}\BibitemShut {NoStop}%
\bibitem [{\citenamefont {Esposito}\ and\ \citenamefont {Gaspard}(2005)}]{PhysRevB.71.214302}%
  \BibitemOpen
  \bibfield  {author} {\bibinfo {author} {\bibfnamefont {M.}~\bibnamefont {Esposito}}\ and\ \bibinfo {author} {\bibfnamefont {P.}~\bibnamefont {Gaspard}},\ }\bibfield  {title} {\bibinfo {title} {Emergence of diffusion in finite quantum systems},\ }\href {https://doi.org/10.1103/PhysRevB.71.214302} {\bibfield  {journal} {\bibinfo  {journal} {Phys. Rev. B}\ }\textbf {\bibinfo {volume} {71}},\ \bibinfo {pages} {214302} (\bibinfo {year} {2005})}\BibitemShut {NoStop}%
\bibitem [{\citenamefont {Bernard}\ and\ \citenamefont {Jin}(2019)}]{PhysRevLett.123.080601}%
  \BibitemOpen
  \bibfield  {author} {\bibinfo {author} {\bibfnamefont {D.}~\bibnamefont {Bernard}}\ and\ \bibinfo {author} {\bibfnamefont {T.}~\bibnamefont {Jin}},\ }\bibfield  {title} {\bibinfo {title} {Open quantum symmetric simple exclusion process},\ }\href {https://doi.org/10.1103/PhysRevLett.123.080601} {\bibfield  {journal} {\bibinfo  {journal} {Phys. Rev. Lett.}\ }\textbf {\bibinfo {volume} {123}},\ \bibinfo {pages} {080601} (\bibinfo {year} {2019})}\BibitemShut {NoStop}%
\bibitem [{\citenamefont {Haken}\ and\ \citenamefont {Reineker}(1972)}]{haken1972coupled}%
  \BibitemOpen
  \bibfield  {author} {\bibinfo {author} {\bibfnamefont {H.}~\bibnamefont {Haken}}\ and\ \bibinfo {author} {\bibfnamefont {P.}~\bibnamefont {Reineker}},\ }\bibfield  {title} {\bibinfo {title} {The coupled coherent and incoherent motion of excitons and its influence on the line shape of optical absorption},\ }\href {https://doi.org/10.1007/BF01400230} {\bibfield  {journal} {\bibinfo  {journal} {Z. Phys.}\ }\textbf {\bibinfo {volume} {249}},\ \bibinfo {pages} {253} (\bibinfo {year} {1972})}\BibitemShut {NoStop}%
\bibitem [{\citenamefont {Medvedyeva}\ \emph {et~al.}(2016)\citenamefont {Medvedyeva}, \citenamefont {Essler},\ and\ \citenamefont {Prosen}}]{PhysRevLett.117.137202}%
  \BibitemOpen
  \bibfield  {author} {\bibinfo {author} {\bibfnamefont {M.~V.}\ \bibnamefont {Medvedyeva}}, \bibinfo {author} {\bibfnamefont {F.~H.~L.}\ \bibnamefont {Essler}},\ and\ \bibinfo {author} {\bibfnamefont {T.}~\bibnamefont {Prosen}},\ }\bibfield  {title} {\bibinfo {title} {Exact {B}ethe ansatz spectrum of a tight-binding chain with dephasing noise},\ }\href {https://doi.org/10.1103/PhysRevLett.117.137202} {\bibfield  {journal} {\bibinfo  {journal} {Phys. Rev. Lett.}\ }\textbf {\bibinfo {volume} {117}},\ \bibinfo {pages} {137202} (\bibinfo {year} {2016})}\BibitemShut {NoStop}%
\bibitem [{\citenamefont {Catalano}\ \emph {et~al.}(2023)\citenamefont {Catalano}, \citenamefont {Mattiotti}, \citenamefont {Dubail}, \citenamefont {Hagenm{\"u}ller}, \citenamefont {Prosen}, \citenamefont {Franchini},\ and\ \citenamefont {Pupillo}}]{PhysRevLett.131.053401}%
  \BibitemOpen
  \bibfield  {author} {\bibinfo {author} {\bibfnamefont {A.~G.}\ \bibnamefont {Catalano}}, \bibinfo {author} {\bibfnamefont {F.}~\bibnamefont {Mattiotti}}, \bibinfo {author} {\bibfnamefont {J.}~\bibnamefont {Dubail}}, \bibinfo {author} {\bibfnamefont {D.}~\bibnamefont {Hagenm{\"u}ller}}, \bibinfo {author} {\bibfnamefont {T.}~\bibnamefont {Prosen}}, \bibinfo {author} {\bibfnamefont {F.}~\bibnamefont {Franchini}},\ and\ \bibinfo {author} {\bibfnamefont {G.}~\bibnamefont {Pupillo}},\ }\bibfield  {title} {\bibinfo {title} {Anomalous diffusion in the long-range {H}aken-{S}trobl-{R}eineker model},\ }\href {https://doi.org/10.1103/PhysRevLett.131.053401} {\bibfield  {journal} {\bibinfo  {journal} {Phys. Rev. Lett.}\ }\textbf {\bibinfo {volume} {131}},\ \bibinfo {pages} {053401} (\bibinfo {year} {2023})}\BibitemShut {NoStop}%
\bibitem [{\citenamefont {Caruso}\ \emph {et~al.}(2009)\citenamefont {Caruso}, \citenamefont {Chin}, \citenamefont {Datta}, \citenamefont {Huelga},\ and\ \citenamefont {Plenio}}]{Caruso2009}%
  \BibitemOpen
  \bibfield  {author} {\bibinfo {author} {\bibfnamefont {F.}~\bibnamefont {Caruso}}, \bibinfo {author} {\bibfnamefont {A.~W.}\ \bibnamefont {Chin}}, \bibinfo {author} {\bibfnamefont {A.}~\bibnamefont {Datta}}, \bibinfo {author} {\bibfnamefont {S.~F.}\ \bibnamefont {Huelga}},\ and\ \bibinfo {author} {\bibfnamefont {M.~B.}\ \bibnamefont {Plenio}},\ }\bibfield  {title} {\bibinfo {title} {Highly efficient energy excitation transfer in light-harvesting complexes: The fundamental role of noise-assisted transport},\ }\href {https://doi.org/10.1063/1.3223548} {\bibfield  {journal} {\bibinfo  {journal} {J. Chem. Phys.}\ }\textbf {\bibinfo {volume} {131}},\ \bibinfo {pages} {105106} (\bibinfo {year} {2009})}\BibitemShut {NoStop}%
\bibitem [{\citenamefont {Sinayskiy}\ \emph {et~al.}(2012)\citenamefont {Sinayskiy}, \citenamefont {Marais}, \citenamefont {Petruccione},\ and\ \citenamefont {Ekert}}]{PhysRevLett.108.020602}%
  \BibitemOpen
  \bibfield  {author} {\bibinfo {author} {\bibfnamefont {I.}~\bibnamefont {Sinayskiy}}, \bibinfo {author} {\bibfnamefont {A.}~\bibnamefont {Marais}}, \bibinfo {author} {\bibfnamefont {F.}~\bibnamefont {Petruccione}},\ and\ \bibinfo {author} {\bibfnamefont {A.}~\bibnamefont {Ekert}},\ }\bibfield  {title} {\bibinfo {title} {Decoherence-assisted transport in a dimer system},\ }\href {https://doi.org/10.1103/PhysRevLett.108.020602} {\bibfield  {journal} {\bibinfo  {journal} {Phys. Rev. Lett.}\ }\textbf {\bibinfo {volume} {108}},\ \bibinfo {pages} {020602} (\bibinfo {year} {2012})}\BibitemShut {NoStop}%
\bibitem [{\citenamefont {Li}\ and\ \citenamefont {Tang}(2026)}]{li2026classificationtopologicalphasesmatter}%
  \BibitemOpen
  \bibfield  {author} {\bibinfo {author} {\bibfnamefont {D.}~\bibnamefont {Li}}\ and\ \bibinfo {author} {\bibfnamefont {E.}~\bibnamefont {Tang}},\ }\href {https://arxiv.org/abs/2609.38171} {\bibinfo {title} {Classification of topological phases of matter in stochastic systems}} (\bibinfo {year} {2026}),\ \Eprint {https://arxiv.org/abs/2609.38171} {arXiv:2609.38171 [cond-mat.stat-mech]} \BibitemShut {NoStop}%
\bibitem [{\citenamefont {Sawada}\ \emph {et~al.}(2024)\citenamefont {Sawada}, \citenamefont {Sone}, \citenamefont {Hamazaki}, \citenamefont {Ashida},\ and\ \citenamefont {Sagawa}}]{PhysRevLett.132.046602}%
  \BibitemOpen
  \bibfield  {author} {\bibinfo {author} {\bibfnamefont {T.}~\bibnamefont {Sawada}}, \bibinfo {author} {\bibfnamefont {K.}~\bibnamefont {Sone}}, \bibinfo {author} {\bibfnamefont {R.}~\bibnamefont {Hamazaki}}, \bibinfo {author} {\bibfnamefont {Y.}~\bibnamefont {Ashida}},\ and\ \bibinfo {author} {\bibfnamefont {T.}~\bibnamefont {Sagawa}},\ }\bibfield  {title} {\bibinfo {title} {Role of topology in relaxation of one-dimensional stochastic processes},\ }\href {https://doi.org/10.1103/PhysRevLett.132.046602} {\bibfield  {journal} {\bibinfo  {journal} {Phys. Rev. Lett.}\ }\textbf {\bibinfo {volume} {132}},\ \bibinfo {pages} {046602} (\bibinfo {year} {2024})}\BibitemShut {NoStop}%
\bibitem [{\citenamefont {Murugan}\ and\ \citenamefont {Vaikuntanathan}(2017)}]{Murugan2017}%
  \BibitemOpen
  \bibfield  {author} {\bibinfo {author} {\bibfnamefont {A.}~\bibnamefont {Murugan}}\ and\ \bibinfo {author} {\bibfnamefont {S.}~\bibnamefont {Vaikuntanathan}},\ }\bibfield  {title} {\bibinfo {title} {Topologically protected modes in non-equilibrium stochastic systems},\ }\href {https://doi.org/10.1038/ncomms13881} {\bibfield  {journal} {\bibinfo  {journal} {Nat. Commun.}\ }\textbf {\bibinfo {volume} {8}},\ \bibinfo {pages} {13881} (\bibinfo {year} {2017})}\BibitemShut {NoStop}%
\bibitem [{\citenamefont {Dasbiswas}\ \emph {et~al.}(2018)\citenamefont {Dasbiswas}, \citenamefont {Mandadapu},\ and\ \citenamefont {Vaikuntanathan}}]{Dasbiswas2018}%
  \BibitemOpen
  \bibfield  {author} {\bibinfo {author} {\bibfnamefont {K.}~\bibnamefont {Dasbiswas}}, \bibinfo {author} {\bibfnamefont {K.~K.}\ \bibnamefont {Mandadapu}},\ and\ \bibinfo {author} {\bibfnamefont {S.}~\bibnamefont {Vaikuntanathan}},\ }\bibfield  {title} {\bibinfo {title} {Topological localization in out-of-equilibrium dissipative systems},\ }\href {https://doi.org/10.1073/pnas.1721096115} {\bibfield  {journal} {\bibinfo  {journal} {Proc. Natl. Acad. Sci. U.S.A.}\ }\textbf {\bibinfo {volume} {115}},\ \bibinfo {pages} {E9031} (\bibinfo {year} {2018})}\BibitemShut {NoStop}%
\bibitem [{\citenamefont {Gou}\ \emph {et~al.}(2020)\citenamefont {Gou}, \citenamefont {Chen}, \citenamefont {Xie}, \citenamefont {Xiao}, \citenamefont {Deng}, \citenamefont {Gadway}, \citenamefont {Yi},\ and\ \citenamefont {Yan}}]{PhysRevLett.124.070402}%
  \BibitemOpen
  \bibfield  {author} {\bibinfo {author} {\bibfnamefont {W.}~\bibnamefont {Gou}}, \bibinfo {author} {\bibfnamefont {T.}~\bibnamefont {Chen}}, \bibinfo {author} {\bibfnamefont {D.}~\bibnamefont {Xie}}, \bibinfo {author} {\bibfnamefont {T.}~\bibnamefont {Xiao}}, \bibinfo {author} {\bibfnamefont {T.-S.}\ \bibnamefont {Deng}}, \bibinfo {author} {\bibfnamefont {B.}~\bibnamefont {Gadway}}, \bibinfo {author} {\bibfnamefont {W.}~\bibnamefont {Yi}},\ and\ \bibinfo {author} {\bibfnamefont {B.}~\bibnamefont {Yan}},\ }\bibfield  {title} {\bibinfo {title} {Tunable nonreciprocal quantum transport through a dissipative {A}haronov-{B}ohm ring in ultracold atoms},\ }\href {https://doi.org/10.1103/PhysRevLett.124.070402} {\bibfield  {journal} {\bibinfo  {journal} {Phys. Rev. Lett.}\ }\textbf {\bibinfo {volume} {124}},\ \bibinfo {pages} {070402} (\bibinfo {year} {2020})}\BibitemShut {NoStop}%
\bibitem [{\citenamefont {Longhi}(2024)}]{Longhi2024}%
  \BibitemOpen
  \bibfield  {author} {\bibinfo {author} {\bibfnamefont {S.}~\bibnamefont {Longhi}},\ }\bibfield  {title} {\bibinfo {title} {Incoherent non-{H}ermitian skin effect in photonic quantum walks},\ }\href {https://doi.org/10.1038/s41377-024-01438-w} {\bibfield  {journal} {\bibinfo  {journal} {Light Sci. Appl.}\ }\textbf {\bibinfo {volume} {13}},\ \bibinfo {pages} {95} (\bibinfo {year} {2024})}\BibitemShut {NoStop}%
\bibitem [{\citenamefont {Yang}\ and\ \citenamefont {Huang}(2026)}]{2f5t-gc98}%
  \BibitemOpen
  \bibfield  {author} {\bibinfo {author} {\bibfnamefont {W.}~\bibnamefont {Yang}}\ and\ \bibinfo {author} {\bibfnamefont {H.}~\bibnamefont {Huang}},\ }\bibfield  {title} {\bibinfo {title} {Noise-induced revival of dynamical skin effects in quasiperiodic non-{H}ermitian systems},\ }\href {https://doi.org/10.1103/2f5t-gc98} {\bibfield  {journal} {\bibinfo  {journal} {Phys. Rev. B}\ }\textbf {\bibinfo {volume} {113}},\ \bibinfo {pages} {174305} (\bibinfo {year} {2026})}\BibitemShut {NoStop}%
\bibitem [{\citenamefont {Li}\ \emph {et~al.}(2026)\citenamefont {Li}, \citenamefont {Cai}, \citenamefont {Liu},\ and\ \citenamefont {Nori}}]{qrh6-vx64}%
  \BibitemOpen
  \bibfield  {author} {\bibinfo {author} {\bibfnamefont {Y.}~\bibnamefont {Li}}, \bibinfo {author} {\bibfnamefont {Z.-F.}\ \bibnamefont {Cai}}, \bibinfo {author} {\bibfnamefont {T.}~\bibnamefont {Liu}},\ and\ \bibinfo {author} {\bibfnamefont {F.}~\bibnamefont {Nori}},\ }\bibfield  {title} {\bibinfo {title} {Dissipation and interaction-controlled non-{H}ermitian skin effects},\ }\href {https://doi.org/10.1103/qrh6-vx64} {\bibfield  {journal} {\bibinfo  {journal} {Phys. Rev. B}\ }\textbf {\bibinfo {volume} {113}},\ \bibinfo {pages} {035444} (\bibinfo {year} {2026})}\BibitemShut {NoStop}%
\bibitem [{\citenamefont {Breuer}\ and\ \citenamefont {Petruccione}(2002)}]{breuer2002theory}%
  \BibitemOpen
  \bibfield  {author} {\bibinfo {author} {\bibfnamefont {H.-P.}\ \bibnamefont {Breuer}}\ and\ \bibinfo {author} {\bibfnamefont {F.}~\bibnamefont {Petruccione}},\ }\href@noop {} {\emph {\bibinfo {title} {The Theory of Open Quantum Systems}}}\ (\bibinfo  {publisher} {Oxford University Press},\ \bibinfo {address} {Oxford},\ \bibinfo {year} {2002})\BibitemShut {NoStop}%
\bibitem [{\citenamefont {Lampo}\ \emph {et~al.}(2016)\citenamefont {Lampo}, \citenamefont {Lim}, \citenamefont {Wehr}, \citenamefont {Massignan},\ and\ \citenamefont {Lewenstein}}]{PhysRevA.94.042123}%
  \BibitemOpen
  \bibfield  {author} {\bibinfo {author} {\bibfnamefont {A.}~\bibnamefont {Lampo}}, \bibinfo {author} {\bibfnamefont {S.~H.}\ \bibnamefont {Lim}}, \bibinfo {author} {\bibfnamefont {J.}~\bibnamefont {Wehr}}, \bibinfo {author} {\bibfnamefont {P.}~\bibnamefont {Massignan}},\ and\ \bibinfo {author} {\bibfnamefont {M.}~\bibnamefont {Lewenstein}},\ }\bibfield  {title} {\bibinfo {title} {{L}indblad model of quantum {B}rownian motion},\ }\href {https://doi.org/10.1103/PhysRevA.94.042123} {\bibfield  {journal} {\bibinfo  {journal} {Phys. Rev. A}\ }\textbf {\bibinfo {volume} {94}},\ \bibinfo {pages} {042123} (\bibinfo {year} {2016})}\BibitemShut {NoStop}%
\bibitem [{\citenamefont {Caldeira}\ and\ \citenamefont {Leggett}(1983)}]{CALDEIRA1983587}%
  \BibitemOpen
  \bibfield  {author} {\bibinfo {author} {\bibfnamefont {A.}~\bibnamefont {Caldeira}}\ and\ \bibinfo {author} {\bibfnamefont {A.}~\bibnamefont {Leggett}},\ }\bibfield  {title} {\bibinfo {title} {Path integral approach to quantum {B}rownian motion},\ }\href {https://doi.org/10.1016/0378-4371(83)90013-4} {\bibfield  {journal} {\bibinfo  {journal} {Physica A}\ }\textbf {\bibinfo {volume} {121}},\ \bibinfo {pages} {587} (\bibinfo {year} {1983})}\BibitemShut {NoStop}%
\bibitem [{\citenamefont {Caldeira}\ and\ \citenamefont {Leggett}(1981)}]{PhysRevLett.46.211}%
  \BibitemOpen
  \bibfield  {author} {\bibinfo {author} {\bibfnamefont {A.~O.}\ \bibnamefont {Caldeira}}\ and\ \bibinfo {author} {\bibfnamefont {A.~J.}\ \bibnamefont {Leggett}},\ }\bibfield  {title} {\bibinfo {title} {Influence of dissipation on quantum tunneling in macroscopic systems},\ }\href {https://doi.org/10.1103/PhysRevLett.46.211} {\bibfield  {journal} {\bibinfo  {journal} {Phys. Rev. Lett.}\ }\textbf {\bibinfo {volume} {46}},\ \bibinfo {pages} {211} (\bibinfo {year} {1981})}\BibitemShut {NoStop}%
\bibitem [{\citenamefont {{\v{Z}}nidari{\v{c}}}(2010{\natexlab{b}})}]{Znidaric_2010_2}%
  \BibitemOpen
  \bibfield  {author} {\bibinfo {author} {\bibfnamefont {M.}~\bibnamefont {{\v{Z}}nidari{\v{c}}}},\ }\bibfield  {title} {\bibinfo {title} {Exact solution for a diffusive nonequilibrium steady state of an open quantum chain},\ }\href {https://doi.org/10.1088/1742-5468/2010/05/L05002} {\bibfield  {journal} {\bibinfo  {journal} {J. Stat. Mech.: Theory Exp.}\ }\textbf {\bibinfo {volume} {2010}}\bibinfo  {number} { (05)},\ \bibinfo {pages} {L05002}}\BibitemShut {NoStop}%
\bibitem [{\citenamefont {Hislop}\ and\ \citenamefont {Sigal}(1996)}]{Hislop1996}%
  \BibitemOpen
\bibfield  {number} {  }\bibfield  {author} {\bibinfo {author} {\bibfnamefont {P.~D.}\ \bibnamefont {Hislop}}\ and\ \bibinfo {author} {\bibfnamefont {I.~M.}\ \bibnamefont {Sigal}},\ }\bibinfo {title} {The essential spectrum: Weyl's criterion},\ in\ \href {https://doi.org/10.1007/978-1-4612-0741-2_7} {\emph {\bibinfo {booktitle} {Introduction to Spectral Theory: With Applications to Schr{\"o}dinger Operators}}}\ (\bibinfo  {publisher} {Springer New York},\ \bibinfo {address} {New York, NY},\ \bibinfo {year} {1996})\ pp.\ \bibinfo {pages} {69--75}\BibitemShut {NoStop}%
\bibitem [{\citenamefont {Daley}(2014)}]{Daley04032014}%
  \BibitemOpen
  \bibfield  {author} {\bibinfo {author} {\bibfnamefont {A.~J.}\ \bibnamefont {Daley}},\ }\bibfield  {title} {\bibinfo {title} {Quantum trajectories and open many-body quantum systems},\ }\href {https://doi.org/10.1080/00018732.2014.933502} {\bibfield  {journal} {\bibinfo  {journal} {Adv. Phys.}\ }\textbf {\bibinfo {volume} {63}},\ \bibinfo {pages} {77} (\bibinfo {year} {2014})}\BibitemShut {NoStop}%
\bibitem [{\citenamefont {Bu{\v{c}}a}\ and\ \citenamefont {Prosen}(2012)}]{Buca_2012}%
  \BibitemOpen
  \bibfield  {author} {\bibinfo {author} {\bibfnamefont {B.}~\bibnamefont {Bu{\v{c}}a}}\ and\ \bibinfo {author} {\bibfnamefont {T.}~\bibnamefont {Prosen}},\ }\bibfield  {title} {\bibinfo {title} {A note on symmetry reductions of the {L}indblad equation: transport in constrained open spin chains},\ }\href {https://doi.org/10.1088/1367-2630/14/7/073007} {\bibfield  {journal} {\bibinfo  {journal} {New J. Phys.}\ }\textbf {\bibinfo {volume} {14}},\ \bibinfo {pages} {073007} (\bibinfo {year} {2012})}\BibitemShut {NoStop}%
\bibitem [{\citenamefont {Albert}\ \emph {et~al.}(2016)\citenamefont {Albert}, \citenamefont {Bradlyn}, \citenamefont {Fraas},\ and\ \citenamefont {Jiang}}]{PhysRevX.6.041031}%
  \BibitemOpen
  \bibfield  {author} {\bibinfo {author} {\bibfnamefont {V.~V.}\ \bibnamefont {Albert}}, \bibinfo {author} {\bibfnamefont {B.}~\bibnamefont {Bradlyn}}, \bibinfo {author} {\bibfnamefont {M.}~\bibnamefont {Fraas}},\ and\ \bibinfo {author} {\bibfnamefont {L.}~\bibnamefont {Jiang}},\ }\bibfield  {title} {\bibinfo {title} {Geometry and response of {L}indbladians},\ }\href {https://doi.org/10.1103/PhysRevX.6.041031} {\bibfield  {journal} {\bibinfo  {journal} {Phys. Rev. X}\ }\textbf {\bibinfo {volume} {6}},\ \bibinfo {pages} {041031} (\bibinfo {year} {2016})}\BibitemShut {NoStop}%
\bibitem [{\citenamefont {Nakagawa}\ and\ \citenamefont {Ueda}(2025)}]{PhysRevX.15.021016}%
  \BibitemOpen
  \bibfield  {author} {\bibinfo {author} {\bibfnamefont {M.}~\bibnamefont {Nakagawa}}\ and\ \bibinfo {author} {\bibfnamefont {M.}~\bibnamefont {Ueda}},\ }\bibfield  {title} {\bibinfo {title} {Topology of discrete quantum feedback control},\ }\href {https://doi.org/10.1103/PhysRevX.15.021016} {\bibfield  {journal} {\bibinfo  {journal} {Phys. Rev. X}\ }\textbf {\bibinfo {volume} {15}},\ \bibinfo {pages} {021016} (\bibinfo {year} {2025})}\BibitemShut {NoStop}%
\bibitem [{\citenamefont {Gardiner}\ and\ \citenamefont {Zoller}(2004)}]{GardinerZoller2004}%
  \BibitemOpen
  \bibfield  {author} {\bibinfo {author} {\bibfnamefont {C.~W.}\ \bibnamefont {Gardiner}}\ and\ \bibinfo {author} {\bibfnamefont {P.}~\bibnamefont {Zoller}},\ }\href@noop {} {\emph {\bibinfo {title} {Quantum Noise: A Handbook of {M}arkovian and {N}on-{M}arkovian Quantum Stochastic Methods with Applications to Quantum Optics}}},\ \bibinfo {edition} {3rd}\ ed.,\ Springer Series in Synergetics\ (\bibinfo  {publisher} {Springer},\ \bibinfo {address} {Berlin, Heidelberg},\ \bibinfo {year} {2004})\BibitemShut {NoStop}%
\bibitem [{\citenamefont {Horstmann}\ \emph {et~al.}(2013)\citenamefont {Horstmann}, \citenamefont {Cirac},\ and\ \citenamefont {Giedke}}]{PhysRevA.87.012108}%
  \BibitemOpen
  \bibfield  {author} {\bibinfo {author} {\bibfnamefont {B.}~\bibnamefont {Horstmann}}, \bibinfo {author} {\bibfnamefont {J.~I.}\ \bibnamefont {Cirac}},\ and\ \bibinfo {author} {\bibfnamefont {G.}~\bibnamefont {Giedke}},\ }\bibfield  {title} {\bibinfo {title} {Noise-driven dynamics and phase transitions in fermionic systems},\ }\href {https://doi.org/10.1103/PhysRevA.87.012108} {\bibfield  {journal} {\bibinfo  {journal} {Phys. Rev. A}\ }\textbf {\bibinfo {volume} {87}},\ \bibinfo {pages} {012108} (\bibinfo {year} {2013})}\BibitemShut {NoStop}%
\bibitem [{\citenamefont {Holstein}(1961)}]{Holstein1961}%
  \BibitemOpen
  \bibfield  {author} {\bibinfo {author} {\bibfnamefont {T.}~\bibnamefont {Holstein}},\ }\bibfield  {title} {\bibinfo {title} {Hall effect in impurity conduction},\ }\href {https://doi.org/10.1103/PhysRev.124.1329} {\bibfield  {journal} {\bibinfo  {journal} {Phys. Rev.}\ }\textbf {\bibinfo {volume} {124}},\ \bibinfo {pages} {1329} (\bibinfo {year} {1961})}\BibitemShut {NoStop}%
\bibitem [{\citenamefont {Chahine}\ \emph {et~al.}(2026)\citenamefont {Chahine}, \citenamefont {Siegl},\ and\ \citenamefont {Buchhold}}]{chahine2026independentdensitycoherenceskin}%
  \BibitemOpen
  \bibfield  {author} {\bibinfo {author} {\bibfnamefont {K.}~\bibnamefont {Chahine}}, \bibinfo {author} {\bibfnamefont {M.}~\bibnamefont {Siegl}},\ and\ \bibinfo {author} {\bibfnamefont {M.}~\bibnamefont {Buchhold}},\ }\href {https://arxiv.org/abs/2609.36017} {\bibinfo {title} {Independent density and coherence skin effects in adaptive fermion circuits}} (\bibinfo {year} {2026}),\ \Eprint {https://arxiv.org/abs/2609.36017} {arXiv:2609.36017 [quant-ph]} \BibitemShut {NoStop}%
\bibitem [{\citenamefont {Chenu}\ \emph {et~al.}(2017)\citenamefont {Chenu}, \citenamefont {Beau}, \citenamefont {Cao},\ and\ \citenamefont {del Campo}}]{PhysRevLett.118.140403}%
  \BibitemOpen
  \bibfield  {author} {\bibinfo {author} {\bibfnamefont {A.}~\bibnamefont {Chenu}}, \bibinfo {author} {\bibfnamefont {M.}~\bibnamefont {Beau}}, \bibinfo {author} {\bibfnamefont {J.}~\bibnamefont {Cao}},\ and\ \bibinfo {author} {\bibfnamefont {A.}~\bibnamefont {del Campo}},\ }\bibfield  {title} {\bibinfo {title} {Quantum simulation of generic many-body open system dynamics using classical noise},\ }\href {https://doi.org/10.1103/PhysRevLett.118.140403} {\bibfield  {journal} {\bibinfo  {journal} {Phys. Rev. Lett.}\ }\textbf {\bibinfo {volume} {118}},\ \bibinfo {pages} {140403} (\bibinfo {year} {2017})}\BibitemShut {NoStop}%
\bibitem [{\citenamefont {Gadway}(2015)}]{PhysRevA.92.043606}%
  \BibitemOpen
  \bibfield  {author} {\bibinfo {author} {\bibfnamefont {B.}~\bibnamefont {Gadway}},\ }\bibfield  {title} {\bibinfo {title} {Atom-optics approach to studying transport phenomena},\ }\href {https://doi.org/10.1103/PhysRevA.92.043606} {\bibfield  {journal} {\bibinfo  {journal} {Phys. Rev. A}\ }\textbf {\bibinfo {volume} {92}},\ \bibinfo {pages} {043606} (\bibinfo {year} {2015})}\BibitemShut {NoStop}%
\bibitem [{\citenamefont {Meier}\ \emph {et~al.}(2016)\citenamefont {Meier}, \citenamefont {An},\ and\ \citenamefont {Gadway}}]{PhysRevA.93.051602}%
  \BibitemOpen
  \bibfield  {author} {\bibinfo {author} {\bibfnamefont {E.~J.}\ \bibnamefont {Meier}}, \bibinfo {author} {\bibfnamefont {F.~A.}\ \bibnamefont {An}},\ and\ \bibinfo {author} {\bibfnamefont {B.}~\bibnamefont {Gadway}},\ }\bibfield  {title} {\bibinfo {title} {Atom-optics simulator of lattice transport phenomena},\ }\href {https://doi.org/10.1103/PhysRevA.93.051602} {\bibfield  {journal} {\bibinfo  {journal} {Phys. Rev. A}\ }\textbf {\bibinfo {volume} {93}},\ \bibinfo {pages} {051602(R)} (\bibinfo {year} {2016})}\BibitemShut {NoStop}%
\bibitem [{\citenamefont {Lapp}\ \emph {et~al.}(2019)\citenamefont {Lapp}, \citenamefont {Ang'ong'a}, \citenamefont {An},\ and\ \citenamefont {Gadway}}]{Lapp2019}%
  \BibitemOpen
  \bibfield  {author} {\bibinfo {author} {\bibfnamefont {S.}~\bibnamefont {Lapp}}, \bibinfo {author} {\bibfnamefont {J.}~\bibnamefont {Ang'ong'a}}, \bibinfo {author} {\bibfnamefont {F.~A.}\ \bibnamefont {An}},\ and\ \bibinfo {author} {\bibfnamefont {B.}~\bibnamefont {Gadway}},\ }\bibfield  {title} {\bibinfo {title} {Engineering tunable local loss in a synthetic lattice of momentum states},\ }\href {https://doi.org/10.1088/1367-2630/ab1147} {\bibfield  {journal} {\bibinfo  {journal} {New J. Phys.}\ }\textbf {\bibinfo {volume} {21}},\ \bibinfo {pages} {045006} (\bibinfo {year} {2019})}\BibitemShut {NoStop}%
\bibitem [{\citenamefont {Karamlou}\ \emph {et~al.}(2022)\citenamefont {Karamlou}, \citenamefont {Braum{\"u}ller}, \citenamefont {Yanay}, \citenamefont {Di~Paolo}, \citenamefont {Harrington}, \citenamefont {Kannan}, \citenamefont {Kim}, \citenamefont {Kjaergaard}, \citenamefont {Melville}, \citenamefont {Muschinske}, \citenamefont {Olsson}, \citenamefont {Secrest}, \citenamefont {Sharma}, \citenamefont {Sung}, \citenamefont {Tarahan}, \citenamefont {Weides}, \citenamefont {Wolchover},\ and\ \citenamefont {Oliver}}]{Karamlou2022}%
  \BibitemOpen
  \bibfield  {author} {\bibinfo {author} {\bibfnamefont {A.~H.}\ \bibnamefont {Karamlou}}, \bibinfo {author} {\bibfnamefont {J.}~\bibnamefont {Braum{\"u}ller}}, \bibinfo {author} {\bibfnamefont {Y.}~\bibnamefont {Yanay}}, \bibinfo {author} {\bibfnamefont {A.}~\bibnamefont {Di~Paolo}}, \bibinfo {author} {\bibfnamefont {P.~M.}\ \bibnamefont {Harrington}}, \bibinfo {author} {\bibfnamefont {B.}~\bibnamefont {Kannan}}, \bibinfo {author} {\bibfnamefont {D.}~\bibnamefont {Kim}}, \bibinfo {author} {\bibfnamefont {M.}~\bibnamefont {Kjaergaard}}, \bibinfo {author} {\bibfnamefont {A.}~\bibnamefont {Melville}}, \bibinfo {author} {\bibfnamefont {S.}~\bibnamefont {Muschinske}}, \bibinfo {author} {\bibfnamefont {E.}~\bibnamefont {Olsson}}, \bibinfo {author} {\bibfnamefont {B.}~\bibnamefont {Secrest}}, \bibinfo {author} {\bibfnamefont {R.}~\bibnamefont {Sharma}}, \bibinfo {author} {\bibfnamefont {J.}~\bibnamefont {Sung}}, \bibinfo {author} {\bibfnamefont {R.}~\bibnamefont {Tarahan}}, \bibinfo {author} {\bibfnamefont
  {M.}~\bibnamefont {Weides}}, \bibinfo {author} {\bibfnamefont {A.}~\bibnamefont {Wolchover}},\ and\ \bibinfo {author} {\bibfnamefont {W.~D.}\ \bibnamefont {Oliver}},\ }\bibfield  {title} {\bibinfo {title} {Quantum transport and localization in 1{D} and 2{D} tight-binding lattices},\ }\href {https://doi.org/10.1038/s41534-022-00528-0} {\bibfield  {journal} {\bibinfo  {journal} {npj Quantum Inf.}\ }\textbf {\bibinfo {volume} {8}},\ \bibinfo {pages} {35} (\bibinfo {year} {2022})}\BibitemShut {NoStop}%
\bibitem [{\citenamefont {Cheuk}\ \emph {et~al.}(2015)\citenamefont {Cheuk}, \citenamefont {Nichols}, \citenamefont {Okan}, \citenamefont {Gersdorf}, \citenamefont {Ramasesh}, \citenamefont {Bakr}, \citenamefont {Lompe},\ and\ \citenamefont {Zwierlein}}]{Cheuk2015}%
  \BibitemOpen
  \bibfield  {author} {\bibinfo {author} {\bibfnamefont {L.~W.}\ \bibnamefont {Cheuk}}, \bibinfo {author} {\bibfnamefont {M.~A.}\ \bibnamefont {Nichols}}, \bibinfo {author} {\bibfnamefont {M.}~\bibnamefont {Okan}}, \bibinfo {author} {\bibfnamefont {T.}~\bibnamefont {Gersdorf}}, \bibinfo {author} {\bibfnamefont {V.~V.}\ \bibnamefont {Ramasesh}}, \bibinfo {author} {\bibfnamefont {W.~S.}\ \bibnamefont {Bakr}}, \bibinfo {author} {\bibfnamefont {T.}~\bibnamefont {Lompe}},\ and\ \bibinfo {author} {\bibfnamefont {M.~W.}\ \bibnamefont {Zwierlein}},\ }\bibfield  {title} {\bibinfo {title} {Quantum-gas microscope for fermionic atoms},\ }\href {https://doi.org/10.1103/PhysRevLett.114.193001} {\bibfield  {journal} {\bibinfo  {journal} {Phys. Rev. Lett.}\ }\textbf {\bibinfo {volume} {114}},\ \bibinfo {pages} {193001} (\bibinfo {year} {2015})}\BibitemShut {NoStop}%
\bibitem [{\citenamefont {Parsons}\ \emph {et~al.}(2015)\citenamefont {Parsons}, \citenamefont {Huber}, \citenamefont {Mazurenko}, \citenamefont {Chiu}, \citenamefont {Setiawan}, \citenamefont {Wooley-Brown}, \citenamefont {Blatt},\ and\ \citenamefont {Greiner}}]{Parsons2015}%
  \BibitemOpen
  \bibfield  {author} {\bibinfo {author} {\bibfnamefont {M.~F.}\ \bibnamefont {Parsons}}, \bibinfo {author} {\bibfnamefont {F.}~\bibnamefont {Huber}}, \bibinfo {author} {\bibfnamefont {A.}~\bibnamefont {Mazurenko}}, \bibinfo {author} {\bibfnamefont {C.~S.}\ \bibnamefont {Chiu}}, \bibinfo {author} {\bibfnamefont {W.}~\bibnamefont {Setiawan}}, \bibinfo {author} {\bibfnamefont {K.}~\bibnamefont {Wooley-Brown}}, \bibinfo {author} {\bibfnamefont {S.}~\bibnamefont {Blatt}},\ and\ \bibinfo {author} {\bibfnamefont {M.}~\bibnamefont {Greiner}},\ }\bibfield  {title} {\bibinfo {title} {Site-resolved imaging of fermionic ${}^{6}\mathrm{Li}$ in an optical lattice},\ }\href {https://doi.org/10.1103/PhysRevLett.114.213002} {\bibfield  {journal} {\bibinfo  {journal} {Phys. Rev. Lett.}\ }\textbf {\bibinfo {volume} {114}},\ \bibinfo {pages} {213002} (\bibinfo {year} {2015})}\BibitemShut {NoStop}%
\bibitem [{\citenamefont {Struck}\ \emph {et~al.}(2012)\citenamefont {Struck}, \citenamefont {{\"O}lschl{\"a}ger}, \citenamefont {Weinberg}, \citenamefont {Hauke}, \citenamefont {Simonet}, \citenamefont {Eckardt}, \citenamefont {Lewenstein}, \citenamefont {Sengstock},\ and\ \citenamefont {Windpassinger}}]{PhysRevLett.108.225304}%
  \BibitemOpen
  \bibfield  {author} {\bibinfo {author} {\bibfnamefont {J.}~\bibnamefont {Struck}}, \bibinfo {author} {\bibfnamefont {C.}~\bibnamefont {{\"O}lschl{\"a}ger}}, \bibinfo {author} {\bibfnamefont {M.}~\bibnamefont {Weinberg}}, \bibinfo {author} {\bibfnamefont {P.}~\bibnamefont {Hauke}}, \bibinfo {author} {\bibfnamefont {J.}~\bibnamefont {Simonet}}, \bibinfo {author} {\bibfnamefont {A.}~\bibnamefont {Eckardt}}, \bibinfo {author} {\bibfnamefont {M.}~\bibnamefont {Lewenstein}}, \bibinfo {author} {\bibfnamefont {K.}~\bibnamefont {Sengstock}},\ and\ \bibinfo {author} {\bibfnamefont {P.}~\bibnamefont {Windpassinger}},\ }\bibfield  {title} {\bibinfo {title} {Tunable gauge potential for neutral and spinless particles in driven optical lattices},\ }\href {https://doi.org/10.1103/PhysRevLett.108.225304} {\bibfield  {journal} {\bibinfo  {journal} {Phys. Rev. Lett.}\ }\textbf {\bibinfo {volume} {108}},\ \bibinfo {pages} {225304} (\bibinfo {year} {2012})}\BibitemShut {NoStop}%
\bibitem [{\citenamefont {Gerbier}\ and\ \citenamefont {Castin}(2010)}]{PhysRevA.82.013615}%
  \BibitemOpen
  \bibfield  {author} {\bibinfo {author} {\bibfnamefont {F.}~\bibnamefont {Gerbier}}\ and\ \bibinfo {author} {\bibfnamefont {Y.}~\bibnamefont {Castin}},\ }\bibfield  {title} {\bibinfo {title} {Heating rates for an atom in a far-detuned optical lattice},\ }\href {https://doi.org/10.1103/PhysRevA.82.013615} {\bibfield  {journal} {\bibinfo  {journal} {Phys. Rev. A}\ }\textbf {\bibinfo {volume} {82}},\ \bibinfo {pages} {013615} (\bibinfo {year} {2010})}\BibitemShut {NoStop}%
\bibitem [{\citenamefont {F{\"u}rst}\ \emph {et~al.}(2013)\citenamefont {F{\"u}rst}, \citenamefont {Mendl},\ and\ \citenamefont {Spohn}}]{PhysRevE.88.012108}%
  \BibitemOpen
  \bibfield  {author} {\bibinfo {author} {\bibfnamefont {M.~L.~R.}\ \bibnamefont {F{\"u}rst}}, \bibinfo {author} {\bibfnamefont {C.~B.}\ \bibnamefont {Mendl}},\ and\ \bibinfo {author} {\bibfnamefont {H.}~\bibnamefont {Spohn}},\ }\bibfield  {title} {\bibinfo {title} {Matrix-valued {B}oltzmann equation for the nonintegrable {H}ubbard chain},\ }\href {https://doi.org/10.1103/PhysRevE.88.012108} {\bibfield  {journal} {\bibinfo  {journal} {Phys. Rev. E}\ }\textbf {\bibinfo {volume} {88}},\ \bibinfo {pages} {012108} (\bibinfo {year} {2013})}\BibitemShut {NoStop}%
\bibitem [{\citenamefont {Faugno}\ and\ \citenamefont {Ozawa}(2022)}]{PhysRevLett.129.180401}%
  \BibitemOpen
  \bibfield  {author} {\bibinfo {author} {\bibfnamefont {W.~N.}\ \bibnamefont {Faugno}}\ and\ \bibinfo {author} {\bibfnamefont {T.}~\bibnamefont {Ozawa}},\ }\bibfield  {title} {\bibinfo {title} {Interaction-induced non-{H}ermitian topological phases from a dynamical gauge field},\ }\href {https://doi.org/10.1103/PhysRevLett.129.180401} {\bibfield  {journal} {\bibinfo  {journal} {Phys. Rev. Lett.}\ }\textbf {\bibinfo {volume} {129}},\ \bibinfo {pages} {180401} (\bibinfo {year} {2022})}\BibitemShut {NoStop}%
\bibitem [{\citenamefont {Metelmann}\ and\ \citenamefont {Clerk}(2015)}]{PhysRevX.5.021025}%
  \BibitemOpen
  \bibfield  {author} {\bibinfo {author} {\bibfnamefont {A.}~\bibnamefont {Metelmann}}\ and\ \bibinfo {author} {\bibfnamefont {A.~A.}\ \bibnamefont {Clerk}},\ }\bibfield  {title} {\bibinfo {title} {Nonreciprocal photon transmission and amplification via reservoir engineering},\ }\href {https://doi.org/10.1103/PhysRevX.5.021025} {\bibfield  {journal} {\bibinfo  {journal} {Phys. Rev. X}\ }\textbf {\bibinfo {volume} {5}},\ \bibinfo {pages} {021025} (\bibinfo {year} {2015})}\BibitemShut {NoStop}%
\bibitem [{\citenamefont {Wanjura}\ \emph {et~al.}(2020)\citenamefont {Wanjura}, \citenamefont {Brunelli},\ and\ \citenamefont {Nunnenkamp}}]{Wanjura2020}%
  \BibitemOpen
  \bibfield  {author} {\bibinfo {author} {\bibfnamefont {C.~C.}\ \bibnamefont {Wanjura}}, \bibinfo {author} {\bibfnamefont {M.}~\bibnamefont {Brunelli}},\ and\ \bibinfo {author} {\bibfnamefont {A.}~\bibnamefont {Nunnenkamp}},\ }\bibfield  {title} {\bibinfo {title} {Topological framework for directional amplification in driven-dissipative cavity arrays},\ }\href {https://doi.org/10.1038/s41467-020-16863-9} {\bibfield  {journal} {\bibinfo  {journal} {Nat. Commun.}\ }\textbf {\bibinfo {volume} {11}},\ \bibinfo {pages} {3149} (\bibinfo {year} {2020})}\BibitemShut {NoStop}%
\bibitem [{\citenamefont {McDonald}\ and\ \citenamefont {Clerk}(2020)}]{mcdonald2020exponentially}%
  \BibitemOpen
  \bibfield  {author} {\bibinfo {author} {\bibfnamefont {A.}~\bibnamefont {McDonald}}\ and\ \bibinfo {author} {\bibfnamefont {A.~A.}\ \bibnamefont {Clerk}},\ }\bibfield  {title} {\bibinfo {title} {Exponentially-enhanced quantum sensing with non-{H}ermitian lattice dynamics},\ }\href {https://doi.org/10.1038/s41467-020-19090-4} {\bibfield  {journal} {\bibinfo  {journal} {Nat. Commun.}\ }\textbf {\bibinfo {volume} {11}},\ \bibinfo {pages} {5382} (\bibinfo {year} {2020})}\BibitemShut {NoStop}%
\bibitem [{\citenamefont {Reulet}\ \emph {et~al.}(2003)\citenamefont {Reulet}, \citenamefont {Senzier},\ and\ \citenamefont {Prober}}]{PhysRevLett.91.196601}%
  \BibitemOpen
  \bibfield  {author} {\bibinfo {author} {\bibfnamefont {B.}~\bibnamefont {Reulet}}, \bibinfo {author} {\bibfnamefont {J.}~\bibnamefont {Senzier}},\ and\ \bibinfo {author} {\bibfnamefont {D.~E.}\ \bibnamefont {Prober}},\ }\bibfield  {title} {\bibinfo {title} {Environmental effects in the third moment of voltage fluctuations in a tunnel junction},\ }\href {https://doi.org/10.1103/PhysRevLett.91.196601} {\bibfield  {journal} {\bibinfo  {journal} {Phys. Rev. Lett.}\ }\textbf {\bibinfo {volume} {91}},\ \bibinfo {pages} {196601} (\bibinfo {year} {2003})}\BibitemShut {NoStop}%
\bibitem [{\citenamefont {Bomze}\ \emph {et~al.}(2005)\citenamefont {Bomze}, \citenamefont {Gershon}, \citenamefont {Shovkun}, \citenamefont {Levitov},\ and\ \citenamefont {Reznikov}}]{PhysRevLett.95.176601}%
  \BibitemOpen
  \bibfield  {author} {\bibinfo {author} {\bibfnamefont {Y.}~\bibnamefont {Bomze}}, \bibinfo {author} {\bibfnamefont {G.}~\bibnamefont {Gershon}}, \bibinfo {author} {\bibfnamefont {D.}~\bibnamefont {Shovkun}}, \bibinfo {author} {\bibfnamefont {L.~S.}\ \bibnamefont {Levitov}},\ and\ \bibinfo {author} {\bibfnamefont {M.}~\bibnamefont {Reznikov}},\ }\bibfield  {title} {\bibinfo {title} {Measurement of counting statistics of electron transport in a tunnel junction},\ }\href {https://doi.org/10.1103/PhysRevLett.95.176601} {\bibfield  {journal} {\bibinfo  {journal} {Phys. Rev. Lett.}\ }\textbf {\bibinfo {volume} {95}},\ \bibinfo {pages} {176601} (\bibinfo {year} {2005})}\BibitemShut {NoStop}%
\bibitem [{\citenamefont {Hargus}\ \emph {et~al.}(2021)\citenamefont {Hargus}, \citenamefont {Epstein},\ and\ \citenamefont {Mandadapu}}]{Hargus2021}%
  \BibitemOpen
  \bibfield  {author} {\bibinfo {author} {\bibfnamefont {C.}~\bibnamefont {Hargus}}, \bibinfo {author} {\bibfnamefont {J.~M.}\ \bibnamefont {Epstein}},\ and\ \bibinfo {author} {\bibfnamefont {K.~K.}\ \bibnamefont {Mandadapu}},\ }\bibfield  {title} {\bibinfo {title} {Odd diffusivity of chiral random motion},\ }\href {https://doi.org/10.1103/PhysRevLett.127.178001} {\bibfield  {journal} {\bibinfo  {journal} {Phys. Rev. Lett.}\ }\textbf {\bibinfo {volume} {127}},\ \bibinfo {pages} {178001} (\bibinfo {year} {2021})}\BibitemShut {NoStop}%
\bibitem [{\citenamefont {Bhatnagar}\ \emph {et~al.}(1954)\citenamefont {Bhatnagar}, \citenamefont {Gross},\ and\ \citenamefont {Krook}}]{BGK1954}%
  \BibitemOpen
  \bibfield  {author} {\bibinfo {author} {\bibfnamefont {P.~L.}\ \bibnamefont {Bhatnagar}}, \bibinfo {author} {\bibfnamefont {E.~P.}\ \bibnamefont {Gross}},\ and\ \bibinfo {author} {\bibfnamefont {M.}~\bibnamefont {Krook}},\ }\bibfield  {title} {\bibinfo {title} {A model for collision processes in gases. {I}. small amplitude processes in charged and neutral one-component systems},\ }\href {https://doi.org/10.1103/PhysRev.94.511} {\bibfield  {journal} {\bibinfo  {journal} {Phys. Rev.}\ }\textbf {\bibinfo {volume} {94}},\ \bibinfo {pages} {511} (\bibinfo {year} {1954})}\BibitemShut {NoStop}%
\bibitem [{\citenamefont {Goldstein}(1951)}]{Goldstein1951}%
  \BibitemOpen
  \bibfield  {author} {\bibinfo {author} {\bibfnamefont {S.}~\bibnamefont {Goldstein}},\ }\bibfield  {title} {\bibinfo {title} {On diffusion by discontinuous movements, and on the telegraph equation},\ }\href {https://doi.org/10.1093/qjmam/4.2.129} {\bibfield  {journal} {\bibinfo  {journal} {Q. J. Mech. Appl. Math.}\ }\textbf {\bibinfo {volume} {4}},\ \bibinfo {pages} {129} (\bibinfo {year} {1951})}\BibitemShut {NoStop}%
\bibitem [{\citenamefont {Kac}(1974)}]{Kac1974}%
  \BibitemOpen
  \bibfield  {author} {\bibinfo {author} {\bibfnamefont {M.}~\bibnamefont {Kac}},\ }\bibfield  {title} {\bibinfo {title} {A stochastic model related to the telegrapher's equation},\ }\href {https://doi.org/10.1216/RMJ-1974-4-3-497} {\bibfield  {journal} {\bibinfo  {journal} {Rocky Mt. J. Math.}\ }\textbf {\bibinfo {volume} {4}},\ \bibinfo {pages} {497} (\bibinfo {year} {1974})}\BibitemShut {NoStop}%
\bibitem [{\citenamefont {Weiss}(2002)}]{Weiss2002}%
  \BibitemOpen
  \bibfield  {author} {\bibinfo {author} {\bibfnamefont {G.~H.}\ \bibnamefont {Weiss}},\ }\bibfield  {title} {\bibinfo {title} {Some applications of persistent random walks and the telegrapher's equation},\ }\href {https://doi.org/10.1016/S0378-4371(02)00805-1} {\bibfield  {journal} {\bibinfo  {journal} {Physica A}\ }\textbf {\bibinfo {volume} {311}},\ \bibinfo {pages} {381} (\bibinfo {year} {2002})}\BibitemShut {NoStop}%
\bibitem [{\citenamefont {Chapman}\ and\ \citenamefont {Cowling}(1970)}]{ChapmanCowling1970}%
  \BibitemOpen
  \bibfield  {author} {\bibinfo {author} {\bibfnamefont {S.}~\bibnamefont {Chapman}}\ and\ \bibinfo {author} {\bibfnamefont {T.~G.}\ \bibnamefont {Cowling}},\ }\href@noop {} {\emph {\bibinfo {title} {The Mathematical Theory of Non-Uniform Gases}}},\ \bibinfo {edition} {3rd}\ ed.\ (\bibinfo  {publisher} {Cambridge University Press},\ \bibinfo {address} {Cambridge},\ \bibinfo {year} {1970})\BibitemShut {NoStop}%
\bibitem [{\citenamefont {Van~Kampen}(2007)}]{vankampen2007stochastic}%
  \BibitemOpen
  \bibfield  {author} {\bibinfo {author} {\bibfnamefont {N.~G.}\ \bibnamefont {Van~Kampen}},\ }\href@noop {} {\emph {\bibinfo {title} {Stochastic Processes in Physics and Chemistry}}},\ \bibinfo {edition} {3rd}\ ed.\ (\bibinfo  {publisher} {Elsevier},\ \bibinfo {address} {Amsterdam},\ \bibinfo {year} {2007})\BibitemShut {NoStop}%
\end{thebibliography}%

\appendix

\begin{widetext}
\section{Correlation-matrix equation of motion}\label{appendix:CM_derivation}
We derive the equation of motion Eq.~\eqref{eq:CM_EOM}, which is the same both for bosons and fermions. Let $K = \sum_{i,j}\mathcal{K}_{ij}\cdag_ic_j$ be a generic Hermitian operator, with $c_i\cdag_j-\zeta \cdag_j c_i=\delta_{ij}$. Here $\zeta=1$ for bosons, and $\zeta=-1$ for fermions. For both kinds, we have the identities
\begin{equation}
\label{eq:comms}
    \begin{aligned}
        \left[\cdag_i c_j, \cdag_l c_n\right] &= \delta_{jl} \cdag_i c_n - \delta_{in}\cdag_lc_j, \\
    \left[\cdag_ic_j, \cdag_l\right] &= \delta_{jl}\cdag_i, \quad \left[c_l, \cdag_ic_j\right] = \delta_{li}c_j.
    \end{aligned}
\end{equation}
These in turn imply
\begin{equation}
    \begin{aligned}
        K \cdag_i &= \cdag_i K + \sum_l\mathcal{K}_{li}\cdag_l, \\
        c_j K &= Kc_j + \sum_l\mathcal{K}_{jl}c_l.
    \end{aligned}
\end{equation}
Using these, we obtain
\begin{equation}
\begin{aligned}
    \cdag_ic_j K &= \cdag_iKc_j + \sum_l\mathcal{K}_{jl}\cdag_ic_l, \\
    K \cdag_ic_j  &= \cdag_iKc_j + \sum_l\mathcal{K}_{li}\cdag_lc_j,    
\end{aligned}
\end{equation}
which give the Hamiltonian part
\begin{equation}
\begin{aligned}
\mathrm{Tr}\left(\cdag_ic_j \left[H,\rho\right]\right) &=\sum_l\mathcal{H}_{jl}\mathrm{Tr}\left(\cdag_ic_l\rho\right) \\&\quad - \sum_l\mathcal{H}_{li}\mathrm{Tr}\left(\cdag_lc_j\rho\right).
\end{aligned}
\end{equation}
Now for the dissipator, we have    
\begin{equation}
\begin{aligned}
    \cdag_ic_j K^2 &= \cdag_iKc_jK + \sum_l\mathcal{K}_{jl}\cdag_ic_lK \\
    &= \cdag_i K^2 c_j + 2\sum_l\mathcal{K}_{jl}\cdag_i K c_l + \sum_n \mathcal{K}^2_{jn}\cdag_ic_n.
\end{aligned}
\end{equation}
Similar expansions give
\begin{equation}
\begin{aligned}
    K^2 \cdag_ic_j &= \cdag_i K^2 c_j + 2\sum_l\mathcal{K}_{li}\cdag_l K c_j + \sum_n \mathcal{K}^2_{ni}\cdag_nc_j,
\end{aligned}
\end{equation}
and
\begin{equation}
\begin{aligned}
    K \cdag_ic_j K &= \cdag_i K^2 c_j + \sum_n \mathcal{K}_{jn}\cdag_i K c_n + \sum_n\mathcal{K}_{ni}\cdag_n K c_j + \sum_{n,l}\mathcal{K}_{ni}\cdag_nc_l\mathcal{K}_{jl}.
\end{aligned}
\end{equation}
Combining all of these (using cyclicity of the trace) gives
\begin{equation}
\begin{aligned}
    \mathrm{Tr}\left(\cdag_ic_j \left[K, \left[K,\rho\right]\right]\right) &= \mathrm{Tr}\left[\left(\sum_n \left(\mathcal{K}^2_{ni}\cdag_nc_j + \mathcal{K}^2_{jn}\cdag_ic_n\right) - 2\sum_{l,n}\mathcal{K}_{li}\cdag_lc_n\mathcal{K}_{jn}\right)\rho \right].
\end{aligned}
\end{equation}
Together, these combine in $\mathrm{Tr}\left[\cdag_ic_j\mathcal{L}(\rho)\right]$ to give Eq.~\eqref{eq:CM_EOM}. 

The derivation only relies on the commutators of Eq.~\eqref{eq:comms}, which are the same for bosons and fermions, since number-conserving bilinears $\cdag_ic_j$ obey the same algebra for both statistics. For Hermitian jumps, the Heisenberg equation of motion, $\dot O=\im[H,O]-\frac12\sum_m[K_m,[K_m,O]]$, involves only commutators with such bilinears, so the equation of motion of $C$ does not depend on the statistics. The same holds for products of bilinears, such as $\langle \cdag_ic_j\cdag_lc_n\rangle$, whose equations of motion also close. Statistics only enters when such products are reordered, for instance into normal-ordered correlators, and through constraints on the allowed states. This is special to Hermitian jumps: for linear jumps describing incoherent gain, for instance, the dissipator is not a double commutator, and the equation of motion of $C$ differs for bosons and fermions. Higher-order equations of motion, e.g. for $\langle \cdag_i c_j \cdag_l c_n \rangle$, are beyond our current study.

\section{Late-time expansions}\label{appendix:late_time_expansions}
We now analytically obtain late-time transport coefficients $c_n, D_n$ [Eq.~\eqref{eq:eigenvalue_expansion} of the main text], for both the dephased flux chain and ladder. The two-site chain of Sec.~\ref{subsec:two_band_chain} is solved numerically. 
\subsection{The general bound state problem}
Consider a general hopping Hamiltonian with entries $\mathcal{H}_{i,i+s}:=h_s$; Hermiticity enforces $h^*_s = h_{-s}$. For example, our dephased flux chain has $h_1 = t, h_2 = t'\emi{\phi}$. Adding on-site dephasing to that gives a topologically trivial $\Theta$. From Eq.~\eqref{eq:CM_EOM}, one can Fourier transform via $C_{k,a} := \sum_i C_{i,i+a}\ei{ki}$ to obtain
\begin{equation}
   \im\dot{C}_{k,a}=\sum_s h_s\left[1-e^{\im ks}\right]C_{k,a+s}-\im\gamma(1-\delta_{a,0})C_{k,a}.
\end{equation}
Defining $\tilde{C}_{k,a}=\ei{ka/2}C_{k,a}$ and dropping the tilde,
\begin{equation}
    \im\dot{C}_{k,a} = \sum_s T_s(k) C_{k,a+s}-\im\gamma(1-\delta_{a,0})C_{k,a},
\end{equation}
where $T_s(k) = -2\im\,h_s\sin\frac{sk}{2}$. This is Eq.~\eqref{eq:impurity_problem} of the main text, which corresponds to a fictitious particle hopping between sites $a$ and $a+s$ with hopping amplitude $T_s$, as well as an impurity at site $0$. We thus expect bound states around site $0$, which is the origin of the point-gapped bands in the main text. 
We now Fourier transform again by defining
\begin{equation}
    C_{k,q} := \sum_{a \in \mathbb{Z}}C_{k,a}\emi{ q a},
\end{equation}
where $q \in \frac{2\pi}{N}\mathbb{Z}$.
Applied on the above, this gives
\begin{equation}\label{eq:kspace_EOM}
    \im \dot{C}_{k,q} + \im\gamma C_{k,q}= \epsilon_{k,q} C_{k,q} + \im\gamma C_{k,a=0}.
\end{equation}
Letting $\Lambda(k):=\lambda(k)+\im\gamma$, consider now the corresponding eigen-equation
\begin{equation}
\Lambda(k)C_{k,q} = \epsilon_{k,q}C_{k,q}+\im\gamma C_{k,a=0},
\end{equation}
where we defined $\epsilon_{k,q} = \sum_s \ei{qs}T_s(k)$. Summing over $q$ and dividing by $C_{k,a=0}$ gives the self-consistency equation
\begin{equation}
    1 = \frac{\im\gamma}{N} \sum_q \frac{1}{\Lambda(k)-\epsilon_{k,q}}.
\end{equation}
We assumed that $C_{k,a=0} \neq 0$, which corresponds to seeking bound-state solutions, since continuum modes have $C_{k,a=0} = 0$, corresponding to $\Lambda(k) = \epsilon_{k,q}$.

The steady state is at $k=0$, corresponding to $\epsilon_{0,q}=0$ for all $q$. Expanding around this gives
\begin{equation}\label{eq:general_coefficients_expansion}
    \lambda(k) = \im\gamma\sum_{n = 1}^{\infty}\frac{\langle\epsilon^n_{k}\rangle}{\Lambda^n(k)},
\end{equation}
where we defined $\langle \epsilon^n_{k} \rangle = \frac{1}{N}\sum_q \epsilon^n_{k,q}.$ This equation is implicit in $\lambda$, since $\Lambda$ contains it as well. We can directly compare with the late-time expansion [Eq.~\eqref{eq:eigenvalue_expansion}] to obtain transport coefficients. For example, we always have $\langle \epsilon_{k}\rangle = 0$, i.e. no contribution at order $n=1$. Generally, if $E(k)$ are the eigenvalues of $\mathcal{H}$, we have $\epsilon_{k,q} = E\big(q-\tfrac k2\big)-E\big(q+\tfrac k2\big)$, giving a Taylor expansion
\begin{equation}
    \epsilon_{k,q} = -\sum_{m\ \text{odd}} \partial^m_q E(q)\,\frac{k^m}{2^{m-1}\,m!}.
\end{equation}
Let the group velocity be $v = \partial_q E(q)$. Since $\langle v \rangle = 0$, we have $c_1 = 0$. Generically, if $n_*$ denotes the order of the first nonzero odd coefficient, we have $c_{n_*} = (-1)^{\frac{n_*-3}{2}}\frac{\langle v^{n_*}\rangle}{\gamma^{n_*-1}}$, with $\langle\cdot\rangle$ the average over $q$. For example, $c_3 = \frac{\langle v^3\rangle}{\gamma^2}$ exactly. Using $E(k) = 2t\cos k + 2t'\cos(2k-\phi)$ for the dephased flux chain gives the coefficients in Eq.~\eqref{eq:analytical_dephased_flux_chain_cumulants} of the main text.

The above can be interpreted via semiclassical kinetic theory. The object $C_{k,q}(\tau)$ is analogous to a lattice Wigner function $W(x,q,\tau)$ of a particle, Fourier-transformed in the center-of-mass coordinate $x$. Then, $q$ is the momentum for the relative coordinate. Its $a=0$ component $C_{k,a=0} = \langle C_{k,q} \rangle_q$ is the density $n(k)$. To leading order in gradients, $\epsilon_{k,q} \approx -v(q)k$, so that the equation of motion becomes
\begin{equation}
\partial_\tau W+v(q)\,\partial_x W=-\gamma\big[W-n\big],
\qquad n=\int\frac{\mathrm{d}q}{2\pi}W,
\end{equation}
which is a Boltzmann equation with a (single relaxation
time) collision term~\cite{BGK1954}. The ``particle'' undergoes a random-walk (velocity-jump) process: it streams with group velocity $v(q)$, and at rate $\gamma$ it samples the momentum $q$ uniformly~\cite{Goldstein1951,Kac1974,Weiss2002}. Thus, the continuum of $\Chat(k)$ at $\mathrm{Im}\lambda = -\gamma$ is made of free streaming kinetic modes. Bound states are hydrodynamic modes, and Eq.~\eqref{eq:general_coefficients_expansion} is the so-called Chapman--Enskog expansion~\cite{ChapmanCowling1970}. For on-site dephasing, this structure is exact, and the jump term evolves the state toward a distribution that is uniform in $q$. For instance, $D_2=2\sum_{s>0} s^2|h_s|^2/\gamma$ is the Haken--Strobl--Reineker
diffusion constant~\cite{haken1973exactly,KenkreReineker1982}. Generically, $\dot\kappa_{n_*}=n_*!\,\langle v^{n_*}\rangle/\gamma^{n_*-1}$ at the first
non-vanishing odd order $n_*$ coincides with the cumulant growth of the classical
velocity-jump process. The higher Taylor terms of
$\epsilon_{k,q}=E(q-\tfrac k2)-E(q+\tfrac k2)$ are quantum corrections. They do not affect $c_3$ for any
on-site-dephased chain, nor $c_5$ for the model of
Appendix~\ref{appendix:higher_orders}. The ``quantum'' ingredient is therefore the
velocity distribution rather than the transport mechanism. A Peierls flux makes
$E(q)\neq E(-q)$, so $v(q)$ acquires odd moments, which dephasing converts into
a hydrodynamic asymmetry. What the point-gap description adds
beyond kinetic theory is threefold. First, a topological invariant fixes that
sign without relying on the above Boltzmann equation structure [Appendix~\ref{appendix:multi_band}],
and so applies to non-on-site jumps such as those of Ref.~\onlinecite{chaduteau2026},
where the collision kernel may be more complicated. Second, it brings a bulk--boundary
correspondence [Appendix~\ref{appendix:OBC}]. Third, as in the main text, the case of multiple sites per unit cell goes beyond this semiclassical kinetic framework.

\subsection{Dephased flux chain}
The correlation-matrix equations of motion for Eq.~\eqref{eq:dephased_flux_chain} are
\begin{equation}
\begin{aligned}
    \im\dot{C}_{i,i+a}+\im\gamma(1-\delta_{a,0})C_{i,i+a} &= t\left[C_{i,i+a+1}+C_{i,i+a-1} - C_{i+1,i+a} - C_{i-1,i+a}\right] \\ 
    &\quad + t'\left[ \ei{\phi}(C_{i,i+a-2}-C_{i+2,i+a}) + \emi{\phi} (C_{i,i+a+2}-C_{i-2,i+a})\right].
\end{aligned}
\end{equation}
We now solve the eigen-equation $\im\dot{C}_{i,i+a} = \lambda C_{i,i+a}$. When Fourier transforming we obtain
\begin{equation}
\begin{aligned}
    \left[\lambda(k) + \im\gamma(1-\delta_{a,0})\right]C_{k,a} &= tu_k \left[\emi{k/2}C_{k,a-1}-\ei{k/2}C_{k,a+1}\right] \\
    &\quad + t' u_{2k}\left[\ei{(\phi-k)}C_{k,a-2}-\emi{(\phi-k)}C_{k,a+2}\right],
\end{aligned}
\end{equation}
where we defined $u_k = 2\im \sin k/2$. 
As in the general method above, we can absorb the complex exponentials $\ei{k/2}, \ei{k}$ by redefining $C_{k,a} \rightarrow \emi{ka/2}C_{k,a}$. By that general method one gets
\begin{equation}
        c_1 = 0, \quad D_2 = \frac{2t^2+8t'^2}{\gamma}, \quad c_3 = -\frac{12t^2t'\sin\phi}{\gamma^2}.
\end{equation}
These exact formulae are used to predict transport, matching numerics in Fig.~\ref{fig:dephased_flux_chain} of the main text. In addition, as discussed in the main text, the asymmetry is amplified in the tails of the density profile. These are governed by the same band at imaginary momentum. With $\psi(\kappa):=-\im\lambda(-\im\kappa)$, which is real by Eq.~\eqref{eq:eigenvalue_expansion}, $\sum_m n_m(\tau)e^{\kappa m}\sim e^{\psi(\kappa)\tau}$, so that the density at velocity $u=x/\tau$ scales as $n_x(\tau)\sim e^{-\tau I(u)}$ with $I(u)=\sup_\kappa[\kappa u-\psi(\kappa)]$~\cite{vankampen2007stochastic}. Hence
\begin{equation}\label{eq:tail_asymmetry}
\ln\frac{n_{x}(\tau)}{n_{-x}(\tau)}\simeq-\tau\left[I(u)-I(-u)\right],
\end{equation}
whose magnitude grows linearly in time at fixed $u$. For small $u$, $I(u)-I(-u)\simeq-c_3u^3/(4D_2^3)$, so its sign is set by $\nu$. We compute $\psi(\kappa)$ as the eigenvalue of $-\im\Chat(-\im\kappa)$ with largest real part, beyond the Taylor expansion. The resulting slopes match the time evolution without fitting parameters, shown in Fig.~\ref{fig:fig_tails}.

\begin{figure}
    \centering
    \includegraphics[width=0.5\linewidth]{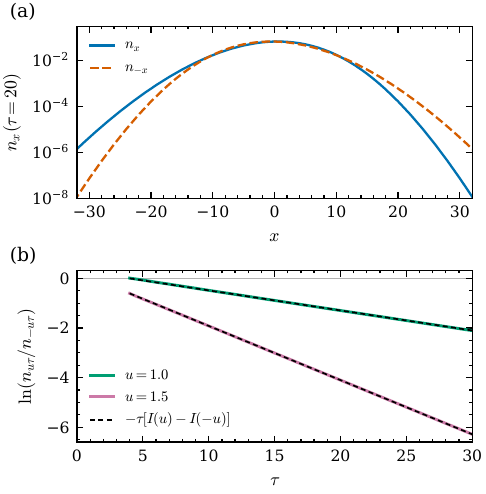}
    \caption{Time evolution of the tail asymmetry of the density profile in the dephased flux chain, with parameters as in Fig.~\ref{fig:dephased_flux_chain}. The initial state is localized on a single site. (a) Density profile at $\tau=20$ on a log scale (solid) and its mirror image (dashed). (b) At fixed velocity $u=x/\tau$, $\ln(n_{u\tau}/n_{-u\tau})$ decreases linearly in time, with slope $-[I(u)-I(-u)]$ (dashed) given by Eq.~\eqref{eq:tail_asymmetry}.}
    \label{fig:fig_tails}
\end{figure}
\subsection{Dephased flux ladder}
We now consider the odd-parity sector of the dephased flux ladder. We let $P_{i,j} = C_{a_i,b_j}$ and $Q_{i,j}=C_{b_i,a_j}$. From Eq.~\eqref{eq:CM_EOM}, one can derive 
\begin{equation}
    \im \dot{P}_{i,j} = t\ei{\phi/2}\left[P_{i,j+1}-P_{i+1,j}\right] + t\emi{\phi/2}\left[P_{i,j-1}-P_{i-1,j}\right] - (\Delta + \im \gamma) P_{i,j} + \im\gamma Q_{i,i}\delta_{ij}.
\end{equation}
The equation for $Q_{i,j}$ is the same up to flipping $\Delta$ and $\phi$. The same procedure as for the dephased flux chain gives
\begin{equation}
\begin{aligned}
    \im\dot P_{k,a}=-2\im t\sin\!\Big(\frac{k-\phi}{2}\Big)(P_{k,a+1}-P_{k,a-1})-(\Delta+\im\gamma)P_{k,a}+\im\gamma\delta_{a,0}Q_{k,0},\\
    \im\dot Q_{k,a}=-2\im t\sin\!\Big(\frac{k+\phi}{2}\Big)(Q_{k,a+1}-Q_{k,a-1})+(\Delta-\im\gamma)Q_{k,a}+\im\gamma\delta_{a,0}P_{k,0}.  
\end{aligned}
\end{equation}
The only term that couples the two chains is the interaction term, again giving an impurity problem at $a=0$. 

With $\Lambda=\lambda+\im\gamma$ as above, let $\tau_\pm(k)=2t\sin\frac{k\pm\phi}{2}$. Fourier
transforming in $a$ as for the chain, and using
$\langle(z-2\tau\sin q)^{-1}\rangle_q=(z^2-4\tau^2)^{-1/2}$, we eliminate
$P_{k,0}$ and $Q_{k,0}$ to obtain the bound-state condition
\begin{equation}\label{eq:ladder_condition}
\sqrt{(\Lambda+\Delta)^2-4\tau_-^2}\;\sqrt{(\Lambda-\Delta)^2-4\tau_+^2}=-\gamma^2 ,
\end{equation}
with branches such that $\sqrt{z^2-4\tau^2}\to z$ for $|z|\to\infty$. At $k=0$,
$\tau_\pm^2=\tau_0^2$ with $\tau_0=2t\sin\frac\phi2$, and the root is purely
imaginary, $\lambda(0)=-\im\Gamma_0$, with decay rate
\begin{equation}\label{eq:ladder_gamma}
\Gamma_0=\gamma-\sqrt{\sqrt{\gamma^4+16\Delta^2\tau_0^2}-\Delta^2-4\tau_0^2}.
\end{equation}
Differentiating Eq.~\eqref{eq:ladder_condition} implicitly at $k=0$ gives the drift
coefficient
\begin{equation}
c_1=-\frac{8t^2\Delta\sin\phi}{\sqrt{\gamma^4+64t^2\Delta^2\sin^2(\phi/2)}},
\end{equation}
matching numerics in Fig.~\ref{fig:dephased_flux_ladder} of the main text.

\section{Open boundary conditions}\label{appendix:OBC}
As is guaranteed by the nonzero winding number in the spectrum of $\Chat$, the non-Hermitian skin effect must occur, meaning exponential localization of these eigenmodes at an edge~\cite{PhysRevLett.124.086801, PhysRevLett.125.126402, chaduteau2026}. We verify this numerically for the dephased flux chain and ladder, with results shown in Fig.~\ref{fig:OBC_results}. PBC modes are completely delocalized. As in Ref.~\onlinecite{chaduteau2026}, the eigenmodes localize in varying degree as a function of their eigenvalue's proximity to the origin. Since the steady state has a uniform density profile in both OBC and PBC, the skin effect can only affect the
relaxation~\cite{PhysRevLett.123.170401}.

\begin{figure}
    \centering
    \includegraphics[width=0.5\linewidth]{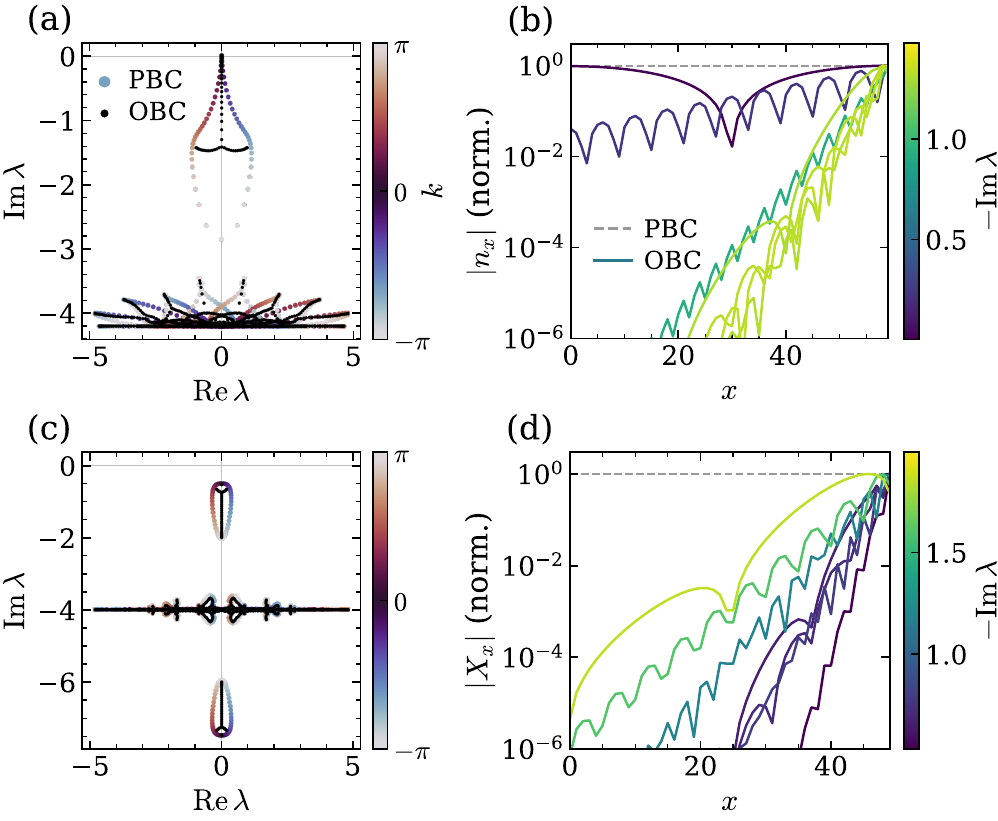}
    \caption{Spectra and eigenmodes in OBC. (a) Spectrum of $\Chat$ in OBC and PBC for the dephased flux chain [Eq.~\eqref{eq:dephased_flux_chain}]. The point gap closes in OBC, signaling the skin effect. (b) Corresponding eigenmode localization of the bound states, which are completely delocalized in PBC (gray dashed line) and exponentially localized in OBC in varying degree as a function of eigenvalue. (c), (d) Same plots for the odd-parity sector of the dephased flux ladder [Eq.~\eqref{eq:dephased_flux_ladder}], with similar behavior.}
    \label{fig:OBC_results}
\end{figure}

\section{Multi-band case}\label{appendix:multi_band}
We recall that $\Chat(k)$ is a matrix in the relative indices $a,b$, spanning the whole lattice~\cite{chaduteau2026}. Explicitly, we have the block-diagonal form
\begin{equation}
    \Chat = \sum_{k,a,b}\Chat_{ab}(k)\ket{k,a}\bra{k,b},
\end{equation}
where $\ket{k,a} := \frac{1}{\sqrt{N}}\sum_{i}\ket{i,i+a}\emi{ki}$. The matrix $\Chat_{ab}(k)$ is infinite-dimensional, with most of the eigenvalues forming the continuum. We write $\Chat(k) = \Chat_{\rm post}(k) + \im \Chat_{\rm Q}(k)$ where $\Chat_{\rm Q}(k)$ is the jump term. It is a finite-rank perturbation to the infinite-dimensional matrix $\Chat_{\rm post}(k)$. Multiple bound-state bands can arise from a sublattice structure, or from jumps $K_m$ which are not on-site. These bands are allowed to merge with the continuum, as in the dephased flux chain for $\gamma < 4t$. This merging happens below the base points we will use to diagnose late-time winding numbers, which does not affect our arguments.

Consider multiple bound-state/late-time bands in $\Chat(k)$, labeled by eigenvalues $\lambda_\alpha(k)$. We let the corresponding right and left eigenvectors be $\ket{R_\alpha(k)}$ and $\bra{L_\alpha(k)}$, satisfying biorthonormality $\bra{L_\alpha(k)}R_\beta(k)\rangle = \delta_{\alpha\beta}$.
Consider an initial profile that is made of densities, $C_{ij}(\tau=0) = \delta_{ij}n_i(\tau=0)$, with $\sum_i n_i(0) \neq 0$. It has momentum components $n(k,0) = \bra{k,0}C(0)\rangle = \frac{1}{\sqrt{N}}\sum_i \ei{ki}n_i(0)$, so that $\ket{C(\tau=0)}=\sum_k n(k,0)\ket{k,0}$. The time-evolved state can be decomposed as
\begin{equation}
\begin{aligned}
    \ket{C(\tau)} &= \sum_k n(k,0)\emi{\Chat(k)\tau}\ket{k,0} \\
    &=\sum_k n(k,0) \sum_\alpha \bra{L_\alpha(k)}k,0\rangle \emi{\lambda_\alpha(k)\tau} \ket{R_\alpha(k)}.
\end{aligned}
\end{equation}
Overlapping with $\bra{k,0}$ gives $n(k,\tau) = n(k,0)S(k,\tau)$, where
\begin{equation}
    S(k,\tau) = \bra{k,0}\emi{\Chat(k)\tau}\ket{k,0} = \sum_\alpha d_\alpha(k) \emi{\lambda_\alpha(k)\tau}.
\end{equation}
We defined $d_\alpha(k) = \bra{k,0}R_\alpha(k)\rangle \bra{L_\alpha(k)}k,0\rangle$. Then the real-space profile can be decomposed as
\begin{equation}
    n_m(\tau) = \sqrt{N}\int_{-\pi}^{\pi}\frac{\mathrm{d}k}{2\pi}\emi{km}n(k,0) S(k,\tau).
\end{equation}
The initial profile is thus convolved with a propagator whose Fourier transform is $S(k,\tau)$. By particle number conservation, we have $S(0,\tau)=1$.
The definition of cumulants in Eq.~\eqref{eq:cumulants_definition} of the main text gives
\begin{equation}\label{eq:cumulants_multi_band}
    \ln S(k,\tau) = \sum_n \kappa_n(\tau) \frac{(\im k)^n}{n!},
\end{equation}
where $\kappa_n$ are the cumulants of the profile minus those of the initial profile. For a unique bound-state band [Eq.~\eqref{eq:cumulants}], this gives a straightforward relation between the Taylor expansion coefficients of $\lambda_1(k)$ and the cumulants. For more bands, the left-hand side is the logarithm of a sum, so the cumulants are no longer linearly related to the Taylor coefficients of $\lambda_\alpha(k)$.

Let us now assume that $\lambda(k=0) = 0$ is a non-degenerate eigenvalue, which we choose to be in band $1$. We assume $\mathrm{Im}\lambda_1(k) < 0$ for any $k \neq 0$. We also assume that the identity is the only eigenvector with $\mathrm{Im}\lambda = 0$, so there are no modes that only oscillate. We further suppose that there exists a $\delta < 0$ such that $\mathrm{Im}\lambda_\alpha(k) \leq \delta$ for all $\alpha > 1$. We know also that $\ket{R_1(0)}=\ket{0,0}, \bra{L_1(0)}=\bra{0,0}$, with $\ket{0,0}=\frac{1}{\sqrt{N}}\sum_{i}\ket{i,i}$ a rescaled version of the vectorized identity. Biorthogonality implies that for all $\alpha \neq 1$, $\bra{0,0}R_\alpha(0)\rangle = \bra{L_\alpha(0)}0,0\rangle = 0$. We can then expand, for $\alpha \neq 1$,
\begin{equation}
    d_\alpha(k) = A_\alpha k^2 + O(k^3),
\end{equation}
where $A_\alpha = \bra{0,0}\partial_k R_\alpha(0)\rangle\bra{\partial_k L_\alpha(0)}0,0\rangle$. Importantly, there is no $O(k)$ term. Now let us factor out the first band,
\begin{equation}
    S(k,\tau) = d_1(k)\emi{\lambda_1(k)\tau}\left[1+\sum_{\alpha \neq 1}\frac{d_\alpha(k)}{d_1(k)}\emi{(\lambda_\alpha(k)-\lambda_1(k))\tau}\right].
\end{equation}
Using $d_1(k) = 1 - \sum_{\alpha \neq 1} d_\alpha(k) = 1 - O (k^2)$, we Taylor expand Eq.~\eqref{eq:cumulants_multi_band} to get
\begin{equation}
    \ln S(k,\tau) = - \im\lambda_1(k)\tau + \sum_{\alpha \neq 1} d_\alpha(k)(\emi{(\lambda_\alpha(k)-\lambda_1(k))\tau} -1)+ O(k^4).
\end{equation}
We used $\ln d_1(k) = -\sum_{\alpha \neq 1}d_\alpha(k) + O(k^4)$. Comparing to the cumulants, we then have
\begin{equation}
    \kappa_n(\tau) =  \dot{\kappa}_n \tau + \delta\kappa_n(\tau),
\end{equation}
where $\dot{\kappa}_n$ is as obtained from the $\lambda_1(k)$ band only [Eq.~\eqref{eq:cumulants} of the main text]. The deviations are, for instance for $n=2$,
\begin{equation}
    \delta\kappa_2=2\sum_{\alpha \neq 1}A_\alpha\left[1-e^{-\im\lambda_\alpha(0)\tau}\right].
\end{equation}
To see that these deviations decay, we let
\begin{equation}
    S(k,\tau) = S_1(k,\tau) + \delta S(k,\tau),
\end{equation}
where $S_1(k,\tau) = d_1(k)\emi{\lambda_1(k)\tau}$ and $\delta S(k,\tau) = \bra{k,0}\emi{\Chat(k)\tau}Q(k)\ket{k,0}$. Here $Q(k) = \id - \ket{R_1(k)}\bra{L_1(k)}$ is the projector onto all other bands, including the continuum. The spectrum of $\Chat$ restricted to the subspace defined by $Q(k)$ satisfies $\mathrm{Im}\lambda(k) \leq \delta$, so each term in the Taylor expansion of $\delta S(k,\tau)$ decays at least by $e^{\delta \tau}$. Thus, in the long-time limit, the deviations $\delta \kappa_n$ tend to constants, which are the cumulants $\kappa^{(1)}_n$ obtained from $d_1(k)$ via $\ln d_1(k) = \sum_n \kappa^{(1)}_n \frac{(\im k)^n}{n!}$. These deviations store the memory of the initial transient. In sum, the growth rates of late-time cumulants are only set by the $\alpha = 1$ band.

Consider initial states now with coherences, which we assume decay with relative index $a$. Then $\ket{C(0)}$ has components in $\ket{k,a}$ where $a \neq 0$; the factor $\bra{L_\alpha(k)}k,0\rangle$ must be replaced by $\bra{L_\alpha(k)} C_k(0)\rangle$. Other bands can now have $O(k)$ coefficients in $d_\alpha(k)$, so that transients appear in odd cumulants as well. However, the decay argument is unchanged, so that rates $\dot{\kappa}_n$ are the same for all initial states with nonzero particle number.
The winding number is defined via the log of a determinant, so it is the sum of winding numbers of each band, $\nu(E) = \sum_\alpha \nu_\alpha(E)$. For the continuum, we use the same truncation procedure as Ref.~\onlinecite{chaduteau2026}. Taking $E=-\im \epsilon$ on the negative imaginary axis, with $0<\epsilon < |\delta|$, the other bands $\alpha \neq 1$ cannot give a nonzero winding number around $E$. So $\nu(E) = \nu_1(E)$.
By the SWAP symmetry, $\mathrm{Re}\lambda_1$ and $\mathrm{Im}\lambda_1$ are odd and even in $k$, respectively. Let $n_*$ be the lowest odd power such that $c_{n_*} \neq 0$ in the $\lambda(k)$ expansion, i.e.
\begin{equation}
    \lambda_1(k) = -\im D_2 k^2 + c_{n_*}k^{n_*} + O(k^{n_* + 1}).
\end{equation}
For example, $n_*=3$ for the dephased flux chain. For our $U(1)$ symmetric case, since the steady state carries no current, the density cannot drift. Our argument nevertheless covers the $n_*=1$ case, beyond $U(1)$-symmetric models. For small $\epsilon$, there are two points $\pm k_\epsilon$ at which $\mathrm{Im}\lambda_1(k)=-\epsilon$, and $\mathrm{Re}\lambda_1(\pm k_\epsilon) = \pm c_{n_*} k_\epsilon^{n_*} + \ldots$, which have opposite signs. As we increase $k$, the spectrum $\lambda_1(k)$ passes over the reference point $E$ only once; the winding number then depends on $c_{n_*}$ via 
\begin{equation}
    \lim_{\epsilon \to 0^+}\nu(-\im\epsilon)=-\operatorname{sgn}c_{n_*}.
\end{equation}
The reference point $E$ must be near enough the origin to be inside the region $|\mathrm{Re}\,E| < |c_{n_*}|(|\mathrm{Im}\,E|/D_2)^{n_*/2}$.
If no odd coefficient is nonzero, then analyticity and the SWAP symmetry give $\mathrm{Re}\lambda_1(k) = 0$ near $k=0$. The arc lies on the imaginary axis, so $\nu(E)$ is undefined, and $\nu(E) = 0$ for other reference points with nonzero real part. In the late-time limit, the rates of the odd cumulants satisfy $\dot\kappa_n=-(-1)^{(n-1)/2}n!\,c_n$, with contribution only from band $1$. Odd cumulants with $n < n_*$ tend to constants, while $\kappa_{n_*}$ grows linearly with
\begin{equation}
    \sgn\dot\kappa_{n_*}=(-1)^{(n_*-1)/2}\nu .
\end{equation}
Hence, all bands considered, the winding number sufficiently near the origin is nonzero if and only if there is an odd cumulant that grows. Its sign fixes the direction of asymmetric transport. This holds for initial states with nonzero particle number.
The above arguments assume that there is a unique $\lambda=0$ eigenvalue of $\Chat$. This is not true for the many-body Lindbladian, which has one steady state per $U(1)$ symmetry sector. However, these all map to a correlation matrix proportional to the identity, $C \propto \id$.

Since $\Chat$ is non-Hermitian, one should also consider the possibility that multiple bands touch at the origin while sharing the same eigenvector there. This would mean that the steady state is an exceptional point of $\Chat(0)$. We argue now that this cannot happen. Expanding the jump term, $\Chat=\hat{\mathcal{H}}-\im\hat{\mathcal{K}}$, with
$\hat{\mathcal{H}}=\mathcal H\otimes\id-\id\otimes\mathcal H^*$ and
$\hat{\mathcal{K}}=\frac12\sum_m(\mathcal K_m\otimes\id-\id\otimes\mathcal K_m^*)^2$. The same decomposition holds for each block
$\Chat(k)$. If $\Chat(k)\ket{C}=\lambda(k) \ket{C}$ with $\lambda(k)$ real, then
$0=\mathrm{Im}\bra{C}\Chat|C\rangle=-\bra{C}\hat{\mathcal{K}}|C\rangle$. As $\hat{\mathcal{K}}$ is positive semi-definite, this gives $||\hat{\mathcal{K}}^{1/2}\ket{C}||=0$, so $\hat{\mathcal{K}}\ket{C}=0$, and hence
$\Chat^\dagger \ket{C}=\hat{\mathcal{H}} \ket{C}=\lambda(k) \ket{C}$. The left and right eigenvectors therefore coincide, whereas at an exceptional point they are orthogonal. Exceptional points can only occur at $\mathrm{Im}\lambda<0$. This also justifies
$\bra{L_1(0)}=\bra{0,0}$ above.
If instead there are more null eigenvectors of $\Chat$ (i.e. more conserved charges), multiple bands pass through the origin and the above can fail. As an example, consider two decoupled flux chains with opposite fluxes, each conserving their own particle number. We have two bands through the origin with opposite
windings, so $\nu=0$, yet a profile initialized on one chain develops a skewness, violating
Eq.~\eqref{eq:winding_sign_rule}.

We numerically test robustness against static on-site disorder $\epsilon_i\in[-W/2,W/2]$ in the
dephased flux chain with $t=1$, $t'=0.5$, $\phi=\pi/2$ and $\gamma=4$. Since
$\Theta=\mathcal H-\im\gamma/2$ still has a real spectrum up to a constant shift, any
point gap of $\Chat$ remains noise-induced. Disorder breaks translation
invariance, so we replace the momentum $k$ by a twist $\theta$: on a ring of $N$
sites, both the ket and bra indices of $C_{ij}$ acquire a phase $e^{\im\theta/2}$
when hopping forward ($c^\dagger_{m+1}c_m$) across one bond,
\begin{equation}
\Chat(\theta)=\Theta(\theta/2)\otimes\id-\id\otimes\Theta^*(-\theta/2)
+\im\sum_m\mathcal K_m\otimes\mathcal K_m^T ,
\end{equation}
and we compute
$\nu(E)=\frac{1}{2\pi\im}\oint\mathrm d\theta\,\partial_\theta\log\det[\Chat(\theta)-E]$~\cite{PhysRevX.8.031079}.
Without disorder, $\Chat(\theta)$ decomposes into the blocks $\Chat(k)$ with
$k=(2\pi n+\theta)/N$, so that $\nu(E)$ reduces to
Eq.~\eqref{eq:winding_number_definition}. For $0\le W\le12$, we find $\nu=+1$ for base
points inside the bound-state loop and $\nu=0$ outside it. The disorder-averaged
skewness rate from time evolution keeps its sign while its magnitude decreases,
from $\dot\kappa_3=-2.25$ in the clean limit to $-2.07$ at $W=2$ and $-0.91$ at $W=8$. Disorder thus reduces the asymmetry without changing its sign.

\section{Higher-order asymmetry}\label{appendix:higher_orders}
We now show how to engineer late-time asymmetries in ever-higher-order cumulants, which open point gaps at appropriate parameters. This can be realized in the dephased flux chain family by having longer-range hoppings. Consider
\begin{equation}\label{eq:model14}
\begin{aligned}
H&=\sum_m\Big[t_1e^{\im\phi_1}c^\dagger_{m+1}c_m+t_4e^{\im\phi_4}c^\dagger_{m+4}c_m+\text{h.c.}\Big], \\
K_m&=\sqrt\gamma\,n_m,
\end{aligned}
\end{equation}
where $t_1, t_4 > 0$. This corresponds now to a flux through a pentagon formed by vertices $m, m+1, \ldots, m+4$. The same method as the above general bound-state problem, with $E(k)=2t_1\cos (k-\phi_1) + 2t_4 \cos(4k - \phi_4)$, now gives 
\begin{equation}
v(k)=-2t_1\sin (k-\phi_1)-8t_4\sin(4k-\phi_4).
\end{equation}
Computing averages gives
$\langle v^2\rangle=2t_1^2+32t_4^2$, $\langle v^3\rangle=0$ and
$\langle v^5\rangle=40\,t_1^4t_4\sin(\phi_4 - 4\phi_1)$. Thus
\begin{equation}\label{eq:alpha5v}
D_2=\frac{\langle v^2\rangle}{\gamma},\qquad
c_5=-\frac{\langle v^5\rangle}{\gamma^4}.
\end{equation}
The asymmetry only comes from the fifth cumulant here, as $c_3=0$. The winding number is $\nu = -\sgn c_5$. The bound-state spectrum has an even flatter cusp. For small $\gamma$, the bound band merges with the continuum. There are straightforward generalizations to longer-range hoppings giving an asymmetry manifesting in arbitrarily high-order cumulants, which are fine tuned and serve as a proof of concept.

\end{widetext}

\end{document}